\documentclass[twocolumn,trackchanges,twocolappendix]{aastex63}

\usepackage{xcolor}

\DeclareRobustCommand{\ers}{\bgroup\markoverwith{\textcolor{red}{\rule[.5ex]{2pt}{0.4pt}}}\ULon}

\received{}
\revised{}
\accepted{}
\submitjournal{ApJ}

\shorttitle{Occurrence and Lifetime of Radio Halos}
\shortauthors{Nishiwaki \& Asano}

\begin{document}

\title{Statistical Properties of Radio Halos in Galaxy Clusters and Origin of Seed Electrons for Reacceleration}

\author{Kosuke Nishiwaki}
\affiliation{Institute for Cosmic Ray Research, The University of Tokyo, 5-1-5 Kashiwanoha, Kashiwa, Chiba 277-8582, Japan}

\author{Katsuaki Asano}
\affiliation{Institute for Cosmic Ray Research, The University of Tokyo, 5-1-5 Kashiwanoha, Kashiwa, Chiba 277-8582, Japan}




\begin{abstract}
One of the most promising mechanisms for producing radio halos (RHs) in galaxy clusters is the reacceleration of cosmic-ray electrons by turbulences. However, the origin of the seed electrons for the reacceleration is still poorly constrained. In the secondary scenario, most of the seed electrons are injected via collision of proton cosmic-rays, while non-thermal electrons are directly injected in the primary scenario. In this paper, we examine the two scenarios for the seed electrons with the observed statistical properties of RHs, combining two methods: following the temporal evolutions of the electron energy and radial distributions in a cluster, and the merger history of clusters. We find that the RH lifetime largely depends on the seed origin, as it could be longer than the cosmological timescale in the secondary scenario. We study the condition for the onset of RHs with the observed RH fraction and the RH lifetime we obtained, and find that long-lived RHs in the secondary scenario should be originated from major mergers with a mass ratio of $\xi\sim0.1$, while the short lifetime in the primary scenario requires more frequent onsets by minor mergers with $\xi\sim0.01$.
Our simple model of the turbulence acceleration can reproduce the observed radio luminosity-mass relation. The RH luminosity functions we obtained suggest that the expected RH number count with the ASKAP survey will detect $\approx10^3$ RHs in both the scenarios.

\end{abstract}
\section{Introduction\label{sec:intro}} 

An increasing number of galaxy clusters are found with diffuse cluster-scale synchrotron emission in the radio band. Giant Radio halos (RHs) are the Mpc-sized emission often found in the central region of massive clusters \citep[see][for a review]{review_vanWeeren}. The progress in the RH observation has been achieved with the NRAO VLA Sky Survey \citep[NVSS,][]{Giovannini_1999}, the Giant Meterwave Radio Telescope \citep[GMRT,][]{Venturi_2007A&A...463..937V} and its extension \citep[][]{2013A&A...557A..99K}. A novel progress in this field is now ongoing with current generation of high-sensitivity radio telescopes, such as LOw Frequency ARay \citep[LOFAR,][]{vanHaarlem_2013A&A...556A...2V} and the the pathfinders for Square Kilometer Array \citep[SKA,][]{Dewdney_2009IEEEP..97.1482D}.\par

A variety of statistical properties of RHs have been reported. For example, the occurrence of RHs correlates with the dynamical disturbance of clusters observed in X-ray \citep[e.g.,][]{Schuecker_2001A&A...378..408S,Cassano_2010_LOFAR,Cassano_2013ApJ...777..141C,2021A&A...647A..50C}, and the radio luminosity is known to correlate with the X-ray luminosity or the mass measured with the Sunyaev-Zel'dovich (SZ) effect \citep[e.g.,][]{Cassano_2013ApJ...777..141C}. In addition, those correlations show bimodalities, as the clusters without RHs fall well below the relations \citep[][]{Brunetti_2007ApJ...670L...5B}. The occurrence of RHs seems to increase with the mass of the cluster \citep[][]{Cuciti_2015A&A...580A..97C,2021A&A...647A..50C}. Those facts indicate a firm connection between the formation process of galaxy clusters and the non-thermal components in the intra-cluster medium (ICM).  \par

During the structure formation of galaxy clusters, a part of the gravitational potential energy would be dissipated into the acceleration of relativistic particles by the merger. In the so-called turbulent reacceleration model, RHs originate from the reacceleration of seed cosmic-ray electrons (CREs) by the merger-induced turbulence \citep[e.g.,][]{Brunetti_2001MNRAS.320..365B,Petrosian_2001ApJ...557..560P,Fujita_2003ApJ...584..190F}. As the turbulence could permeates the most of the cluster volume, this model can explain the large extension of the radio emission. This model is supported by the observational signature of the turbulence acceleration such as the steep spectral index ($F_\nu\propto\nu^{-\alpha_{\rm syn}}$ with $\alpha_{\rm syn}\approx1.5$) or the break feature in the radio spectrum \citep[e.g.,][]{Brunetti_2008Natur.455..944B,Macario_2010A&A...517A..43M,Wilber_2018MNRAS.473.3536W,DiGennaro_2021NatAs...5..268D}. However, the origin of the seed electrons for the reacceleration has not been revealed yet \citep[e.g.,][]{Brunetti2014}.  \par

There are two possible origins of the seed CREs. One is the {\it primary} scenario, where CREs are directly injected by accelerators like large-scale accretion shocks, or escaped from radio jets launched from the central active galactic nuclei (AGNs) \citep[e.g.,][]{Kang_2012ApJ...756...97K,Volk_1999APh....11...73V,Gitti_2002A&A...386..456G,Heinz_2006MNRAS.373L..65H}. The other is the {\it secondary} scenario, where the primary CRE injection is dominated by the secondary injection from the pp collisions between cosmic-ray protons (CRPs) and thermal protons in the ICM \citep[e.g.,][]{Brunetti2005,Brunetti2011,Pinzke2017}. Recently, by modeling the well-studied RH in the Coma cluster, we have found that the non-thermal properties, such as the CR energy density and the location of the primary CR accelerators, could be significantly different between those two scenarios, which could be a clue to reveal the seed origin \citep[][]{paperI}. In this study, we further discuss possible differences appearing in the statistics of RHs. \par


The luminosity function of RHs has been calculated with the scaling relation between the radio power and the mass or X-ray luminosity \citep[e.g.,][]{Ensslin_2002}. Several theoretical attempts have been made to investigate the occurrence of RHs. In the context of the reacceleration scenario, \citet{Cassano_2005} constructed a self-consistent statistical model for the turbulent injection and the RH generation during the cluster evolution. Their model is constrained by the observed occurrence of 30\% around 1.4 GHz, and predicts a much smaller  RH fraction ($\le 5\%$) in less massive clusters. The increased occurrence at lower frequencies ($\approx100~{\rm MHz}$) is predicted in \citet{Cassano_2010_LOFAR}. The bimodality in the RH luminosity distribution may be explained with the rapid CR streaming or diffusion rather than the reacceleration. \citet{Cassano_2016} pointed out that the lifetime of RHs and the critical mass ratio for the reacceleration onset can be constrained from the observed fraction of radio-loud clusters with the merger rate of the dark matter halo obtained in N-body simulations.
\citet{Zandanel2014b} and \citet{Cassano_2012} proposed the combined scenario, where the emission in the core region is produced by the secondary CREs, while the outer parts are dominated by primary CREs.\par 




The main objective of this paper is to investigate how the statistical properties found in the RH observations provide constrains on those different scenarios for the seed population. In this paper, we combine following two methods; the time-dependent calculation of the CR spectral evolution solving the Fokker-Planck equation, and the Monte Carlo (MC) procedure to simulate the macroscopic cluster evolution through mergers and the mass accretion. In the former calculation, we consider the reacceleration of CRs and compare the two scenarios, secondary or primary, of the CRE injection (Section~\ref{sec:CR}). Those models are constrained by the observed spectrum and spatial profile of the Coma RH (Section~\ref{subsec:Coma}). With this method, we can estimate how different the lifetime of the radio emission is between the secondary and primary scenarios (Section~\ref{sec:lifetime}). \par

In the latter part, we construct the MC merger tree from the halo merger rate based on N-body simulations (Section~\ref{sec:MT}). Considering the lifetime obtained from the former part, we investigate consistent conditions for the onset of RHs with the observed statistics for both the two scenarios (Section~\ref{sec:Statis}). Our attempt requires that the peak radio luminosity after a merger is related to the cluster mass scale. The required correlation is partially justified in Section~\ref{sec:origin_of_PM}, where we again solve the evolution of CRs in RHs with various masses, assuming a scaling between the acceleration efficiency and the cluster mass. Finally, the detectability of low-mass RHs with upcoming radio observations is discussed in Section~\ref{sec:ASKAP}.




Throughout this paper, we assume the $\Lambda$CDM cosmology and adopt the parameters from \citet{Planck18}, where $H_0=67~{\rm km/s/Mpc}$ ($h=0.67$ or $h_{70}=0.96$), $\Omega_m h^2=0.143$ and $\Omega_{\Lambda}=0.68$. \par

\section{Time evolution of cosmic-ray distribution \label{sec:CR}}
First, we study the emission lifetime of RHs in the turbulent reacceleration model. For this purpose, we integrate the Fokker-Planck (FP) equation for the CR distribution function and follow the time evolution of the synchrotron spectrum. We mainly discuss the comparison of the two scenarios for the origin of seed CREs, i.e., the primary scenario and the secondary scenario. Those scenarios are distinguished by the injection rate of primary cosmic-ray protons (CRPs) relative to that of CREs. Our model here is basically the same as \citet{paperI}, which includes the reacceleration of CRs. We adopt the observed property of the Coma RH, for which we have well-examined data sets, to constrain the model parameters.  \par


Our model is one-dimensional in space, which enable us to follow the evolution of brightness profile in addition to the spectral evolution. In our previous study, the large extension of the RH was explained with the radial dependence in the CR injection with a constant reacceleration efficiency in space. In this paper, we investigate another possibility: the radial dependence in the reacceleration efficiency for a fixed injection profile. While both the models well reproduce the Coma RH, the spectral evolution or the lifetime is not largely affected by the injection or reacceleration efficiency profile. Our model includes relatively less parameters and it is easy to apply to other RHs with different masses (Section~\ref{sec:origin_of_PM}).  \par

\subsection{The Fokker--Planck equation \label{subsec:FP}}
We solve the isotropic one-dimensional Fokker-Planck (FP) equation to follow the time evolution of the distribution function in the phase space, $N_{s}(r,p,t)=(4\pi r^2)(4\pi p^2) f_{s}(r,p,t)$, where $f_{s}(r,p,t)$ is the number density of particle species $s$ of momentum $p$ at radial position $r$, and time $t$ in the phase space. 

For CRPs, the FP equation can be written as \citep[e.g.,][]{Brunetti2005,Fujita_2003ApJ...584..190F,Pinzke2017,paperI}

\begin{eqnarray}
\frac{\partial N_\mathrm{p}}{\partial t}&=&\frac{\partial}{\partial p}\left[N_\mathrm{p}\left(b^{(p)}_\mathrm{C}-\frac{1}{p^2}\frac{\partial}{\partial p}\left(p^2D_{pp}\right)\right)\right] \nonumber\\
& &+\frac{\partial^2}{\partial p^2}[D_{pp} N_\mathrm{p}]+Q_\mathrm{p}(r,p)-\frac{N_\mathrm{p}}{\tau_{pp}(r,p)},
\label{eq:Np}
\end{eqnarray}
where $N^\mathrm{tot}_p(r,p,t)$ is the distribution function of CRPs, $b^{(p)}_\mathrm{C}$ is the momentum loss rate ($b\equiv -dp/dt$) due to the Coulomb collisions, $D_{pp}$ is the momentum diffusion coefficient due to interactions with turbulence, and $Q_\mathrm{p}(r,p)$ denotes the injection of primary CRPs. The factor $\tau_{pp}$ denotes the $pp$ collision timescale. For simplicity, we ignore the effect of repeated collisions of a CRP, so we do not follow an energy loss per collision. The cooling due to the $pp$ collision is expressed like escape as $-N_\mathrm{p}/\tau_{pp}$, where the loss timescale includes the inelasticity coefficient of $\kappa_{pp}=0.5$. The details of $\tau_{pp}$ and $b^{(p)}_\mathrm{C}$ are shown in \citet{paperI}.\par
Throughout this paper, we focus on the synchrotron emission only below $\sim$GHz and the gamma-ray emission below $\sim100$GeV. The spatial diffusion of low-energy particles responsible for such emission is negligible, so we omit the spatial diffusion terms, $D_{rr}$. \par


For CREs, the FP equation is expressed as
\begin{eqnarray}
\frac{\partial N_\mathrm{e}}{\partial t}&=&\frac{\partial}{\partial p}\left[N_\mathrm{e}\left(b_{\mathrm{rad}}+b^{(e)}_\mathrm{C}-\frac{1}{p^2}\frac{\partial}{\partial p}\left(p^2D_{pp}\right)\right)\right] \nonumber\\
& &+\frac{\partial^2}{\partial p^2}[D_{pp} N_\mathrm{e}]+Q_\mathrm{e}(r,p;N_\mathrm{p}),
\label{eq:Ne}
\end{eqnarray}
where $b^{(e)}_\mathrm{C}$ is the energy loss rate due to CRE-e collisions, $b_\mathrm{rad}$ denotes the radiative cooling, and $Q_\mathrm{e}(r,p;N_\mathrm{p})$ is the source term, which includes both the secondary and primary injections: $Q_\mathrm{e}(r,p;N_\mathrm{p})=Q^{\mathrm{sec}}_\mathrm{e}(r,p,;N_{\rm p})+Q_\mathrm{e}^{\mathrm{pri}}(r,p)$. We use the same formulae for the momentum loss term $b$, $b^{(e)}_\mathrm{C}$ and $b$ as those given in \citet{paperI}. We do not solve the evolution of high-energy CREs of $p > 10^7 m_{\rm e} c$ to save the computation time. \par

The radial profile of the thermal electron density is expressed with the beta-model profile as
\begin{eqnarray}
n_{\mathrm{th}}(r)=n_{\mathrm{th}}(0)\left\{1+\left(\frac{r}{r_\mathrm{c}}\right)^2\right\}^{-\frac{3}{2}\beta}.
\label{eq:beta-model}
\end{eqnarray}
The parameters for the Coma cluster are \citep{1992A&A...259L..31B}, $n_{\mathrm{th}}(0)=3.42\times10^{-3}\:\mathrm{cm}^{-3}$, $\beta=0.75$, and $r_{\rm c} = 290$ kpc. We also use the same temperature profile as that in \citet{paperI}. For simplicity, the ICM is a pure hydrogen plasma, and we use Eq.~(\ref{eq:beta-model}) for the density of thermal protons in our model.  \par

Relativistic electrons can be produced through the decay of pions produced by the inelastic collisions between CRPs and thermal protons ($p+p\rightarrow \pi^{0,\pm} + X$). We adopt the same method as that in \citet{paperI} to calculate the secondary injection from $pp$ collisions. \par

\subsection{Injection of primary cosmic-rays \label{subsec:primary}}


Primary CRPs can be accelerated through the shocks caused by the structure formation process or injected in the ICM from the internal sources like AGNs. We assume a single power-law spectrum with an exponential cutoff for the injection spectrum of primary CRs; 
\begin{eqnarray} 
Q_\mathrm{p} (r,p)=C^\mathrm{inj}_\mathrm{p}p^{-\alpha_{\rm inj}}\exp\left[-\frac{E_\mathrm{p}}{E_\mathrm{p}^{\mathrm{max}}}\right]{\cal Q}(r),
\label{eq:primary_injection}
\end{eqnarray}
where $C^\mathrm{inj}_\mathrm{p}$ is the normalization factor, and ${\cal Q}(r)$ represents the radial dependence of the injection\footnote{${\cal Q}(r)$ is equivalent to $K(r)$ in \citet{paperI}}. Those two quantities can be constrained by the observed flux and the brightness profile of the RH, respectively. The maximum energy of primary CRPs is taken to be  $E^\mathrm{max}_\mathrm{p}=100$ PeV. The minimum momentum of CRPs is taken to be ten times larger than that of the thermal protons. Following \citet{Brunetti2017}, we adopt the injection index of $\alpha_{\rm inj}=2.45$.\par

We also take into account the injection of primary CREs. The number ratio of primary CREs to primary CRPs at the injection is expressed with a parameter $f_\mathrm{ep}$;
\begin{eqnarray}
Q_\mathrm{e}^{\mathrm{pri}} (r,p) = f_{\rm ep} Q_\mathrm{p}(r,p).
\label{eq:Qepri}
\end{eqnarray}
Eq.~(\ref{eq:primary_injection}), which is defined only above the minimum momentum of CRPs, $p=30$ MeV$/c$, is extrapolated to the minimum momentum of CREs, $p_\mathrm{e}=0.3m_\mathrm{e}c=150$ keV$/c$, where the strong Coulomb cooling inhibits the reacceleration.


As we have mentioned, we compare the two scenarios: the secondary scenario ($f_\mathrm{ep}=0$), where all of CREs are produced through the $pp$ collision, and the primary scenario ($f_\mathrm{ep}=0.01$), where the primary CRE injection dominates the injection from the hadronic interaction. The adopted ratio of $f_\mathrm{ep}=0.01$ is close to the observed CRE to CRP ratio in our galaxy \citep[e.g.,][]{SchlickiBuch}. We neglect the possible radial dependence of $f_\mathrm{ep}$ \citep[e.g.,][]{2008MNRAS.385.1211P} for simplicity. In Sect.~\ref{sec:lifetime}, we show that the expected lifetime of the radio emission is significantly different between these two scenarios.
\par

The extended profile of RHs indicates a fair amount of CREs outside the thermal core of the cluster. This challenges the secondary scenario, since the injection of secondary CREs is more effective at smaller radii where the timescale of the $pp$ interaction is shorter. \citet{paperI} tested the case where the injection profile $K(r)$ has a peak at $r\approx1$ Mpc to reproduce the surface brightness profile of the Coma RH. However, even though CRPs are injected outside the core, high-energy CRPs selectively diffuse into the core, where the radio emissivity is higher due to the stronger magnetic field.
As a result, the spectral hardening is caused by this radial diffusion of parental CRPs, and this makes it difficult to fit the observed spectrum.
This difficulty is expected to be relaxed if the reacceleration is more efficient at larger radii. \par

In this paper, we assume that CRs in the ICM are mainly injected from internal sources, such as AGN activities and other outflows from member galaxies. We adopt the injection profile of primary CRs proportional to the thermal density profile (Eq.~(\ref{eq:beta-model})):
\begin{eqnarray}
{\cal Q}(r) &=& \left\{1+\left(\frac{r}{r_\mathrm{c}}\right)^2\right\}^{-\frac{3}{2}\beta},
\label{eq:Q(r)}
\end{eqnarray}
where we use the same $r_{\rm c}$ and $\beta$ as Eq.~(\ref{eq:beta-model}). We fix this injection for both the primary and secondary scenarios. 
As will be explained in Sect.~\ref{subsec:acc_diff}, the observed brightness profile of the RH can be modeled by tuning the radial profile of $D_{pp}$. \par


\subsection{Turbulent reacceleration \label{subsec:acc_diff}}

Recent studies show that the observed $\sim$Mpc extension of the Coma RH with the cored profile of the magnetic field \citep[][]{Bonafede2010} requires the CR profile flatter than the ICM profile \citep[e.g.,][]{Brunetti2017,Pinzke2017}. In this study, we consider that the extended profile of radio-emitting CREs is reproduced by the radial dependence of the reacceleration coefficient, $D_{pp}$, rather than that of the CR injection. This assumption is supported by the 
turbulent profile often seen in the numerical simulations of the cluster formation \citep[e.g.,][]{Nelson2014,Vazza_2017MNRAS.464..210V,Angelinelli_2020MNRAS.495..864A}. In the following, we discuss how the profile of $D_{pp}$ is related to the profile of the turbulence, using the expression derived from the quasi-linear theory. As a result, $D_{pp}(r,p)$ will be expressed with two parameters; one is for the radial dependence, and the other is for the normalization. \par


We assume that the (re)acceleration of CRs occurs through the transit-time damping (TTD) with the compressible turbulence in the ICM \citep[e.g.,][]{Brunetti2007, 2019ApJ...877...71T}. In this case, the momentum diffusion coefficient becomes hard-sphere type ($D_{pp}\propto p^2$), which implies that the acceleration timescale is independent of the particle momentum $t_{\rm acc}=p^2/(4D_{pp})\propto p^0$. We focus on only the TTD acceleration in this paper, though other possibilities for the reacceleration  mechanism have been proposed, such as the reacceleration by the incompressible turbulence\citep[e.g.,][]{Brunetti_2016MNRAS.458.2584B} or Alfv\'{e}n waves\citep[e.g.,][]{Fujita_2003ApJ...584..190F}.

We assume that the fast mode turbulence in the ICM is well described  with the Iroshnikov-Kraichnan (IK) scaling \citep[e.g.,][]{Brunetti2011}. Isotropic cascades of fast mode turbulences with the IK spectrum are seen in numerical simulations \citep[e.g.,][]{2003MNRAS.345..325C}. 




Using the turbulent energy spectrum ${\cal W}(k)\propto k^{-3/2}$, $D_{pp}$ can be written as \citep[e.g.,][]{Miniati_2015ApJ...800...60M}
\begin{eqnarray}\label{eq:Dpp_integral}
D_{pp}(r,p) = \frac{p^2\pi I_\theta(x)}{8c}\int_{k_L}^{k_{\rm cut}}dk k{\cal W}(k),
\end{eqnarray}
where $k_{L}$ corresponds to the injection scale, while $k_{\rm cut}$ represents the cut-off scale (see Appendix \ref{app:Dpp}). Here, we have assumed that the turbulence is isotropic with respect to the background magnetic field. The average over the angle $\theta$ between the direction of the wave and the background magnetic field is represented by a dimensionless function, $I_\theta(x)\equiv\int^{\arccos(x)}_0d\theta \frac{\sin^3\theta}{|\cos\theta|}\left[1-\left(\frac{x}{\cos\theta}\right)^2\right]^2\approx5$, where $x=v_{\rm ph}/c$ and $v_{\rm ph}$ is the phase velocity of the wave. Because the TTD is the interaction with the fast mode waves and the ICM is a high-beta plasma ($\beta_{\rm pl}\gtrsim100$ for $\sim1\mu{\rm G}$ magnetic field), we adopt the sound velocity $c_{\rm s}$ for $v_{\rm ph}$. \par






The diffusion coefficient $D_{pp}$ could depend on the radial coordinate $r$ through the profile of the turbulent energy density. Instead of specifying the parameters such as $k_L$ and the normalization of ${\cal W}(k)$, we use another parameter, $t_{\rm acc}(r_{\rm c})$, that represents the acceleration timescale at $r=r_{\rm c}$, and discuss constraints on this parameter from the observed spectrum. Introducing cut-off factors for both the maximum and the minimum energies \citep{paperI}, the diffusion coefficient may be written as
\begin{eqnarray}
D_{pp} (r,p) &=& {\cal D}(r)\frac{p^2}{4t_\mathrm{acc}(r_{\rm c})} \nonumber \\
&&\times\exp\left[-\frac{E(p)}{E^{\mathrm{max}}_c(r)}\right]\exp\left(-\frac{m_sc}{p}\right),
\label{eq:Dpp}
\end{eqnarray}
where ${\cal D}(r)$ is a dimensionless factor representing the radial dependence, and $m_s$ is the mass of particle species $s$. We adopt the Hillas limit \citep{1984ARA&A..22..425H} for the maximum energy of CRs, $E^{\mathrm{max}}_\mathrm{c} (r) = qB(r) l^\mathrm{F}_\mathrm{c} \sim 9\times 10^{19} (B(r)/1~\mu G)(l^\mathrm{F}_\mathrm{c}/0.1 \mathrm{Mpc})$ eV, where $l^\mathrm{F}_\mathrm{c}$ is the maximum size of the turbulent eddy of compressible turbulence.
We assume $l^\mathrm{F}_\mathrm{c} = 0.1$ Mpc as a reference, which is compatible with the constraint from the thermal Sunyaev-Zel'dovich effect (SZ) observation of the Coma cluster \citep{2012MNRAS.421.1123C}.



Following \citet{Pinzke2017}, the turbulent energy density, $\varepsilon_{\rm turb}\approx \rho V_L^2$, is assumed to scale with the thermal energy density $\varepsilon_{\rm th}$ as
\begin{eqnarray}\label{eq:eps_tu}
\frac{\varepsilon_{\rm turb}}{\varepsilon_{\rm th}}\propto\varepsilon_{\rm th}^{\frac{\alpha_{\rm turb}-1}{2}}(r)T^{-\frac{1}{4}}(r)\exp\left[-\left(\frac{r}{R_{500}}\right)^\frac{1}{2}\right],
\end{eqnarray}
where $R_{500}$ is a parameter $R_\Delta$ with $\Delta=500$.
The radius $R_\Delta$ is defined as the radius inside which the average of the total density becomes $\Delta$ times the critical density of the Universe, $\rho_{\rm cr}$:
\begin{eqnarray}\label{eq:R_vir}
R_\Delta  = \left[\frac{3M_\Delta}{4\pi\Delta\rho_{\rm cr}(z)}\right]^{1/3}. 
\end{eqnarray}

We adopt $R_{500}=47~{\rm arcmin}$ for the Coma cluster \citep[][]{Planck2013_Coma}, which corresponds to $R_{500}=1.35~{\rm Mpc}$ under the cosmology we adopted. The need for the cut-off in Eq.~(\ref{eq:eps_tu}) is discussed in Section~\ref{subsec:Coma}. 

\par

From Eq.~(\ref{eq:Dpp_integral}) and the Mach number for the turbulent velocity at the injection scale ${\cal M}_{\rm s}=V_L/c_{\rm s}$, we obtain $D_{pp}(r,p)\propto {\cal M}_{\rm s}^4c^2_{\rm s}k_L \propto \frac{c_{\rm s}(r)}{c}\frac{I_L(r)}{\rho(r)c_{\rm s}^2(r)}$, where $I_L$ is the volumetric injection rate of turbulence energy. For simplicity, the injection scale $k_L$ is assumed to be a constant. With those assumptions concerning $\varepsilon_{\rm turb}$ and $k_L$, the radial dependence factor ${\cal D}(r)$ in Eq.~(\ref{eq:Dpp}) can be expressed with one parameter, $\alpha_{\rm turb}$:
\begin{eqnarray}\label{eq:Dpp_r}
{\cal D}(r) = {\cal D}_0 \varepsilon_{\rm th}^{\alpha_{\rm turb}-1}(r)\sqrt{T_{\rm e}(r)}\exp\left[-\left(\frac{r}{R_{500}}\right)\right],
\end{eqnarray}
where the normalization is given at $r = r_{\rm c}$, i.e., ${\cal D}_0^{-1}= \varepsilon_{\rm th}^{\alpha_{\rm turb}-1}(r_{\rm c})\sqrt{T_{\rm e}(r_{\rm c})}\exp\left[-\left(\frac{r_{\rm c}}{R_{500}}\right)\right]$. \par

Using Eq.~(\ref{eq:Dpp_integral}) and neglecting the cut-off terms in Eq.~(\ref{eq:Dpp}), $t_{\rm acc}(r_{\rm c})$ can be obtained as a function of the ratio of the turbulent energy density to the thermal energy density:
\begin{eqnarray} \label{eq:tacc_eturb}
t_{\rm acc}(r) \approx 300~{\rm Myr}\left(\frac{\varepsilon_{\rm turb}/\varepsilon_{\rm th}(r)}{0.5}\right)^{-2}\left(\frac{L}{500~{\rm kpc}}\right) \nonumber \\
\times \left(\frac{I_\theta(x)}{0.5}\right)^{-1}\left(\frac{c_{\rm s}(r)}{10^3~{\rm km/s}}\right)^{-2},
\end{eqnarray}
where $x=c_{\rm s}(r)/c$.

\subsection{Model outline}\label{subsec:outline}
In our modeling, the history of the RH evolution is divided into following three phases:
\begin{itemize}
    \item {\it Injection phase}. In this phase, we follow the long-term evolution of the CR distribution before the major merger event. We integrate the FP equations (Eqs.~(\ref{eq:Np}) and (\ref{eq:Ne})) with $D_{pp}\equiv0$ for 4 Gyr until the the CRE spectrum settles in a steady shape due to the balance between the constant injection and the cooling processes. \par
    
    \item {\it Reacceleration phase}.
    This phase starts just after a major merger event.
    The merger-induced turbulence reaccelerates CRs. The initial condition of this phase is the final state of the injection phase. We follow the evolution including the reacceleration expressed with the diffusion coefficient of Eq.~(\ref{eq:Dpp}). The Coma RH is assumed to be in this phase and the elapsed time in this phase, $t_{\rm R}$, is treated as a model parameter. We do not consider the time evolution of the turbulence, so the parameters, $\alpha_{\rm turb}$ and $t_{\rm acc}$ are not time dependent. \par
    
    \item {\it Cooling phase}.
    After the reacceleration ceases, the emission decays with time. This phase is modeled with $D_{pp}=0$, and the emission lifetime will be investigated in Section~\ref{sec:lifetime}.
    
    
\end{itemize}

The CR injection spectral index is fixed to be $\alpha_{\rm inj} = 2.45$, which can model the typical radio spectral index of RHs, $\alpha_{\rm syn}\approx1.2$. Concerning the CR injection, we have one free parameter, $L_{\rm p}^{\rm inj}$, which is the injection luminosity (in unit of [erg/s]) of relativistic ($p> m_{\rm p} c$) protons integrated over the cluster volume, i.e., $L_{\rm p}^{\rm inj}=\int dr\int dp E_{\rm p}(p)Q_{\rm p}(r,p)$, where $E_{\rm p}(p)=\sqrt{m^2_{\rm p}c^4+p^2 c^2}$. The injection rate of primary CREs is determined by Eq~(\ref{eq:Qepri}).  \par

For a given magnetic field, our model includes four parameters: the CR injection luminosity, $L_{\rm p}^{\rm inj}$, the power-law index for the radial dependence of $D_{pp}$, $\alpha_{\rm turb}$, the timescale of the reacceleration at a given radius, $t_{\rm acc}(r_{\rm c})$, and the elapsed time after the reacceleration onset, $t_{\rm R}$.

\subsection{Modeling the Coma RH \label{subsec:Coma}}
In this section, we apply the model explained above to the Coma RH and discuss the constraints on the model parameters. The radial dependence of $D_{pp}$ is constrained by the observed brightness profile, while $t_{\rm R}$, $t_{\rm acc}$, and $L_{\rm p}^{\rm inj}$ are by the spectral shape and the flux at a given frequency. With those parameters, we study the emission lifetime in Section~\ref{sec:lifetime}. \par

The magnetic field in the Coma cluster is well studied with the rotation measure (RM) measurements. Here we use the following scaling of the magnetic field strength with cluster thermal density: 
\begin{eqnarray}
B(r)=B_0\left(\frac{n_{\mathrm{th}}(r)}{n_{\mathrm{th}}(0)}\right)^{\eta_B},
\label{eq:B}
\end{eqnarray}
where the best fit values are $B_0=4.7\:\mu \mathrm{G}$ and $\eta_B=0.5$ \citep{Bonafede2010}, and see Eq. (\ref{eq:beta-model}) for $n_{\mathrm{th}}(r)$. The uncertainty in the magnetic field and its impact on the multi-wavelength spectrum are discussed in \citet{paperI}. \par

\begin{figure*}
    \centering
    \plottwo{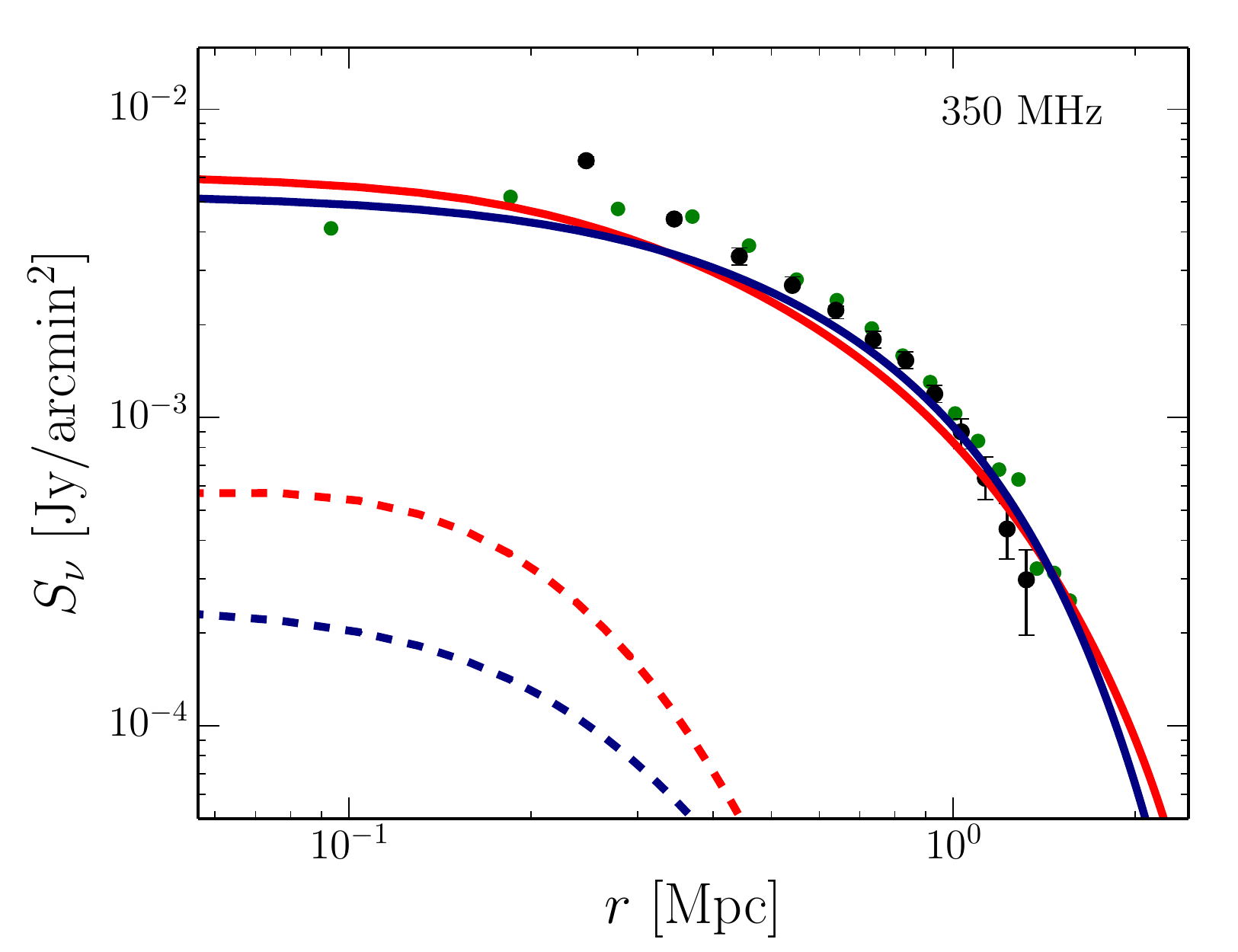}{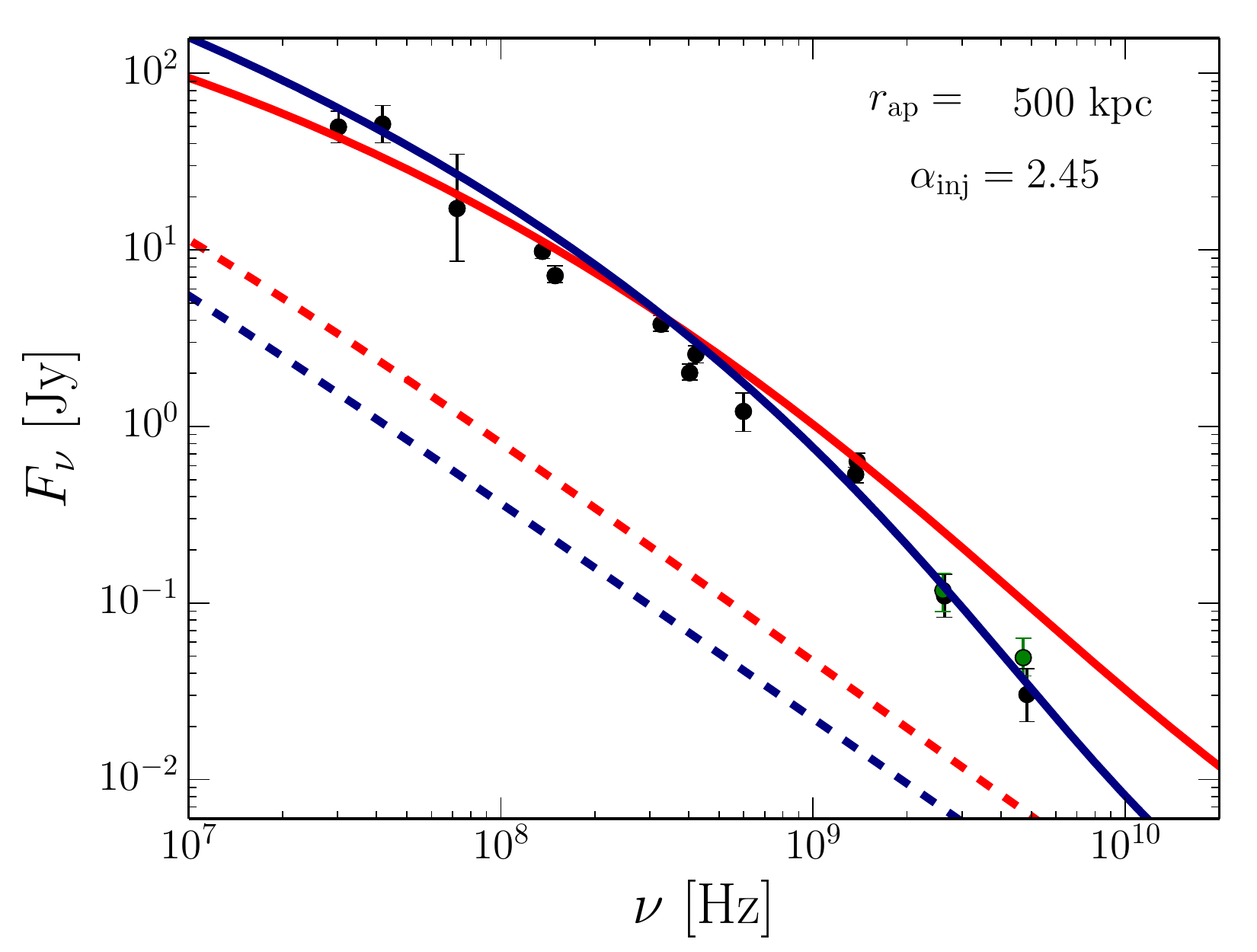}
    \caption{Left panel: Brightness profile of the Coma RH at 350 MHz. Right panel: The spectrum is calculated by integrating the intensity within the aperture radius of $r_{\rm ap} = 500~{\rm kpc}$. The results for secondary and primary scenarios are shown with red and blue lines, respectively. The dashed lines show the emission before the reacceleration. Data points are taken from \citet{Pizzo2010,Pinzke2017,Brunetti2017}.}
    \label{fig:Coma_RH}
\end{figure*}

For a fixed injection of primary CRPs, the profile of the turbulence $\alpha_{\rm turb}$ can be constrained from the brightness profile of the RH. Before the reacceleration, the injection with profile of Eq.~(\ref{eq:primary_injection}) produces too steep brightness profile in both primary and secondary scenarios (dashed lines in Figure~\ref{fig:Coma_RH}). This steepness is further enhanced in the secondary model because of the density dependence of the electron injection.\par

\begin{figure*}
    \centering
    \plottwo{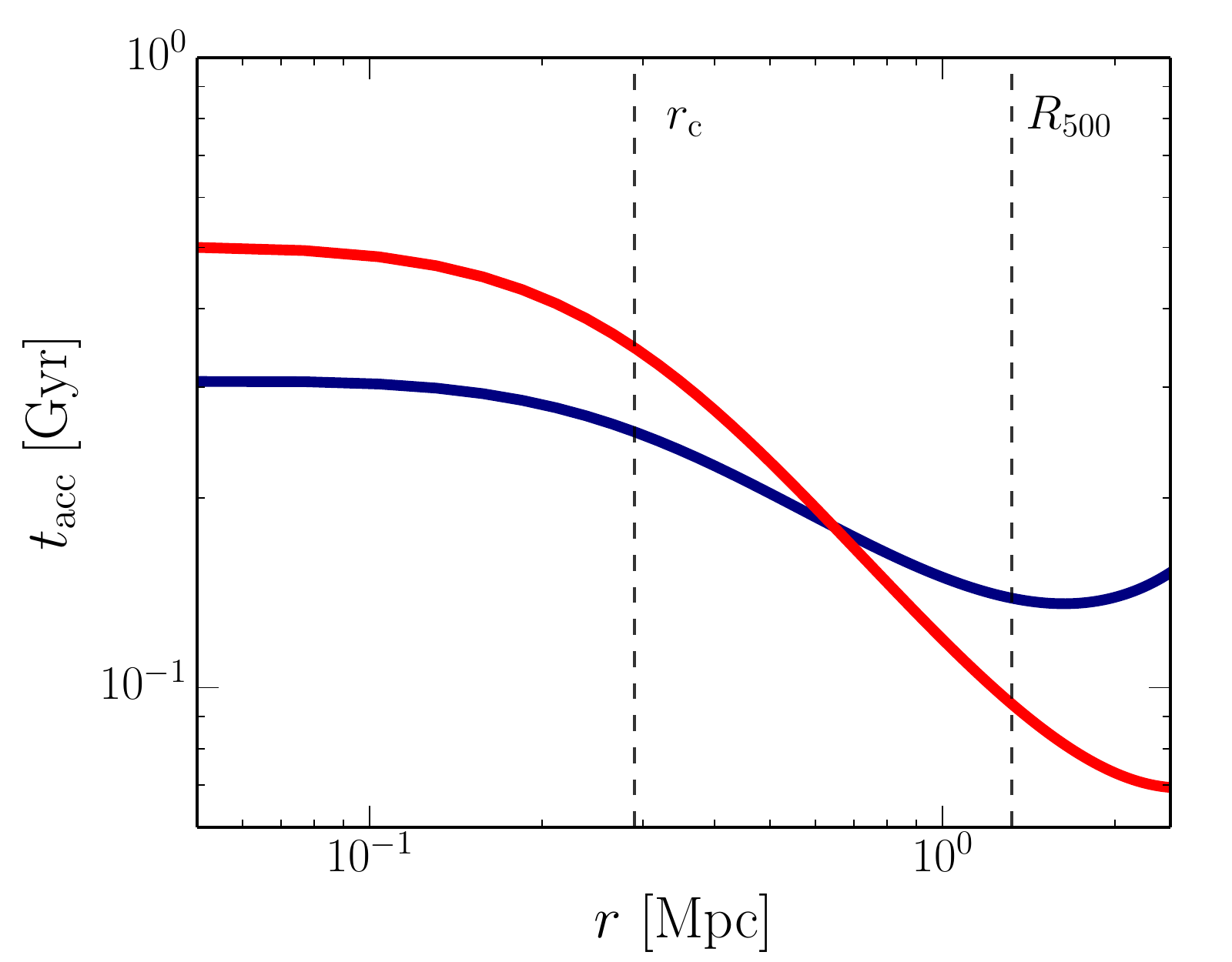}{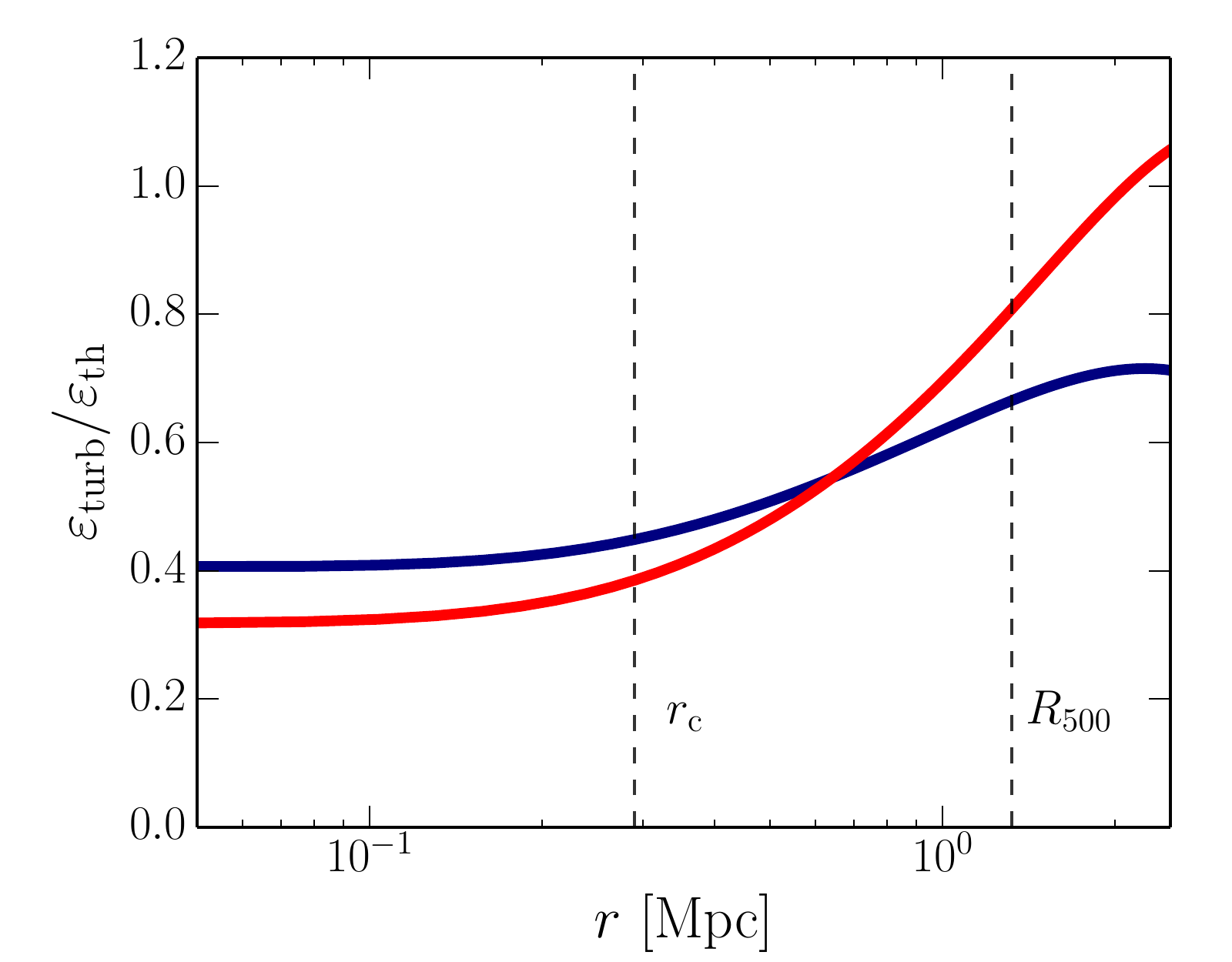}
    \caption{Radial profile of $t_{\rm acc}$ (Left) and $\varepsilon_{\rm turb}/\varepsilon_{\rm th}$ (Right) in the same models in Fig. \ref{fig:Coma_RH}. The red and blue lines show results of the secondary ($\alpha_{\rm turb}=0.28$) and the primary ($\alpha_{\rm turb}=0.52$) scenarios, respectively. The turbulent energy density is calculated from Eq.~(\ref{eq:tacc_eturb}) with $L=500~{\rm kpc}$. The vertical dashed lines show the radius correspond to $r_{\rm c}$ and $R_{500}$ of the Coma cluster.}
    \label{fig:tacc_eps_tu}
\end{figure*}

As seen from Eq.~(\ref{eq:Dpp}), the turbulent profile with $\alpha_{\rm turb}<1$ enhances the acceleration at larger radii ($r>r_{\rm c}$), and the brightness profile would be modified to fit the observation. As shown in Figure~\ref{fig:Coma_RH}, we find that $\alpha_{\rm turb} = 0.28$ and $\alpha_{\rm turb} = 0.52$ provide the best fit to the secondary and the primary scenarios, respectively. The other model parameters are summarized in Table \ref{tab:params}. The radial dependence of the acceleration timescale $t_{\rm acc}(r)\equiv p^2/(4D_{pp}(r,p))$ is shown in Figure~\ref{fig:tacc_eps_tu}. Adopting Eq.~(\ref{eq:tacc_eturb}) with constant $L=500~{\rm kpc}$, we estimate the turbulent energy density from $t_{\rm acc}$ as shown in Figure~\ref{fig:tacc_eps_tu}. \par


In the secondary scenario, the reacceleration timescale at the edge of the RH is required to be almost 2 times shorter than inside the core region. The decline of $t_{\rm acc}$ towards large radii is suppressed by the artificial cut-off of the turbulent profile at $R_{500}$ (Eq.~(\ref{eq:Dpp_r})), with which the observed decline of the brightness around 1 Mpc is well explained. 


If we omit the cut-off term in Eq.~(\ref{eq:Dpp_r}), the acceleration efficiency becomes unrealistically high at the periphery of the cluster. Although a model without the cutoff can explain the synchrotron spectrum within $r_{\rm ap}=500~{\rm kpc}$, the gamma-ray flux calculated within $r<R_{200}$ exceeds the upper-limit given by Fermi-LAT \citep[][]{Ackermann2016} in the secondary scenario. This problem can be circumvented by introducing the cut-off. \par

The elapsed time, $t_{\rm R}$, and the timescale of reacceleration, $t_{\rm acc}$, can be constrained from the spectral shape of the RH \citep[see][for the detail]{paperI}, especially the spectral steepening around 1 GHz. We find that $t_{\rm acc}(r_{\rm c})\approx300$ Myr is favorable in both the scenarios.\par

The log-scaling in the cross section of the pp collision \citep[e.g.,][]{Kamae2006} causes the hardening in the secondary CRE specrtum, and our secondary model predicts a larger flux above $1~{\rm GHz}$ (Figure~\ref{fig:Coma_RH}). A possible prescription for that tension is adopting a steeper injection index. The spectra for different $\alpha_{\rm inj}$ are shown in Appendix \ref{app:ComaAlpha}, where we change $L_{\rm p}^{\rm inj}$ and $t_{\rm R}$ for each $\alpha_{\rm inj}$ to fit the data, while the other parameters, $\alpha_{\rm turb}$ and $t_{\rm acc}$ are fixed. We find that the $\alpha_{\rm inj} = 3.0$ model provides a better fit to the higher frequencies. Since the purpose of this paper is the discussion about the RH statistics, we do not excessively tune the parameters to match with the Coma RH spectrum. Throughout this paper, we fix $\alpha_{\rm inj} = 2.45$ as a fiducial value \citep[e.g.,][]{Pinzke2010,Brunetti2017}. \par



In each scenario, The expected flux of gamma-ray becomes smaller than the upper limit by Fermi-LAT observation \citep[][]{Ackermann2016}. Note that possible gamma-ray signals are reported in the direction of Coma \citep[e.g.,][]{Xi_2018PhRvD..98f3006X,2021A&A...648A..60A}. The predicted flux in our secondary scenario is comparable to the data reported in those papers, while the GeV gamma-ray flux becomes almost two order of magnitude lower in the primary model. \par

\begin{table}[hbt]
    \caption{Model parameters for the Coma radio halo}
    \centering
    \begin{tabular}{ccccc}
    \hline
model & $\alpha_{\rm turb}$ & $t_{\rm acc}$ (at $r=r_{\rm c}$)  & $t_{\rm R}$ & $L_{\rm p}^{\rm inj}$\\ 
& & [Myr]  &  [Myr] & [erg s$^{-1}$]\\ \hline
secondary & 0.28 & 330 & 340 & $9.3\times10^{43}$\\
primary & 0.52 & 255 &  300 & $7.3\times10^{42}$\\
\hline
    \end{tabular}
   \tablecomments{The other fixed parameters are $4~{\rm Gyr}$ for the duration of the injection phase, $\alpha_{\rm inj} = 2.45$ for the slope of the injection spectrum, and the cutoff $R_{500}$ of the turbulent profile.}
    \label{tab:params}
\end{table}

As shown in Table \ref{tab:params}, the acceleration timescale $t_{\rm acc}$ are similar to those in the previous studies \citep[e.g.,][]{Brunetti2017,paperI}. However, the primary CR injection luminosity in the primary scenario is one order of magnitude larger than that in \citet{paperI}. This is attributed to the different best fit values of $t_{\rm R}/t_{\rm acc}$. In our previous model, $t_{\rm R}/t_{\rm acc}\gtrsim3.0$ is required to match the brightness at outer radii ($r>500~{\rm kpc}$), while this requirement is relaxed considering the radially increasing turbulent profile. The relatively small $t_{\rm R}/t_{\rm acc}$ in the current model reduces the energy contribution from the reacceleration, which in turn increases the contribution from the primary CR injection. The injection power of primary CREs ($p_{\rm e}c \geq 150$ keV) is smaller than that of CRPs ($pc \geq 30$ MeV) by a factor of $\approx5$ in the primary scenario with $\alpha_{\rm inj} = 2.45$. \par 

In our previous study, we modeled the brightness profile of Coma with a fine-tuned injection profile with larger number of model parameters, under the assumption of spatially homogeneous acceleration efficiency. In this study, however, we succeeded in the modeling with much simpler injection profile (Eq.~(\ref{eq:primary_injection})), adopting the turbulent profile shown in Figure~\ref{fig:tacc_eps_tu}. Note that both the models are compatible with the current radio observations. The direct measurement of turbulent velocity with future X-ray telescopes featured with improved spectral resolution and effective area is desired to distinguish those two models. A similar turbulent profile for a Coma-like halo is obtained in \citet{Pinzke2017}. The best-fit value for their {\it M-turbulence} model is $\alpha_{\rm turb}=0.67$, which is slightly larger than that in our primary scenario. Our model is in good agreement with that model concerning the profile of $t_{\rm acc}$, although the overall normalization is larger in their {\it M-turbulence} model by a factor $\sim$2.







\section{Lifetime of radio halos \label{sec:lifetime}} 
In this Section, we discuss the lifetime of RHs in the cooling phase (Section~\ref{subsec:outline}). As an example case, we present the time evolution of a Coma-like cluster after the reacceleration ends. In this cooling phase, we solve the FP equation with $D_{pp}=0$. In our method, we do not follow the evolution of the turbulent spectrum, so the acceleration timescale is assumed to be constant within the reacceleration phase and it suddenly turns to be zero at the transition to the cooling phase. Two competing effects govern the evolution of the CR spectra: cooling processes and injections (primary and secondary). \par

The model is basically the same as that explained in Section~\ref{sec:CR}. We adopt the same magnetic field, ICM profile, CR injection profile, slope of the injection spectrum, and redshift. We adopt the model parameters, such as $L_{\rm p}^{\rm inj}$, $t_{\rm acc}$, and $\alpha_{\rm turb}$, from the obtained values in Section~\ref{sec:CR}. However, we change $t_{\rm R}$ as explained in the following. \par

The initial condition for the cooling phase corresponds to the final state of the reacceleration phase. To set the initial condition, the maximum duration of the reacceleration phase needs to be specified. In the case of the Coma RH, the present stage does not necessarily correspond to the end of the reacceleration phase, i.e., the peak of the luminosity evolution. We assume that the Coma RH is in the midst of the reacceleration phase. Considering that there exit some RHs with masses similar to the mass of Coma and much brighter radio luminosities as $P_{1.4}\approx10^{25}~{\rm W/Hz}$, such as RXCJ~0510.7-0801 and Abell 2744 \citep[][]{2021A&A...647A..51C}, we assume that a Coma-like cluster could achieve similar luminosity at $t_{\rm R}=t_{\rm R}^{\rm peak}$. Our model suggests that the luminosity of a Coma-like halo reaches that value around $t_{\rm R}\approx500~{\rm Myr}$ (about 200 Myrs from the present stage of Coma). Thus, we set the initial condition for the cooling phase by the state at $t_{\rm R}^{\rm peak}=500~{\rm Myr}$ \par



In the following, we compare the two scenarios for the seed population: the secondary scenario (Section~\ref{subsec:lifetime_secondary}) and the primary scenario (Section~\ref{subsec:lifetime_primary}). We define the emission lifetime as the duration for which the flux at 1.4 GHz, $F_{1.4}$, satisfies the condition, $F_{1.4}\gtrsim0.1F_{1.4}^{\rm peak}$, where $F_{1.4}^{\rm peak}$ is its peak value.


\begin{figure*}
    \centering
    \plottwo{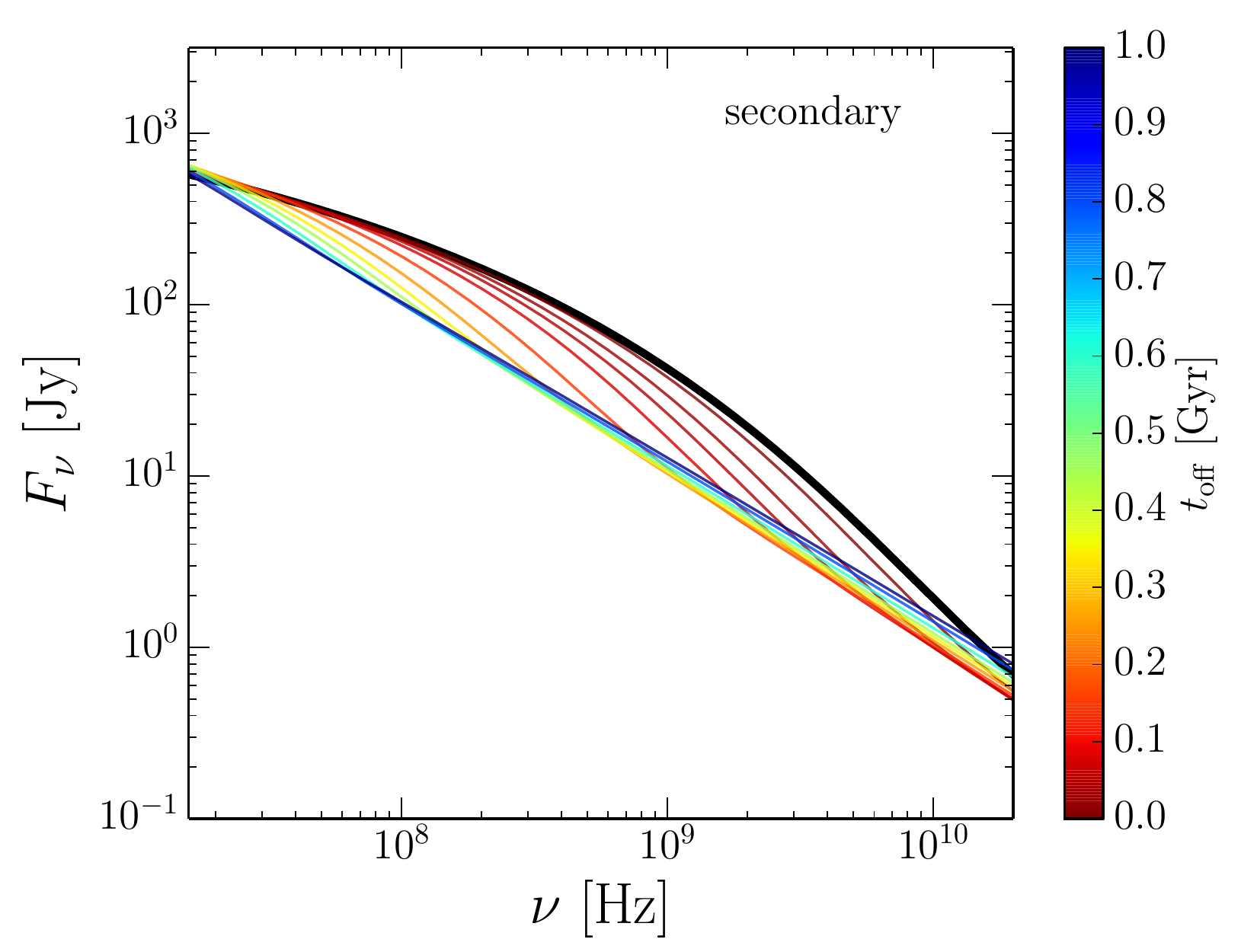}{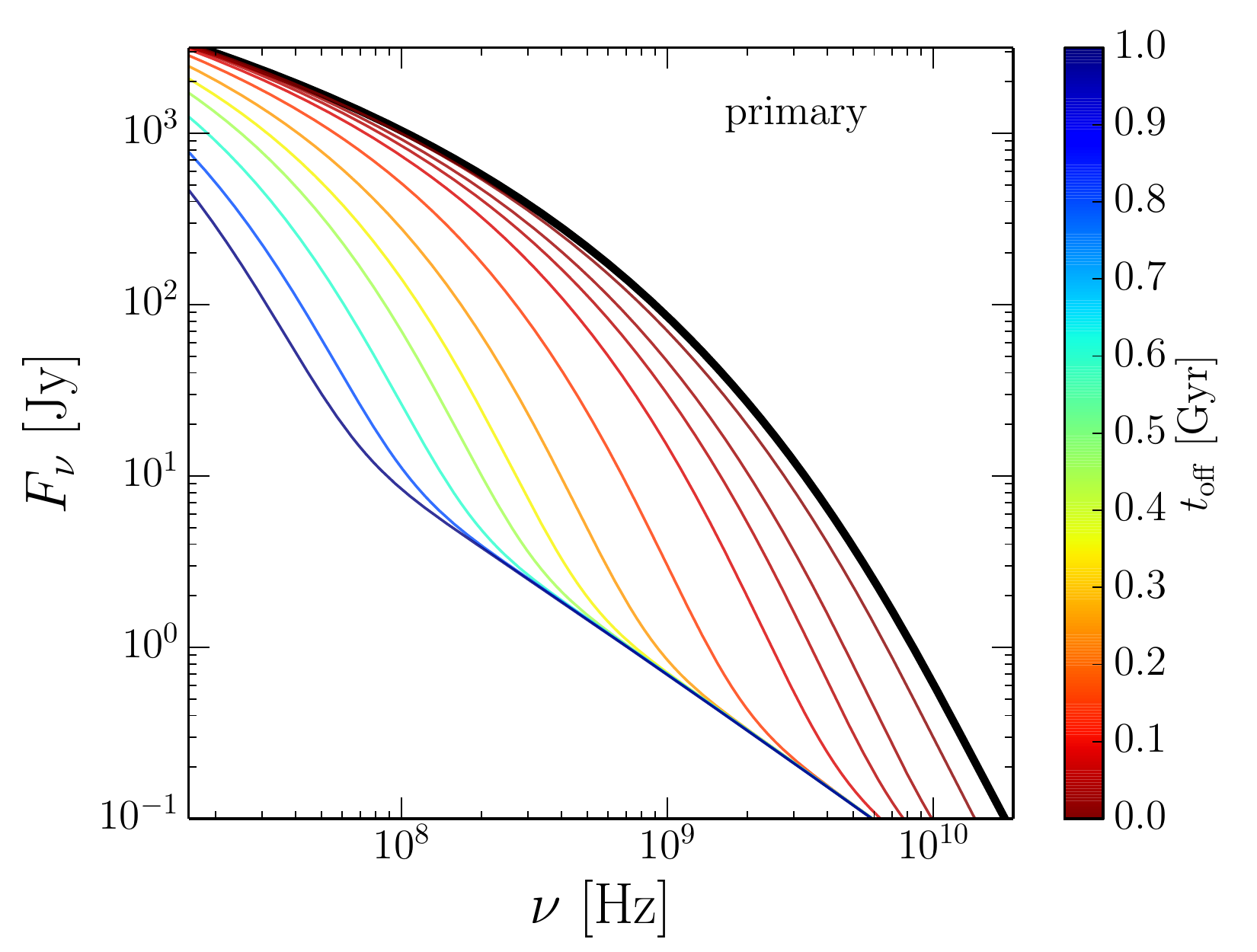}
    \caption{Time evolutions of the synchrotron flux of a Coma-like RH at the cooling phase in the secondary (Left) and primary (Right) scenarios. The elapsed time (distinguished with different colors) is measured from the peak time, which corresponds to the time when the reacceleration ends. The thick black line shows the spectrum at the peak time.}
    \label{fig:radio_flux_off}
\end{figure*}

\subsection{Secondary scenario \label{subsec:lifetime_secondary}}
In the left panel of Figure~\ref{fig:radio_flux_off}, we show the time evolution of synchrotron spectrum of our Coma-like cluster in the secondary scenario. We adopt the model parameters, $\alpha_{\rm turb}=0.28$, $t_{\rm acc} = 330~{\rm Myr}$ (at $r=r_{\rm c}$), and $t_{\rm R}=500~{\rm Myr}$. The aperture radius is fixed to $r_{\rm ap}=500~{\rm kpc}$. The thick black line shows the flux at the end of the reacceleration phase, or the beginning of the cooling phase. The flux at the each epoch in the cooling phase is distinguished by colors, as it evolves from red to blue. After several 100 Myrs, the spectrum approaches a power-law shape due to the balance between the electron injection and the cooling. The flux decrement in the cooling phase is at most one order of magnitude and most highlighted at $\approx300~{\rm MHz}$, where the cooling and acceleration balance in the reacceleration phase. 
At higher frequencies ($\nu>1~{\rm GHz}$), the peak flux is maintained by a high injection rate of secondary CREs from reaccelerated CRPs. Since parent CRPs do not significantly suffer from cooling, the flux at $>1$ GHz is maintained even after the reacceleration phase. Therefore, the lifetime of the RH can be much longer than the cooling time of CREs especially at high frequencies. \par


Our calculation demonstrates that not only the reacceleration of the pre-deposited secondary CREs but also the flesh electrons injected from reaccelerated CRPs make significant contributions to the radio emission. The fractional contribution of those two components could depend on the parameters, $t_{\rm acc}$ and $t_{\rm R}^{\rm peak}$. As shown in \citet{paperI}, in the reacceleration phase, the evolution of the radio flux around 300~MHz is relatively faster than that of the gamma-ray flux originated from $\pi^0$ decay. This suggests that as $t_{\rm R}/t_{\rm acc}$ increases, in the radio flux, the contribution of the reaccelerated CREs dominates that of the enhanced electron injection from reaccelerated CRPs. Even at 1.4~GHz, the flux decay would be more prominent, if the reacceleration is more efficient ($t_{\rm acc}\ll200~{\rm Myr}$) or the reacceleration lasts longer ($t_{\rm R}^{\rm peak}\gg500~{\rm Myr}$). However, such a model causes a tension with the observed spectral feature of the Coma RH \citep[e.g.,][]{paperI} or the radio luminosity becomes far brighter than any observed RHs. \par

Note that such long-living emission is achieved only under the assumption of hard-sphere type reacceleration ($D_{pp}\propto p^2$), where the reacceleration timescale is the same between CRPs and CREs. When the CRP reacceleration is less efficient compared to that for CREs, e.g., in the case of the Kolmogorov type reacceleration \citep{paperI}, the emission would decline within a few times the cooling timescale. \par

As the radiative cooling proceeds, the shape of the spectrum approaches to a single power-law shape. The concave shape caused by the reacceleration disappears within a few times the cooling timescale. If the observed break feature around 1.4 GHz is considered to be significant, the RH in Coma should still be in the reacceleration phase or within $\approx200$ Myrs after the reacceleration stops. \par

\begin{figure*}
    \centering
    \plottwo{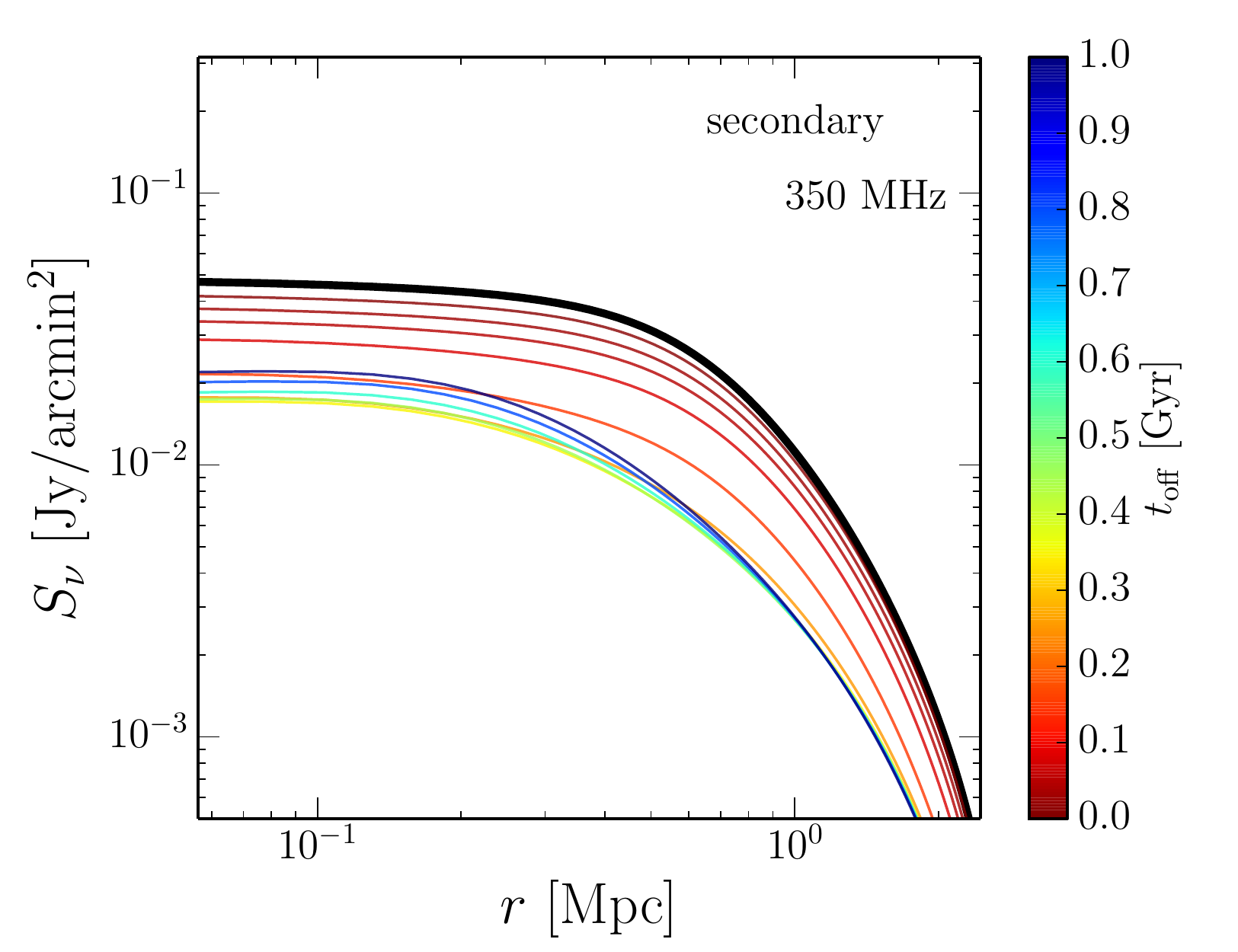}{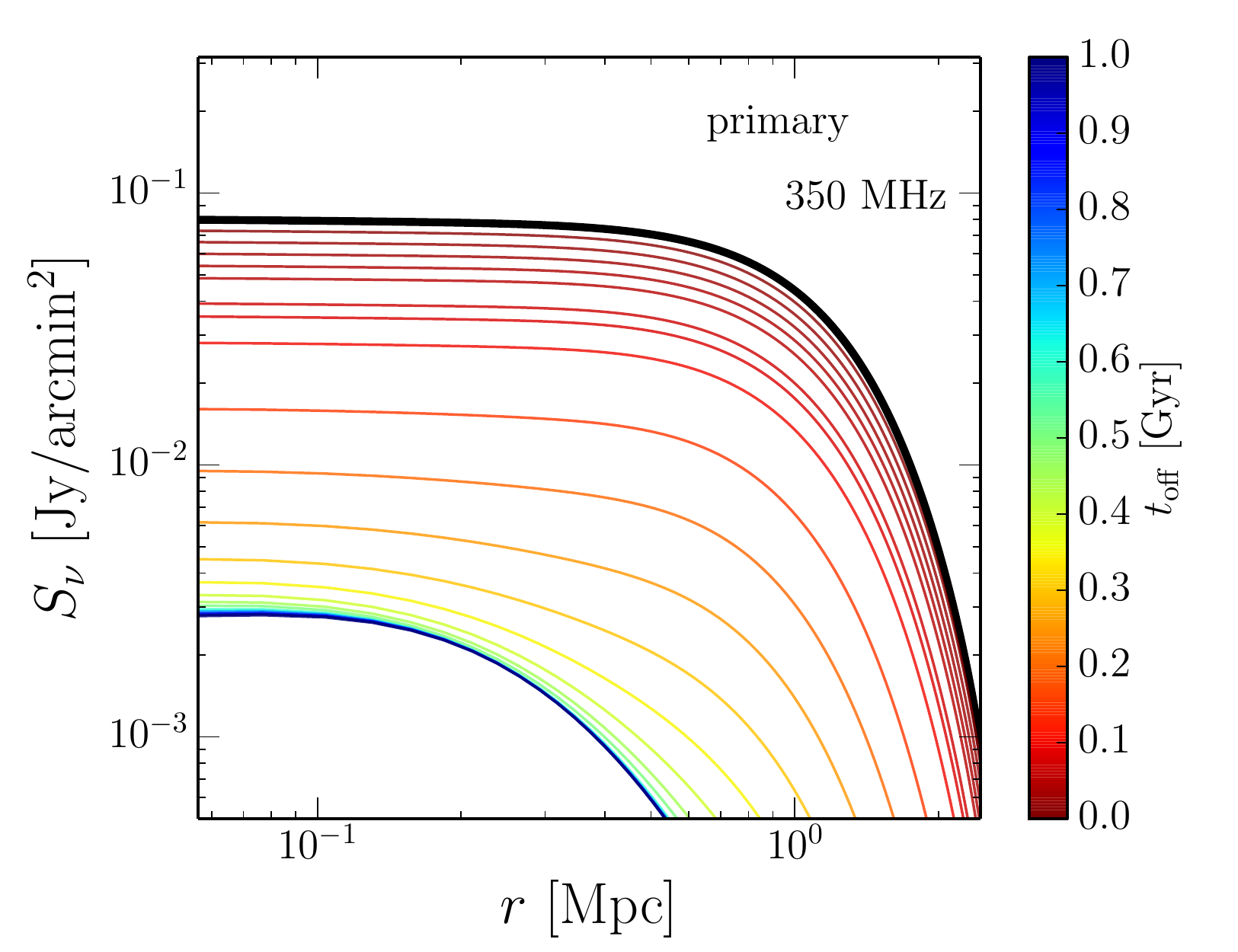}
    \caption{Time evolution of the brightness profile of a Coma-like RH at 350 MHz in the cooling phase. The color code is the same as that in Figure~\ref{fig:radio_flux_off}. }
    \label{fig:radio_profile_off}
\end{figure*}

In Figure~\ref{fig:radio_profile_off}, we show the evolution of the brightness profile at 350 MHz. The decline of the brightness is within one order of magnitude, consistent with the spectral evolution. Given a threshold for $S_\nu$, the size of the RH shrinks with time. For example, considering an observational threshold around $S_\nu\approx10^{-2}~[{\rm Jy/arcmin}^2]$, the apparent RH size evolves from $1~{\rm Mpc}$ to $0.5~{\rm Mpc}$ in the cooling phase. A slight {\it increase} of the brightness in the core region is seen at the later stage. This is a transitional phenomenon while the CRE distribution is evolving towards the equilibrium state regulated by the balance between the cooling and the enhanced injection from reaccelerated CRPs. \par 



\subsection{Primary scenario \label{subsec:lifetime_primary}}
The right panel of Figure~\ref{fig:radio_flux_off} shows the spectral evolution in the primary scenario. The model parameters are, $\alpha_{\rm turb}=0.52$, $t_{\rm acc} = 255~{\rm Myr}$ (at $r=r_{\rm c}$), and $t_{\rm R}=500~{\rm Myr}$. In this case, the evolution of the secondary injection rate is negligible, and the radio flux in lower frequencies ($\nu<1~{\rm GHz}$) decays almost one order of magnitude within $\approx500$ Myr after the end of the reacceleration phase. The lifetime of the RH at 1.4 GHz is comparable to the cooling timescale of responsible electrons, or the reacceleration timescale, as shown in the previous studies \citep[][]{Cassano_2005}. \par


The evolution of the brightness profile is shown in the right panel of Figure~\ref{fig:radio_profile_off}. Adopting the same threshold for $S_\nu$ as in Sect.~\ref{subsec:lifetime_secondary}, we find that the RH disappear in 400 Myr after the peak time. In this scenario, not only the flux but also the size of the RH show rapid increase and decrease. Considering both the reacceleration phase and the cooling phase, one can approximate the lifetime of the giant RH as $\approx$500 Myr in total. \par


The spectral profile at the late stage of the cooling phase, $t_{\rm off}>1~{\rm Gyr}$, is not equivalent to the pre-reacceleration profile (Figure~\ref{fig:Coma_RH} dashed line). This difference is originated from the assumption of $f_{\rm ep}= 0.01$, i.e., the finite injection rate of CRPs even in the primary scenario. The off-state emission is supported by the enhanced injection from reaccelerated CRPs, although the synchrotron flux becomes smaller by one order of magnitude than the peak state.  \par



\subsection{Discussion \label{subsec:caveats}}
In this Section, we have discussed the lifetime of the radio emission around 1.4 GHz. As seen from Figure~\ref{fig:radio_flux_off}, the lifetime depends on the frequency. A longer time is required for lower frequencies to reach the steady flux due to the balance between the cooling and the injection. In addition, the cooling timescale depends on the magnetic field and redshift. Especially for the primary scenario, one can expect a much longer lifetime for a weaker magnetic field and lower frequencies.  \par 


While we have adopted $\alpha_{\rm inj}=2.45$, as demonstrated in Figure 7 in \citet{paperI}, a smaller $\alpha_{\rm inj}$ in the secondary model results in a harder and brighter flux around 10 GHz, where the emission is mostly powered by the injection from reaccelerated CRPs. Then, the decay of the emission seen in secondary scenario (Figure~\ref{fig:radio_flux_off} left) becomes less important when $\alpha_{\rm inj} < 2.45$.  \par



We have estimated the RH lifetime under the ``step-function approximation" for $D_{pp}$. In reality, the transition into the cooling phase would be gradual. If its transition timescale is much longer than the cooling timescale, the lifetime in the primary scenario can be longer than our estimate in the previous section. \par




\section{Merger tree of dark matter halos \label{sec:MT}}
In the previous section, we have studied the lifetime of RHs. Our next steps is to calculate the population of RHs induced by cluster mergers, which can be compared with the observed statistic of RHs. In this Section, we explain our method to follow the cosmological evolution of galaxy clusters by constructing merger trees.\par

Merger trees describe the mass evolution of a dark matter (DM) halo. In the extended Press-Shechter (EPS) formalism \citep[e.g.,][]{PressSchechter_1974ApJ...187..425P,Bond_1991ApJ...379..440B,LaceyCole_1993MNRAS.262..627L}, the mass growth rate of a halo, or the conditional mass function, is given consistently with the Press-Shechter (PS) mass function. However, the EPS formalism tends to underestimate the number density of massive halos compared to cosmological N-body simulations \citep[e.g.,][]{ShethTormen_1999MNRAS.308..119S,Tinker_2008ApJ...688..709}. We introduce a Monte Carlo algorithm to build the merger trees compatible with the mass function given in those simulations. 

\subsection{Mean merger rate per halo \label{method:MergerRate}}

In our picture, DM halos evolve through stochastical mergers and continuous accretion of matters. Each halo experiences both the continuous mass evolution due to the accretion and the sudden mass increase due to the mergers. We employ the fitting formulae for the merger rate and the mass accretion rate (MAR) from N-body simulations \citep[][]{Fakhouri_2010MNRAS.406.2267F}. However, as explained in the following, the separation between accretion and mergers is not definitive, so we need to define the minimum mass ratio for mergers, on which the estimate of the continuous mass accretion depends. \par


We treat a merger as a two-body event. The merger between $n\geq 3$ halos can be regarded as a sequence of binary mergers within a sufficiently small time interval. For a merger between two halos of masses $M_1$ and $M_2$, we call these two halos ``progenitor" halos, and the one produced by the event ``descendant" halo. \par


We use the dimensionless merger rate per descendant halo given by \citet{Fakhouri_2010MNRAS.406.2267F};
\begin{eqnarray}
\frac{dN_{\rm m}}{d\xi dz}(M_0,\xi,z) = &&A\left(\frac{M_0}{10^{12}~{\rm M}_\odot}\right)^{k_1}\xi^{k_2}\exp\left[\left(\frac{\xi}{\tilde{\xi}}\right)^{k_3}\right]\nonumber\\
&&\times(1+z)^{k_4},
\label{eq:Mergerrate}
\end{eqnarray}
where $M_0=M_1+M_2$ is the mass of a descendant halo, $\xi=M_2/M_1 \leq 1$ is the progenitor mass ratio, and the best fit parameters are $(k_1,k_2,k_3,k_4)=(0.133,-1.995,0.263,0.0993)$ and $(A,\tilde{\xi})=(0.0104,9.72\times10^{-3})$. \par

This expression diverges at $\xi=0$, so we introduce a parameter, $\xi_{\rm min}$, below which such frequent minor mergers are treated as a continuous accretion.


In the following sections, we assume that RHs are induced by ``mergers" and discuss the condition for the RH onset. Thus, the value of $\xi_{\rm min}$ needs to be smaller than the threshold mass ratio for the RH onset, $\xi_{\rm RH}$ (Section~\ref{subsec:criterion}). In this paper, we set $\xi_{\rm min}=10^{-3}$ to make the above condition satisfied with saving the computational cost. This value is larger than the minimum $\xi$ used in \citet{Fakhouri_2010MNRAS.406.2267F}. The number of mergers for given intervals $\Delta z$ and $\Delta\xi$ is expressed as

\begin{eqnarray}
{\cal P}(M_0,\xi,z)=\frac{dN_{\rm m}}{d\xi dz}\Delta\xi\Delta z.
\label{eq:P_mer}
\end{eqnarray}

In our computation, we adopt sufficiently small values for $\Delta z$ and $\Delta \xi$ so that ${\cal P}(M_0,\xi,z)<10^{-3}$ for $\xi\ge\xi_{\rm min}$, ensuring the two-body merger approximation. \par



%

\subsection{Mass accretion rate\label{method:MAR}}
In the limit of $\xi_{\rm min}\rightarrow0$, the mass evolution rate due to mergers obtained by integrating Eq. (\ref{eq:Mergerrate}) with respect to $\xi$ becomes infinity. This indicates that the mass evolution is dominated by numerous minor mergers with small $\xi$, rather than rare major mergers. 
For an appropriate time interval, most of halos with similar masses would evolve with similar rates through minor mergers and the accretion, while a small fraction of halos experience abrupt growth due to major mergers. To express the continuous accretion, we adopt the median value of the total MAR in N-body simulations, in which the contribution of major mergers may be negligible. \par

The total MAR is defined as $\frac{dM}{dz}= -(M_0-M_1)/\Delta z$, where $M_0$ is the descendant mass at $z$ and $M_1$ is the mass of its most massive progenitor at $z+\Delta z$. The process of the mass increase, mergers or accretion, is not specified here. According to \citet{Fakhouri_2010MNRAS.406.2267F}, the median values of MAR in the two Millennium simulations is well fitted with the following formula;
\begin{eqnarray}
\left\langle\frac{dM}{dz}\right\rangle_{\rm median} &=& - 3.58\times10^{12}~{\rm M}_\odot\left(\frac{h}{0.7}\right)^{-1} \nonumber\\
&&\times\left(\frac{M}{10^{12}M_\odot}\right)^{1.1} \left(\frac{1+1.65z}{1+z}\right). \label{eq:medianMAR}
\end{eqnarray}
As discussed in \citet{Fakhouri_2010MNRAS.406.2267F}, the mean MAR is overall larger than than median MAR, because the probability distribution of $\frac{dM}{dz}$ has a long tail in the high-accretion rate region. As the long tail, which boosts up the mean value, is mainly due to major mergers, the median value would be appropriate as the accretion rate in our definition. \par



\subsection{Halo mass function\label{method:HMF}}
The merger tree calculation needs to reproduce
the halo mass function (HMF), which describes the number of DM halos per unit comoving volume per unit mass. This can be written as, 
\begin{eqnarray}\label{eq:dndlnM}
\frac{dn}{d\ln M} = \frac{\rho_0}{M}f(\sigma,z)\left|\frac{d\ln\sigma}{d\ln M}\right|,
\end{eqnarray}
where $\rho_0$ is the mean mass density of the Universe, $\sigma$ is the rms mass variance, and the function $f(\sigma,z)$ is called multiplicity function, whose functional form is determined analytically \citep[][]{PressSchechter_1974ApJ...187..425P} or empirically from the fit to the mass function in N-body simulations \citep[e.g.,][]{ShethTormen_1999MNRAS.308..119S,Jenkins_2001MNRAS.321..372J,Tinker_2008ApJ...688..709,Watson_2013MNRAS.433.1230W}. 


The mass variance $\sigma$ is calculated with a top-hat filter on a scale $R_M \equiv \left(\frac{3M}{4\pi\rho_0}\right)^{1/3}$ as
\begin{eqnarray}\label{eq:sigma}
\sigma^2 = \int\frac{dk}{2\pi}k^2P(k)|\hat{W}(kR_M)|^2,
\end{eqnarray}
where $P(k)$ is the linear matter power spectrum and $\hat{W}(\chi)$ is the Fourier transform of the top-hat window function, which is expressed as
\begin{eqnarray}\label{hat_W}
\hat{W}(\chi) = \frac{3[\sin(\chi)-\chi\cos(\chi)]}{\chi^3}.
\end{eqnarray}
The radius $R_M$ is larger than the physical radius of galaxy clusters $R_\Delta$ (Eq.~(\ref{eq:R_vir})), since the mass density in clusters is generally larger than the mean mass density of the Universe.\par 

The matter power spectrum, $P(k)$, in Eq~(\ref{eq:sigma}) is conventionally expressed as
\begin{eqnarray}\label{eq:Pk}
P(k) = A_8 k^{n_s}T^2(k),
\end{eqnarray}
where $T(k)$ is the transfer function, and $n_{s}$ is the slope of the primordial power spectrum. The deformation from the primordial power spectrum at each scale is expressed by $T(k)$. The calculation of $T(k)$ includes technical challenges, because it is affected by, for example, various damping processes due to interaction between matters and relativistic particles. In this study, we adopt $T(k)$ generated with the public code CAMB.py \citep[][]{Lewis_2000ApJ...538..473}. The normalization $A_8$ is calculated from the cosmological parameter, $\sigma_8$, which is the mass variance on a scale of $8h^{-1}~{\rm Mpc}$. We adopt $n_{\rm s}=0.97$ and $\sigma_8=0.81$ from \citet{Planck18}.\par 

From Eq.~(\ref{eq:sigma}), we can calculate the rightmost factor of Eq.~(\ref{eq:dndlnM});
\begin{eqnarray}
\frac{d\ln\sigma}{d\ln M} = \frac{3}{2\sigma^2\pi^2R_M^4}\int_0^\infty\frac{d\hat{W}^2}{dM}\frac{P(k)}{k^2}dk.
\end{eqnarray}

Under the PS formalism, the multiplicity function, $f(\sigma,z)$, appearing in Eq~(\ref{eq:dndlnM}) becomes universal to the changes in redshift and cosmological parameters and has an expression of $f_{\rm PS}(\sigma)=\sqrt{\frac{2}{\pi}}\frac{\delta_{\rm c}}{\sigma}\exp\left[-\frac{\delta_{\rm c}^2}{2\sigma^2}\right]$, where $\delta_{\rm c}=1.69$ is the threshold parameter for the density fluctuation. However, we adopt the empirical formula for the multiplicity function given by \citet{Tinker_2008ApJ...688..709}, which agrees with the results in N-body simulations, as 
\begin{eqnarray}\label{eq:TinkerMF}
f_{\rm T}(\sigma,z) = A_{\rm T}\left(\left(\frac{b}{\sigma}\right)+1\right)\exp\left[-\frac{c}{\sigma^2}\right],
\end{eqnarray}
with best-fit parameters, $A_{\rm T}=0.186(1+z)$, $a=1.47(1+z)^{-0.06}$, $b=0.3(1+z)^{-\zeta}$, $c=1.19$, and $\zeta=\exp\left[-\left(\frac{0.75}{\Delta_{\rm vir}/75}\right)\right]$, where $\Delta_{\rm vir}$ is the overdensity with respect to $\rho_{0}$ within the sphere of radius $R_M$. We adopt $\Delta = 500$ to compare with the mass of the sample clusters in radio observations (Sect.~\ref{sec:lifetime}). \par

\subsection{Monte Carlo Simulation \label{method:MCS}}
For the initial state, we prepare $N=4,000$ halos at $z=0$ in the mass range of $[10^{13}~M_\odot,10^{16.5}~M_\odot]$.  To realize the mass distribution consistent with the HMF of \citet{Tinker_2008ApJ...688..709}, each halo is weighted with the HMF. \par

In our MC algorithm, the halo mass decreases through the mergers and the accretion as the time goes back to higher redshifts. The occurrence of a merger with each $\xi$ in each time step is simulated with a random uniform number ${\cal R}$ in the interval 0 to 1. When ${\cal R}<{\cal P}(M_0,\xi,z)$, the descendant halo with mass $M_0$ split into two progenitor halos with masses of $M_1 = M_0/(1+\xi)$ and $M_2 = \xi M_0/(1+\xi)$. This criterion is adopted for each $\xi$ in each time step. When ${\cal R}\ge{\cal P}(M_0,\xi,z)$ for all $\xi$, the mass decreases only through the accretion. \par


In very rare cases, two or more mergers occur in one time step with different $\xi$. In this case, we choose $M_1$ as the mass of the most massive progenitor and $M_2$ as the sum of the rest, and redefine the mass ratio of the event as $\xi=M_2/M_1$. For example, in an event where a descendant halo with a mass of $M_0$ split into three haloes with $M_{\rm a}>M_{\rm b}>M_{\rm c}$,
we set $M_1=M_{\rm a}$ and $M_2=M_{\rm b}+M_{\rm c}$.
The ratio $\xi=M_2/M_1$ can be larger than unity in this case.
\par


The boundary between the merger and accretion is ambiguous, so that the simple summation of $\Delta M = M_2+\Delta M_{\rm median}$, where $\Delta M_{\rm median}=\langle dM/dz \rangle_{\rm median}\times\Delta z$, would overestimate the mass evolution in each time step in out computation.
Thus, we set $\Delta M = \Delta M_{\rm median}$ for $M_2<\Delta M_{\rm median}$, considering that the contribution of frequent minor mergers are effectively included in $\Delta M_{\rm median}$. When $M_2>\Delta M_{\rm median}$, the halo mass at $z+\Delta z$ is chosen to be $M_1=M_0-M_2$ to account for the abrupt growth due to the major merger. Our code also follows the evolution of sub-progenitors with masses larger than $10^{13}~h^{-1}M_\odot$, although those trees are not used in the following analysis. When multiple mergers occur in one time step, although such event hardly occurs, two or more sub-progenitors are produced in that step.


\begin{figure*}
    \centering
    \plottwo{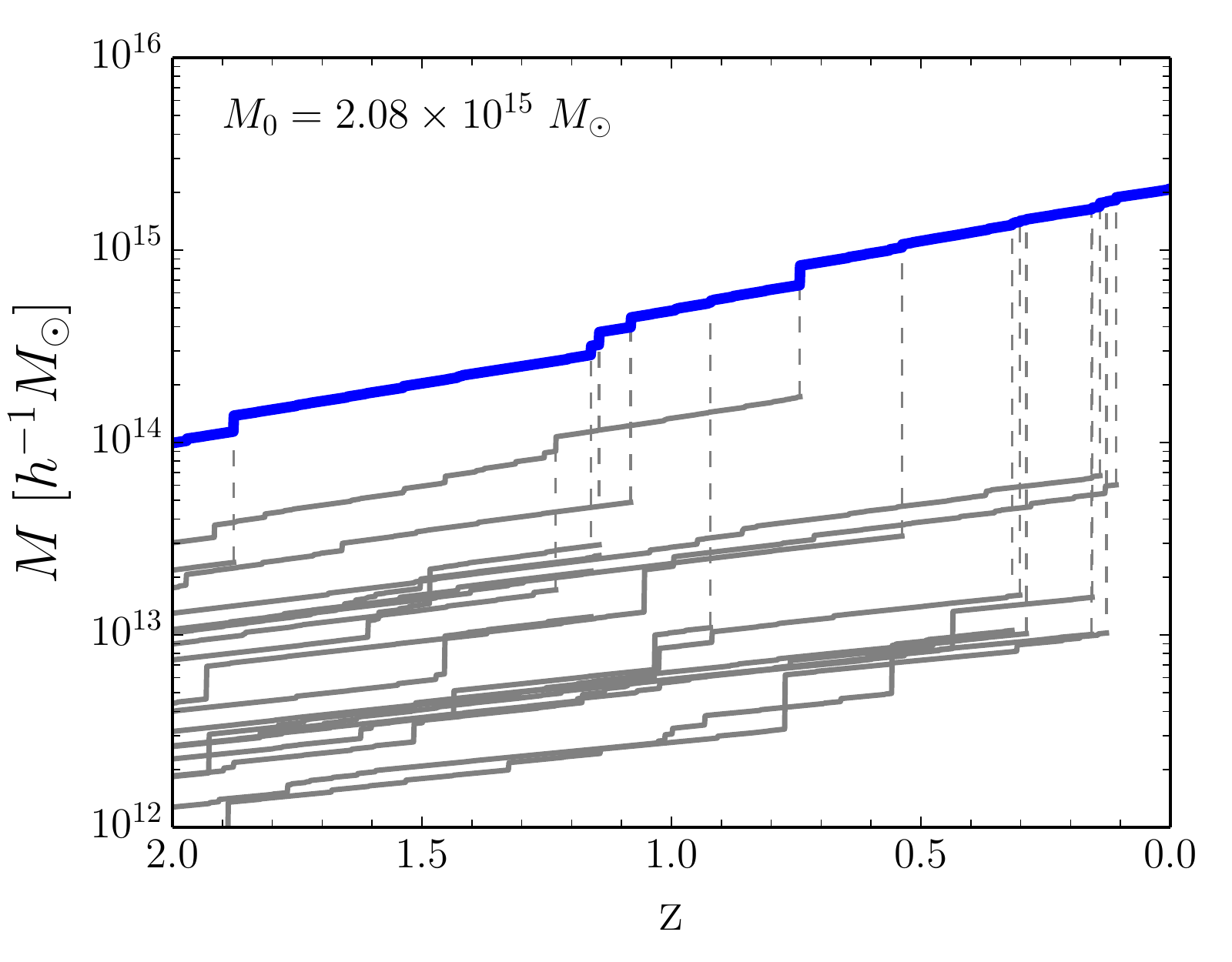}{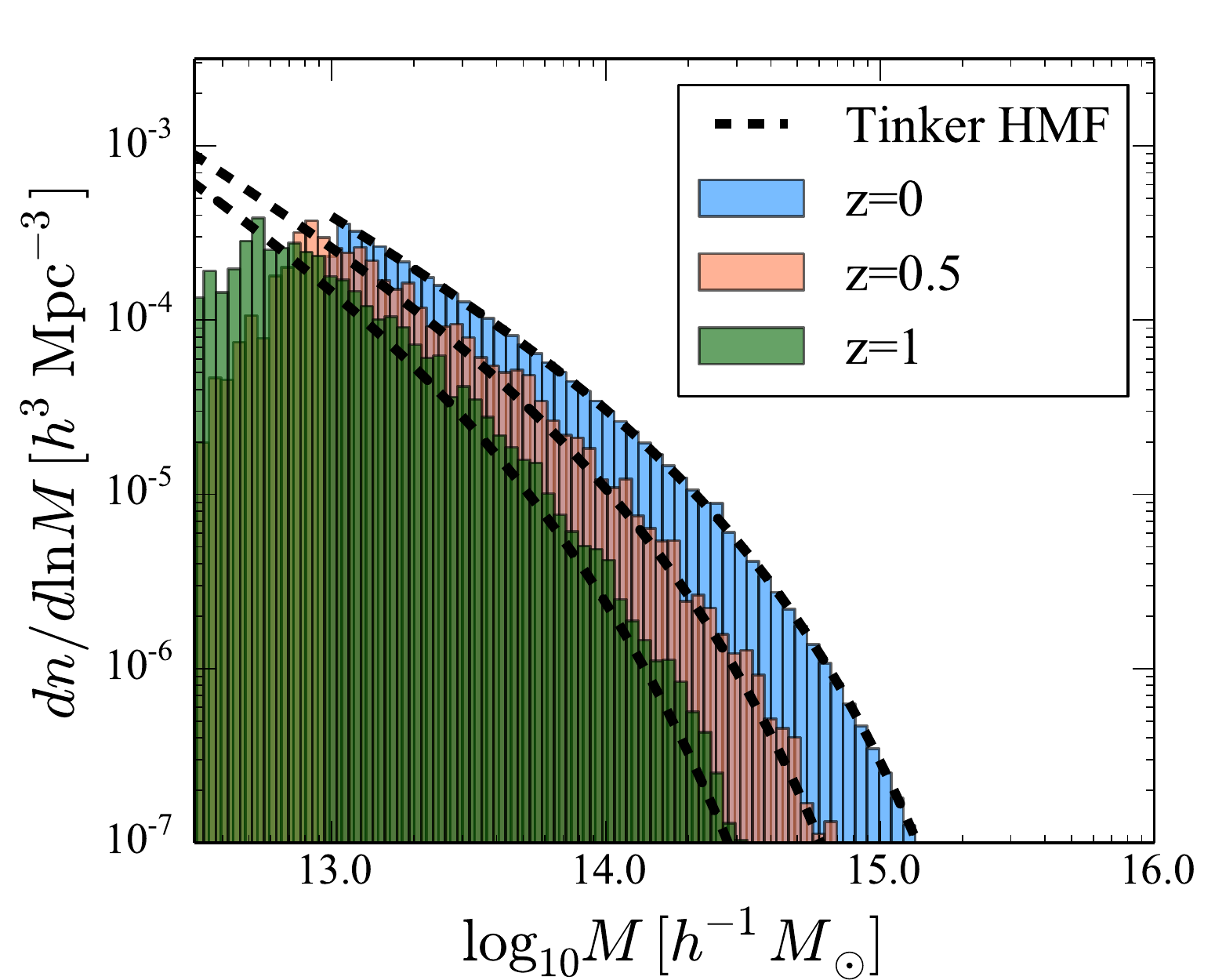}
    \caption{Left: An example of MC merger tree. The blue line shows the evolution of the most massive progenitor, while the gray lines correspond to the sub-progenitors. The merger between progenitors are expressed with the dashed lines. Right: HMFs constructed with the Monte Carlo merger trees (histogram) compared with the fitting function by \citet{Tinker_2008ApJ...688..709} (dashed lines). The HMF at $z =$0.0 (blue), 1.0 (orange), 2.0 (green) are shown.}
    \label{fig:HMF}
\end{figure*}

We follow the mass evolution of each halo up to $z=2$. An example of the MC merger tree is shown in Figure~\ref{fig:HMF} (left). 
Figure~\ref{fig:HMF} (right) compares the HMF obtained from our MC merger tree with the fitting form of \citet{Tinker_2008ApJ...688..709}.
Starting from the HMF given by \citet{Tinker_2008ApJ...688..709} at $z=0$ (blue), the HMF evolution shows a good agreement with the fitting formula at higher redshifts. Our method slightly (a factor of $\lesssim2$) overproduce halos at a higher redshift ($z>1.0$). However, since galaxy clusters above $M>10^{14}M_\odot$ are mostly formed at a low redshift, that difference does not affect our discussion in the following sections. Although better match at higher redshifts would be achieved by tuning the parameters appearing in Eqs.~(\ref{eq:Mergerrate}) and (\ref{eq:medianMAR}), we adopt the same parametrization as \citet{Fakhouri_2010MNRAS.406.2267F} for simplicity. \par

\section{Statistical study on the Merger-RH connection \label{sec:Statis}}
In this section, we study the connection between the halo merger and the occurrence of RHs, using the MC merger tree (Section~\ref{sec:MT}). The RH survey observations have revealed that only a fraction of clusters host RHs, and they are preferentially found in merging systems \citep[e.g.,][]{Buote_2001,Cassano2010}. \par

We assume two conditions for the RH onset: mass-ratio condition and break-frequency condition (Section~\ref{subsec:criterion}). Based on the RH lifetime obtained in Section~\ref{sec:lifetime}, we model the mass dependence and time evolution of the radio power in Section~\ref{subsec:radio_power}. We then calculate the luminosity evolution of each cluster between $0\leq z\leq 2$. A cluster is categorized as a RH only when the radio power at the observation epoch is larger than the minimum value for the detection (Section~\ref{subsec:obs_limit}). In Section~\ref{subsec:occurrence}, we discuss the constraints on model parameters by calculating the fraction of clusters with RHs, $f_{\rm RH}$, and comparing it with the observation \citep[][]{2021A&A...647A..50C}.

The mass of clusters in those observations is defined with the over-density with respect to the cosmic critical density, while our MC merger tree refer to the mean mass density. We multiply a factor of $\Omega_{\rm m}^{1/2}$ to convert the mass in merger tree into the observational mass \citep[see also][]{Ensslin_2002}. \par 

\subsection{Criteria for the onset \label{subsec:criterion}}
We introduce two conditions for the onset of the radio emission in the ICM. The first one is called {\it mass-ratio condition}, which is a simplified approach using a threshold for the merger mass ratio, $\xi_{\rm RH}$, above which mergers can ignite the reacceleration. We treat $\xi_{\rm RH}$ as a model parameter and do not include its dependence on the total mass or redshift. \par


The purpose of this simplified model is to discuss the relation between the RH lifetime (Section~\ref{sec:lifetime}) and the typical mass ratio for the onset that can explain the observed fraction of RHs $f_{\rm RH}$. Since the merger rate (Eq.~(\ref{eq:Mergerrate})) decreases with $\xi$, we can expect that the model with longer lifetime requires larger $\xi_{\rm RH}$ \citep[][]{Cassano_2016}. \par






The other condition, the {\it break-frequency condition}, is more sophisticated approach based on the reacceleration model. The reacceleration of CREs balances with the radiative cooling at a certain energy scale, which causes a spectral steepening at the break frequency $\nu_{\rm b}$. That steepening is actually seen in both the secondary and primary scenarios (Figure~\ref{fig:radio_flux_off}). In this {\it break-frequency condition}, $\nu_{\rm b}$ need to be large enough to reproduce typical power and spectral index of RHs. \par


To calculate the efficiency of the reacceleration, we introduce a parameter $\eta_{\rm t}$, which is the fraction of the turbulent energy to the merger kinetic energy. The formalization in Appendix~\ref{app:Dpp} is used to calculate the reacceleration efficiency, or its timescale, $t_{\rm acc}$, of each merger event as a function of $\eta_{\rm t}$, $M$ and $\xi$.\par



In this model, we firstly set $\nu_{\rm b}$ around 1.4 GHz. The case with different $\nu_{\rm b}$ will be discussed in Section~\ref{subsec:break-frequency_condition}. A merger can ignite the emission, whose spectral feature matches with the typical observed one, only when the acceleration is efficient enough to balance or overcome the cooling at a given $\nu_{\rm b}$, i.e., $t_{\rm acc}\le t_{\rm cool}|_{\nu=\nu_{\rm b}}$. The cooling time scale due to synchrotron and IC radiation in the Thomson limit is approximated as \citep[e.g.,][]{Rybicki:847173,Brunetti2014}
\begin{eqnarray}\label{eq:tcool}
t_{\rm cool} &=& \gamma_{\rm b}m_ec^2\left[\left(\frac{2}{3}\right)^2\sigma_{\rm T}(\gamma_{\rm b}-1)^2\left(u_B+u_{\rm CMB}\right)\right]^{-1},\nonumber\\
&\approx& 360~{\rm Myr}\left[\left(\frac{\gamma_{\rm b}}{10^4}\right)\left\{\left(\frac{B}{3.2~\mu{\rm G}}\right)^2+(1+z)^4\right\}\right]^{-1},
\end{eqnarray}
where $\sigma_{\rm T}$ is the Thomson cross section, $u_{B}=B^2/8\pi$ and $u_{\rm CMB}=0.262~{\rm eV/cm^3}(T_{\rm CMB}/2.73~{\rm K})^4(1+z)^4$ are the energy densities of the magnetic field and the cosmic microwave background (CMB), respectively, and the Lorentz factor $\gamma_{\rm b}$ is related to $\nu_{\rm b}$ as $\nu_{\rm b} = \frac{3eB}{4\pi m_{\rm e}c}\gamma_{\rm b}^2$. For simplicity, we fix $B=3.2~\mu{\rm G}$ and $t_{\rm cool}$ is treated as a function of $z$ and $\nu_{\rm b}$. The acceleration timescale becomes shorter with a larger $\eta_{\rm t}$, which leads to a larger $f_{\rm RH}$ for a fixed $\nu_{\rm b}$. \par

For the secondary scenario, we do not take into account the {\it break-frequency condition}. The spectrum of CRPs has no cooling break in low-energy region. As seen in Section~\ref{subsec:lifetime_secondary} (left panel of Figure~\ref{fig:radio_flux_off}), a significant part of the radio emission at 1.4 GHz is powered by the enhanced electron injection from CRPs. This makes the break feature less significant compared to the primary scenario. To simplify the following discussion, we apply {\it break-frequency condition} only to the primary scenario. \par

\subsection{Power of the radio halo \label{subsec:radio_power}}
One of the main objectives of this paper is to calculate the luminosity function of RHs (RHLF). Combining the HMF and the criteria proposed in Section~\ref{subsec:criterion}, one can obtain the mass function of RHs. \citet{Ensslin_2002} used the empirical relation between the radio power and the cluster mass to transform the HMF into the RHLF. \par

The important improvement in our method is to include the time evolution of the radio luminosity discussed in Section~\ref{sec:lifetime}. Although the calculation in Section~\ref{sec:lifetime} was done only for the Coma-like RH, we have adopted a similar luminosity evolution for different masses. We assume a simple power-law relation between the maximum radio power at the end of the reacceleration period and $M_{500}$ like the previous study in \citet[][]{Ensslin_2002}. \par




With those assumptions, the radio luminosity at 1.4 GHz is expressed as a function of both mass and time. We write it as $P_{1.4}(M_{500},t_{\rm R})$, where $t_{\rm R}$ is the time interval between the RH onset and the observation. We do not consider the possible dependence on $\xi$ for $P_{1.4}$, although this parameter could affect the efficiency of the reacceleration \citep[][]{Cassano_2005}. This simplification may be justified, if the merger rate is a steep function with respect to $\xi$ and a large fraction of RHs arise from mergers with a value close to the lower limit $\xi_{\rm RH}$. \par


Since both the mass and the luminosity evolve with time, it is important to distinguish $P_{1.4}-M_{500}$ relation at the onset from that at the observation. The timescale of mass increment can be evaluated from Eq.~(\ref{eq:medianMAR}), and it is about 5 Gyr for $M_{500}\approx10^{15}M_\odot$. If the RH lifetime is much shorter than this timescale, as in the primary scenario, the mass evolution after the onset is almost negligible. However, that can be not the case for the secondary scenario, since the emission can last for the cosmological timescale ($\approx 10~{\rm Gyr}$). We 
adopt a power-law relation between the peak luminosity and the mass at the onset as $P_{1.4}(t_{\rm R}^{\rm peak})\propto M_{500}^{\alpha_M}$, which could satisfy the observed $P_{1.4}-M_{500}$ relation, neglecting the mass evolution during the reacceleration phase ($\approx 500~{\rm Myr}$). Thus, $P_{1.4}(M_{500},t_{\rm R})$ can be expressed as
\begin{eqnarray}\label{eq:P14_M}
\left(\frac{P_{1.4}(M_{500},t_{\rm R})}{10^{24.5}~{\rm W/Hz}}\right) =10^{A_{\rm RH}}\left(\frac{M_{500}}{10^{14.9}M_\odot}\right)^{\alpha_{M}}f(t_{\rm R}),
\end{eqnarray}
where the function $f(t_{\rm R})$ satisfies $f(t_{\rm R}^{\rm peak})=1$, so that the factor $10^{A_{\rm RH}}\left(M_{500}/10^{14.9}M_\odot \right)^{\alpha_{M}}$ corresponds to the peak luminosity for the descendant mass $M_{500}$. The overall normalization $A_{\rm RH}$ is considered to be a constant, and treated as a model parameter.
\par

\begin{figure*}
    \centering
    \plottwo{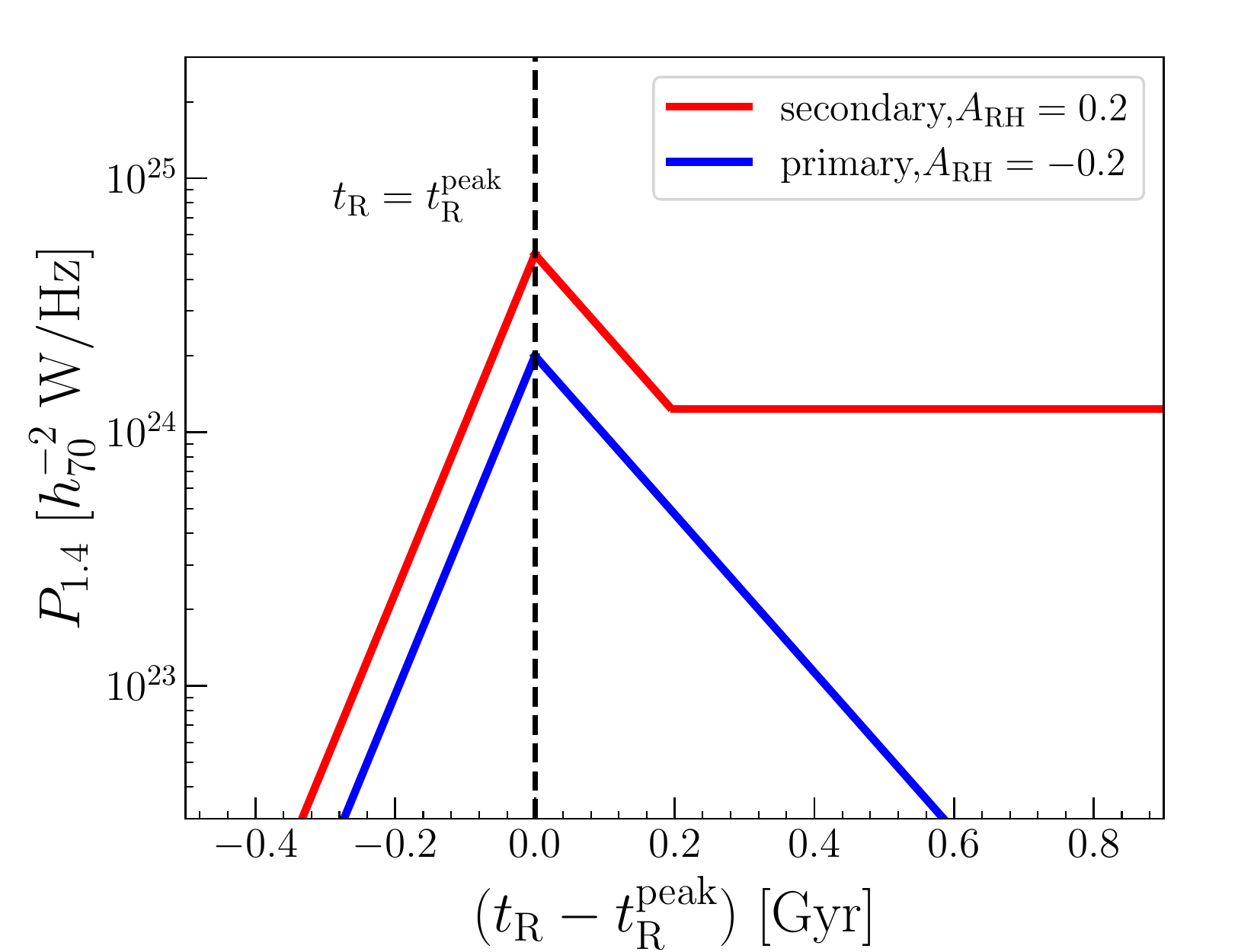}{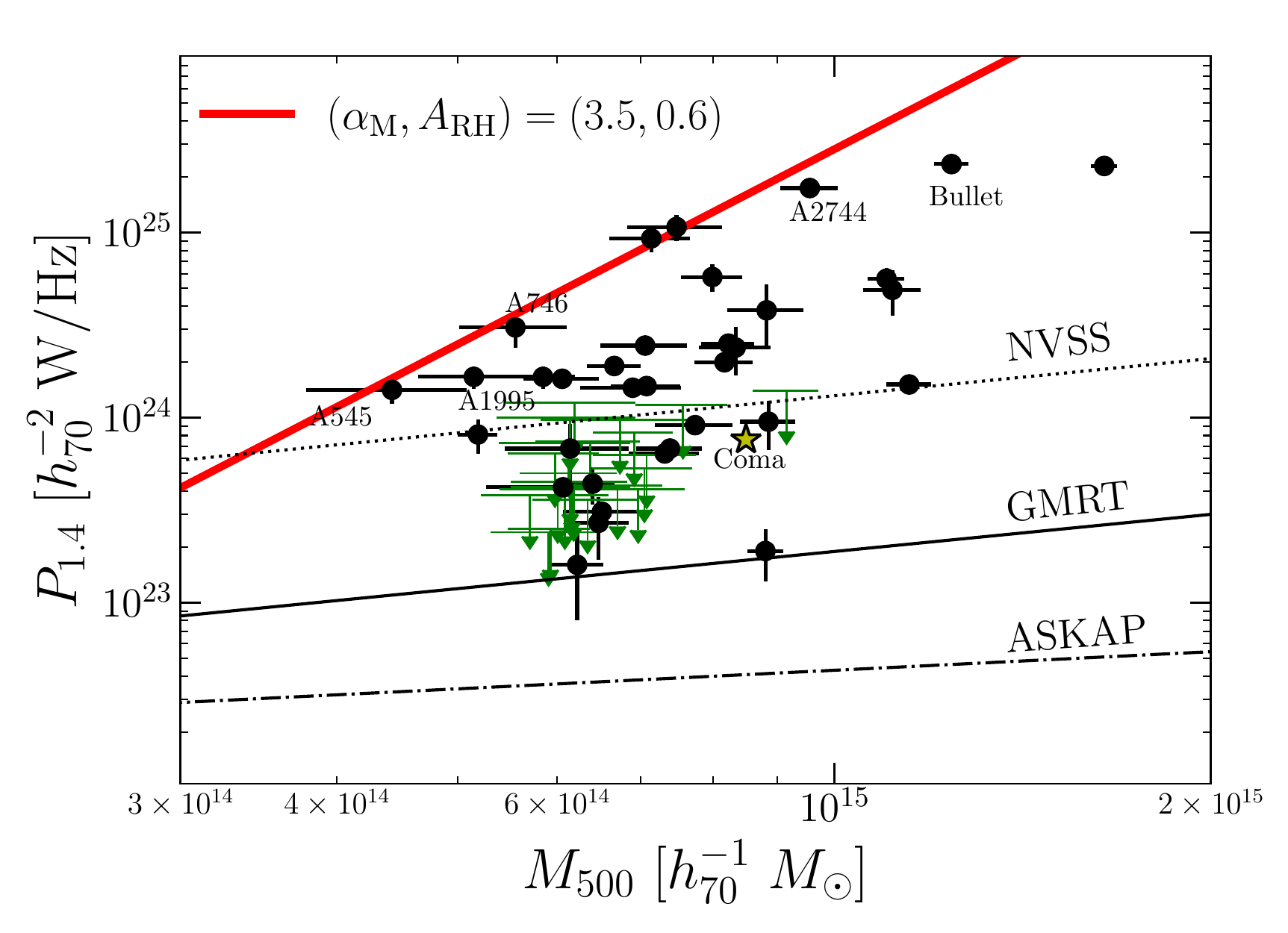}
    \caption{Mass and time dependence of the RH luminosity modeled in Section~\ref{subsec:occurrence}.  Left: The solid lines show the evolution model of the RH luminosity, which are expressed with Eqs.~(\ref{eq:P14_sec}) and (\ref{eq:P14_pri}) for the secondary (red) and the primary (blue) scenarios, respectively. We assumed $M_{500}=10^{14.9}~M_\odot$ for both the lines, while different values for $A_{\rm RH}$ are adopted to improve the visibility. The vertical dashed line shows the peak time.
    Right: The red solid line shows the mass dependence of the peak luminosity. The black lines show the sensitivity limits of observations. The mean redshift $z=0.2$ is adopted to convert the flux limit into the luminosity limit. The observational limits for NVSS and GMRT (dotted and solid) are calculated with Eq.~(\ref{eq:fmin}). The dot-dashed line shows the sensitivity of ASKAP survey (Eq.~(\ref{eq:fmin_2})). Data points are taken from \citet{2021A&A...647A..50C}. The Coma RH is shown with the yellow star.}
    \label{fig:P_M_onset}
\end{figure*}





As shown in Figure \ref{fig:P_M_onset}, we adopt the temporal evolution of RHs extrapolated from the modeling of the Coma-like RH. We have assumed in Section~\ref{sec:lifetime} that the Coma-like RH is in the midst of the reacceleration phase and the peak luminosity has not been achieved. The maximum duration of the reacceleration phase was assumed to be $t_{\rm R}^{\rm peak}\approx500~{\rm Myr}$, which is close to the timescale of the turbulent cascade at the $\approx~500{\rm kpc}$ scale (Section~\ref{sec:origin_of_PM}). The peak luminosity of the Coma-like RH becomes almost two orders of magnitude larger than the luminosity at the pre-acceleration state.

In the secondary scenario, such a high luminosity is sustained even in the cooling phase (Figure~\ref{fig:radio_flux_off}). We take into account the decline (a factor of $\approx4$ at 1.4 GHz) in the cooling phase. The time evolution of the luminosity, $f(t_{\rm R})$, can be modeled as
\begin{eqnarray}\label{eq:P14_sec}
f(t_{\rm R})=
\left\{
\begin{array}{cc}
    10^{2.0\left(\frac{t_{\rm R}-t_{\rm R}^{\rm peak}}{t_{\rm R}^{\rm peak}}\right)} & (t_{\rm R}-t_{\rm R}^{\rm peak}<0), \\
    \exp\left[-\frac{t_{\rm R}-t_{\rm R}^{\rm peak}}{t_{\rm cool}(z)}\right] & (0<t_{\rm R}-t_{\rm R}^{\rm peak}<t_{1}), \\
    \exp\left[-\frac{t_1-t_{\rm R}^{\rm peak}}{t_{\rm cool}(z)}\right] & (t_1<t_{\rm R}-t_{\rm R}^{\rm peak}),
\end{array}
\right.
\end{eqnarray}
where $t_{\rm cool}(z)$ is the cooling timescale at 1.4 GHz (Eq.~(\ref{eq:tcool})), and $t_1=1.4t_{\rm cool}(z)$ corresponds to the time at which the luminosity becomes $\approx4$ times smaller than the peak value, i.e., $f(t_{1})=0.25$. Since the Coulomb cooling is negligible at higher energies ($\gamma>10^3$), $t_{\rm cool}$ is calculated from Eq.~(\ref{eq:tcool}), for which we substitute the Lorentz factor $\gamma_{\rm b}$ corresponding to 1.4 GHz. As done in Section~\ref{subsec:criterion}, we fix $B=3.2~\mu{\rm G}$ for the magnetic field. In this case, $t_{\rm cool}$ only depends on $z$. The luminosity is constant after $t_{\rm R}-t_{\rm R}^{\rm peak}>t_1$. \par

In the primary scenario, the RH would revert to the radio-quiet state within a few times the cooling timescale.
We assume an exponential decay after the peak time as
\begin{eqnarray}\label{eq:P14_pri}
f(t_{\rm R})=
\left\{
\begin{array}{cc}
    10^{2.0\left(\frac{t_{\rm R}-t_{\rm R}^{\rm peak}}{t_{\rm R}^{\rm peak}}\right)} & (t_{\rm R}-t_{\rm R}^{\rm peak}<0), \\
    \exp\left[-\frac{t_{\rm R}-t_{\rm R}^{\rm peak}}{t_{\rm cool}(z)}\right] & (0<t_{\rm R}-t_{\rm R}^{\rm peak}), 
\end{array}
\right.
\end{eqnarray}

When another merger occurs after one merger event, we compare the two values of luminosity induced by each merger and choose the larger one as the luminosity at a given epoch.\par

The relation between the peak ($t_{\rm R} =t_{\rm R}^{\rm peak}$) luminosity and the mass is shown in the left panel of Figure~\ref{fig:P_M_onset}. As a reference, we adopted $(\alpha_{M},A_{\rm RH})=(3.5,0.6)$. Those values are constrained from the observed fraction of RHs, and their number counts per observed flux (see Section~\ref{subsec:occurrence}). The observed power of RHs and the upper limits listed in the extended sample of \citet{2021A&A...647A..50C} are plotted with black points and green arrows, respectively. The black lines are the sensitivity limits for NVSS and ASKAP at $z=0.2$ (Section~\ref{subsec:obs_limit}). The luminosities of some RHs, such as Abell 545, Abell 1995, and Abell 2744 are close to the assumed peak luminosity for their mass, indicating that they are in the transition from the reacceleration phase to the cooling phase. \par

In the right panel of Figure~\ref{fig:P_M_onset}, we plot the assumed evolution of Eqs.~(\ref{eq:P14_sec}) and (\ref{eq:P14_pri}). We have adopted $t_{\rm cool}$ at $z=0$ for the primary scenario. \par

As discussed in Section~\ref{subsec:lifetime_primary}, the CR injection rate does not revert to that at the pre-reacceleration state, when a certain amount of CRPs are injected as primaries. However, we have neglected this effect in Eq.~(\ref{eq:P14_pri}) in order to treat each merger event independently.

\subsection{Observational limit \label{subsec:obs_limit}}
To estimate the RH number expected in the previous and future radio surveys, we need the minimum flux detectable in those surveys.
We adopt the brightness-based criterion of 
\citet{Cassano_2010_LOFAR} for the observational limit in the NVSS survey. In this case, the minimum flux is estimated as,
\begin{eqnarray}\label{eq:fmin}
f_{\rm min} =&& 1.2\times10^{-4}~{\rm mJy}~\zeta_1 \left(\frac{F_{\rm rms}}{{\rm 10\mu Jy}}\right) \nonumber \\
&&\times \left(\frac
{10~{\rm arcsec}}{\theta_{b}}\right)^{2}\left(\frac{\theta_{\rm H}}{{\rm arcsec}}\right)^2,
\end{eqnarray}
where $\theta_{\rm b}$ and $\theta_{\rm H}$ are the angular size of the beam and the source, respectively, and $F_{\rm rms}$ is the rms noise per beam. Following \citet{Cassano_2012}, we assume $\zeta_1\approx3$, and $(F_{\rm rms}, \theta_{\rm b})=(0.45~{\rm mJy}, 45~{\rm arcsec})$ for the rms noise and the beam size of the NVSS survey. We adopt $\theta_{\rm H}=0.35\theta_{500}$ as the typical size of RHs, where $\theta_{500}$ is the angular size corresponding to $R_{500}$. \par

In Section~\ref{subsec:occurrence}, we compare our calculation to the data given by \citet[][]{2021A&A...647A..51C}. The data includes the RHs observed with GMRT at 330 MHz and 610 MHz. Although our method only follows the luminosity at 1.4 GHz, we also consider the sensitivity of those low frequency observations. In this study, we simply scale the sensitivity at 610 MHz to 1.4 GHz, assuming a spectral index of $\alpha_{\rm syn}=-1.2$. We adopt $F_{\rm rms}$ and $\theta_{\rm b}$ for GMRT 610 MHz provided in \citet[][]{Brunett_Venturi_2007ApJ...670L...5B}, which leads to ($F_{\rm rms}$, $\theta_{\rm b}$)=($19~\mu{\rm Jy}$, $25~{\rm arcsec}$) at 1.4 GHz. Note that the value of $F_{\rm rms}$ in \citet[][]{Brunett_Venturi_2007ApJ...670L...5B} is similar to the ones in the actual GMRT survey \citep[][]{Venturi_2007A&A...463..937V,2013A&A...557A..99K}.  \par




The flux of each RH is evaluated from $P_{1.4}$ as $f_{1.4}=(1+z)^{1-\alpha_{\rm syn}}P_{1.4}/(4\pi D_L(z)^2)$, where $D_L(z)$ is the luminosity distance at redshift $z$ and $\alpha_{\rm syn}$ is the spectral index of the RH (Section~\ref{sec:intro}). For simplicity, we fix $\alpha_{\rm syn} = -1.2$ for all RHs in our merger tree. \par

In the right panel of Figure~\ref{fig:P_M_onset}, the observational limits for NVSS and GMRT are shown with the dotted and solid lines. The mean redshift of the sample of \citet{2021A&A...647A..51C}, $z=0.2$, is adopted to convert the flux into luminosity. The figure shows that the NVSS limit is comparable to the upper limits in the sample (green arrows), while the GMRT limit is lower than the upper limits and the power of detected halos (black points). Note that the redshift was fixed only for the visualization purpose. In the following calculation, the flux of each RH is calculated depending on its redshift. \par







In Section~\ref{sec:ASKAP}, we use the flux-based criterion to estimate the number count in the ASKAP survey. In this case, $f_{\rm min}$ can be written as
\begin{eqnarray}\label{eq:fmin_2}
f_{\rm min} =1.43\times10^{-3}~{\rm mJy}~\zeta_2\left(\frac{F_{\rm rms}}{10~{\rm \mu~Jy}}\right)\nonumber\\
\times \left(\frac{10~{\rm arcsec}}{\theta_{\rm b}}\right)\left(\frac{\theta_{H}}{\rm arcsec}\right).
\end{eqnarray}
Following \citet{Cassano_2012}, we adopt $\zeta_2\approx10$ and $(F_{\rm rms},\theta_{\rm b})=(10~\mu{\rm Jy},25~{\rm arcsec})$ for the ASKAP survey. The apparent halo size is assumed to be $\theta_{\rm H}=0.35\theta_{500}$. The radio power at $z=0.2$ corresponding this limit is shown with the dash-dotted line in Figure~\ref{fig:P_M_onset}.

\subsection{Occurrence of RHs \label{subsec:occurrence}}
\subsubsection{mass-ratio condition \label{subsec:mass-ratio_condition}}
First, we discuss the case where the onset is triggered by the condition $\xi>\xi_{\rm RH}$. In this model, we have three model parameters: $\xi_{\rm RH}$, $\alpha_{M}$, and $A_{\rm RH}$. The observed fraction of RHs and the luminosity function (Section~\ref{subsec:RHLF}) give constraints on those parameters. \citet{2021A&A...647A..50C} reported that the fraction of RHs, $f_{\rm RH}$, is $\approx0.7$ in the high-mass (HM) bin ($8.0\times10^{14}<M_{500}<12.0\times10^{14}$), while it drops to $\sim0.35$ in the low-mass (LM) bin ($5.7\times10^{14}<M_{500}<8.0\times10^{14}$) in a sample of clusters from the {\it Planck} SZ catalogue. When the small halos and the ultra-steep spectrum radio halos (USSRHs) \citep[e.g.,][]{Brunetti_2008Natur.455..944B} are excluded from the statistic, $f_{\rm RH}\approx0.33$ in the HM bin and $f_{\rm RH}\approx0.23$ in the LM bin. \par

We adopt a Monte Carlo procedure to calculate the fraction $f_{\rm RH}$ and its statistical error in our MC merger tree. We randomly extract a sub-sample of clusters from our merger tree, and calculate $f_{\rm RH}$ as a ratio between radio-loud and radio-quiet cluster. The size of the sub-sample is taken to be the same as the observation \citep[][]{2021A&A...647A..51C}, i.e., 60 clusters for the LM bin and 15 clusters for the HM bin. Those sub-samples are parts of the population with masses in the LM or HM bin at some time within the sampling range of redshift ($0.088<z<0.33$). If the radio flux of a halo, calculated with Eq.~(\ref{eq:P14_sec}) or Eq.~(\ref{eq:P14_pri}), is larger than the observational limit (Eq.~(\ref{eq:fmin})) at a given redshift, the halo is counted as a radio-loud one, while halos with a flux below the above criteria are regarded as radio-quiet one. The observation redshift ($0.08<z<0.33$) is divided into 50 bins, and we calculate the mean value of $f_{\rm RH}$, weighting each redshift bin equally. Note that we have calculated the luminosity evolution over the whole redshift range of the merger tree ($0\leq z\leq 2$). The time interval $t_{\rm R}$ in Eq.~(\ref{eq:P14_sec}) or (\ref{eq:P14_pri}) is measured from the most recent merger that satisfies the condition $\xi>\xi_{\rm RH}$ before the observation.  \par


The observed fraction could be reproduced in our merger tree by tuning the model parameters. Since the merger rate is larger for smaller $\xi$ (Eq.~(\ref{eq:Mergerrate})), one can expect more RHs for smaller $\xi_{\rm RH}$. 
The parameter $A_{\rm RH}$, which regulates the typical radio flux, also affects the RH fraction. \par

\begin{figure*}
    \centering
    \plottwo{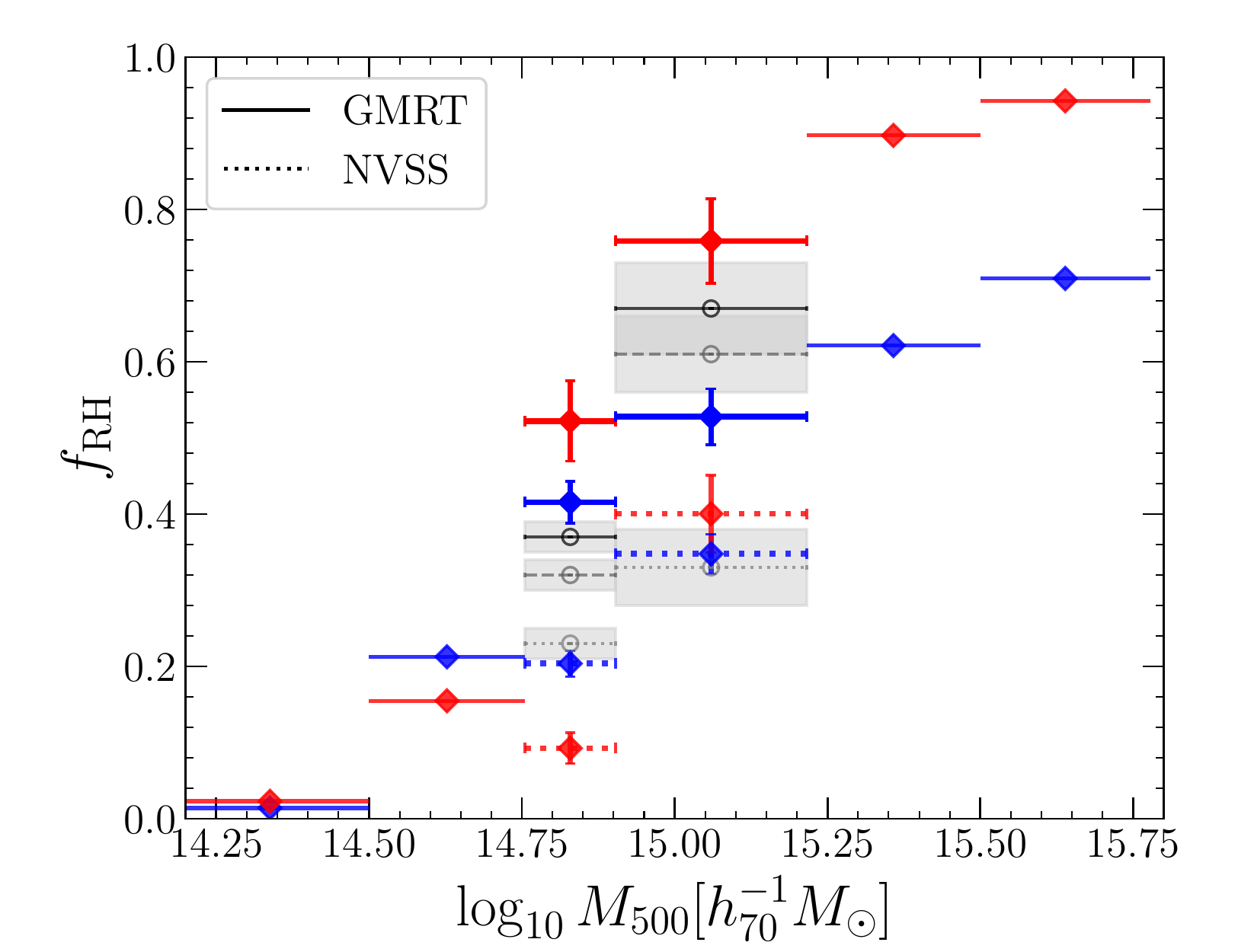}{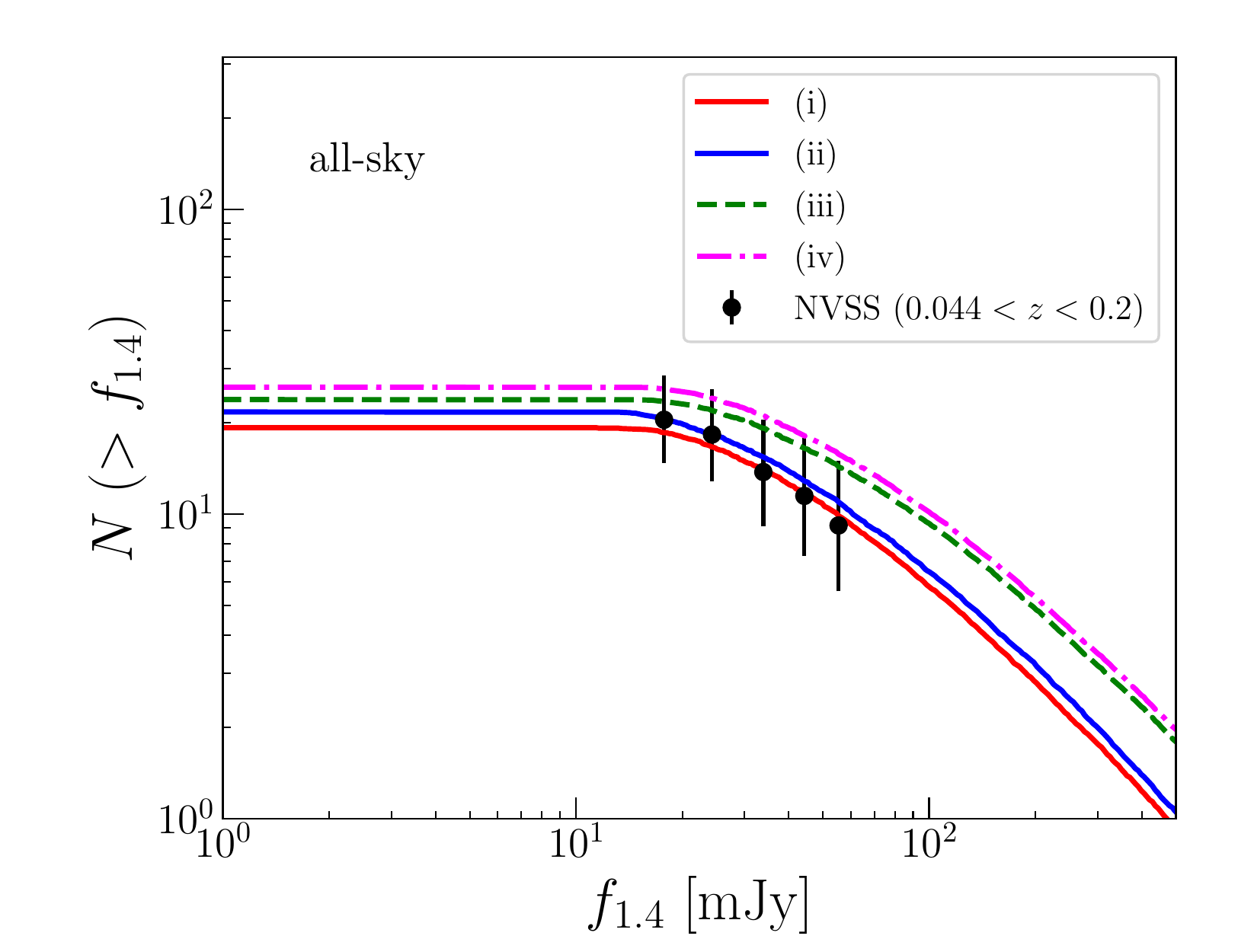}
    \caption{Left: The fraction of RH clusters in six mass bins averaged within $0.088<z<0.33$ (red: secondary scenario, blue: primary scenario). The errors in the models ($1\sigma$) due to statistical fluctuation are given by $N=100$ trials of the MC sampling. The gray lines show the observed fraction for three cases: small halos and USSRHs are counted as RHs (solid), USSRHs are considered as RHs but small haloes are considered as non-RHs (dashed), and both USSRHs and small halos are considered as non-RHs (dotted), adopted from \citet{2021A&A...647A..50C}. The shaded region show the 1$\sigma$ error. See the text for further explanations. Right: Cumulative number count of RHs observable with the NVSS sensitivity for the different models listed in Table~\ref{tab:fRH_params}. The black points show the NVSS number count within $0.04<z<0.2$ for the all-sky, adopted from \citet{Cassano_2012}. }
    \label{fig:fRH}
\end{figure*}

For any $\xi_{\rm RH}$, we can find a parameter sets of $(\alpha_{M}, A_{\rm RH})$ to reproduce the observed $f_{\rm RH}$ and its mass dependence. However, there should exit a typical mass for a given luminosity of RHs, considering the observed RHLF and the MF of clusters. For example, in the models with $A_{\rm RH}\ge1.0$ and $\alpha_{M}=3.5$, even low-mass RHs with $M_{500}\le5\times10^{14}~M_\odot$ can be observable with the NVSS sensitivity, and such a model overproduces the RH number count (Figure~\ref{fig:fRH}). Thus, the parameter $A_{\rm RH}$ is constrained to be $A_{\rm RH}\approx0.5$.\footnote{A larger value of $A_{\rm RH}$ is preferable for a larger $\alpha_{M}$.} \par

If the mass dependence of $f_{\rm min}$ Eq.~(\ref{eq:fmin}) is significantly weak, the assumed peak luminosity-mass relation naturally leads to the positive mass dependence of $f_{\rm RH}$ (see Figure~\ref{fig:P_M_onset}). However, the observed mass dependence is not significant enough to give a definite constraint on the parameters. We temporary adopt $\alpha_{M}$ similar to the one adopted from the fitting to the observed RH, i.e. $\alpha_M=2.5-4.0$. This ensures a reasonable fit to the observed number count (Section~\ref{subsec:RHLF}). In Section~\ref{sec:origin_of_PM}, we show that a similar mass dependence can be reproduced by calculating the peak luminosity for clusters with various masses. On the other hand, the threshold $\xi_{\rm RH}$ does not largely affect the mass dependence of $f_{\rm RH}$, since the merger rate has a weak mass dependence. \par

\begin{table}[hbt]
    \centering
    \caption{Model parameters compatible with the observed RH fraction}
    \begin{tabular}{lcccc}
    \hline
    model  &(i)&(ii)&(iii)&(iv) \\
    scenario&secondary&primary& primary & primary \\ \hline
     $(\alpha_{M},A_{\rm RH})$& (3.5,0.6) & (3.5,0.6) & (4.0, 0.2) & (4.0, 0.2)\\
    $\xi_{\rm RH}$ & 0.12  & 0.01 & --- & --- \\
    $\nu_{\rm b}$ & --- & --- & 1.4 GHz & 500 MHz \\
    $\eta_{\rm t}$ & --- & --- & 0.65  & 0.5 \\
    $\langle f_{\rm RH}\rangle$\tablenotemark{a} & 0.57 & 0.44 & 0.21  & 0.22 \\
    \hline
    \end{tabular}
    \flushleft
   \tablecomments{$^a$The fraction $f_{\rm RH}$ for the GMRT sensitivity in LM and HM bins are weighted with the sample size of each bin.}
    \label{tab:fRH_params}
\end{table}

We search for the best fit value of $\xi_{\rm RH}$ for given $(\alpha_{M},A_{\rm RH})$ in the range discussed above, and find $\xi_{\rm RH}\approx0.1$ and $\xi_{\rm RH}\approx0.01$ for the secondary and the primary models, respectively. The values of the three parameters, constrained from both the observed $f_{\rm RH}$ and the number count (Figure~\ref{fig:fRH}), are listed in Table~\ref{tab:fRH_params}. Here, we refer the secondary and the primary models as model (i) and (ii), respectively. To study the impact of different $\xi_{\rm RH}$ and $P_{1.4}(M_{500},t_{\rm R})$ on the observable quantities, we adopt the same $ (\alpha_{M},A_{\rm RH})=(3.5,0.6)$ for both the models. Possible constraints from the correlation between the RH occurrence and the X-ray morphological disturbance will be discussed in Section~\ref{sec:X_dynamics}.\par

In Figure~\ref{fig:fRH}, we show $f_{\rm RH}$ for models (i) and (ii). The errors in the models due to statistical fluctuation are given by $N=100$ trials of the MC sampling. The gray points and the shaded region show the observed fraction and its 1$\sigma$ uncertainty associated with the statistical error on the masses, adopted from \citet{2021A&A...647A..50C}. 

The calculation was carried out for the two different sensitivities: NVSS  ($F_{\rm rms}=0.45~{\rm mJy}$, $\theta_{\rm b}=45~{\rm arcsec}$) and GMRT ($F_{\rm rms}=19~\mu{\rm Jy}$, $\theta_{\rm b}=25~{\rm arcsec}$). With the NVSS sensitivity (points with dotted error bars), the fraction $f_{\rm RH}$ is about $10-30\%$, while it becomes $40-70\%$ with the GMRT sensitivity (points with solid error bars). The mean value of $f_{\rm RH}$ at the GMRT sensitivity within LM and HM bins are shown in Table~\ref{tab:fRH_params}. 
Both the models result in $f_{\rm RH}$ close to the observation including small halos and USSRHs, $\langle f_{\rm RH}\rangle=0.44$ \citep[][]{2021A&A...647A..50C}.  \par


The difference in the best fit values for $\xi_{\rm RH}$ between the secondary and primary scenarios (models (i) and (ii)) is caused by the difference in the RH lifetimes. The longer lifetime in the secondary scenario requires less frequent onsets and thus a larger value for $\xi_{\rm RH}$. In this scenario, RHs appears only after major mergers. This condition supports the expectation that only major mergers $\xi\approx0.1$ can cause disturbances over the scale of giant RHs \citep[][]{Cassano_2016}. On the other hand, the short lifetime in the primary scenario requires a much smaller $\xi_{\rm RH}$ for the same $(\alpha_{M},A_{\rm RH})$. Thus, RHs need to be powered by frequent minor mergers. \par 
%


Since the lifetime in model (i) is longer than the timescale of the mass evolution, a RH observed with a certain mass, $M_{\rm obs}$, can have the low radio luminosity corresponding to a smaller mass, $M_{\rm onset}<M_{\rm obs}$. Since the relative margin between the peak luminosity and the observational limit is enough for larger masses (see Figure~\ref{fig:P_M_onset}), more ``aged" RHs (larger $t_{\rm R}$), can contribute to the number count in higher mass bins. In other words, almost all of the RHs in lower mass bins are driven by recent ($t_{\rm R}\lesssim1~{\rm Gyr}$) mergers, while the count in the higher mass bins can include RHs with larger $t_{\rm R}$ ($>{\rm Gyr}$). On the other hand, in model (ii), ages of RHs should be similar in all mass bin. 



\subsubsection{Break-frequency condition \label{subsec:break-frequency_condition}}
In this section, we discuss the {\it break-frequency condition} in the primary model. In the models (iii) and (iv), we observe the competition between the acceleration and the cooling (Section~\ref{subsec:criterion}). These models are basically compatible with model (ii) but described with different parameters, $\nu_{\rm b}$ and $\eta_{\rm t}$. As long as the same luminosity evolution is assumed, mergers with $\xi\gtrsim0.01$ are required for the RH onset to explain the observed $f_{\rm RH}$ and number count.  \par


We adopt a slightly steeper $P_{1.4}-M_{500}$ relation for models (iii) and (iv), as it provides a better fit to the mass dependence of $f_{\rm RH}$ for the primary scenario. Because we have not considered the redshift dependence in the parameters, while $t_{\rm cool}$ decreases with $z$ (Eq~(\ref{eq:tcool})), this model tends to predict a smaller $f_{\rm RH}$ at higher redshifts. That makes it difficult to fit simultaneously both the observed $f_{\rm RH}$ and the RH number count (Section~\ref{subsec:RHLF}), which is measured in lower redshifts. We find that the model is improved by the combination of a larger $\eta_{\rm t}$ and a smaller $A_{\rm RH}$, rather than a smaller $\eta_{\rm t}$ and a large $A_{\rm RH}$. Here, we adopt a slightly smaller $A_{\rm RH}$ compared to model (ii). \par

For $\nu_{\rm b}=1.4~{\rm GHz}$, $\eta_{\rm t}\approx0.6$ is required. More than a half of the kinetic energy need to be dissipated into the compressible turbulence to achieve an efficient acceleration in mergers with $\xi\approx0.01$. A smaller value of $\eta_{\rm t}$ can be allowed for a smaller $\nu_{\rm b}$ (model (iv)). Even with those large $\eta_{\rm t}$, models (iii) and (iv) predict a slightly small $f_{\rm RH}$ compared to the observed value, while the RH number count is well reproduced (Figure~\ref{fig:fRH} right). A better model would be available with an even larger $\eta_{\rm t}$ (smaller $A_{\rm RH}$). Therefore, if a large $\eta_{\rm t}>0.5$ or small $\nu_{\rm b}$ is allowed, the observed RH statistics are reproduced even by minor mergers with $\xi\gtrsim0.01$.

In the right panel of Figure \ref{fig:fRH}, we show the all-sky cumulative number counts of RHs observable with NVSS sensitivity and compare them with the number count in the NVSS follow-up \citep[][]{Giovannini_1999} of the XBACs sample \citep[][]{Ebeling_1996}. In this figure, we take into account not only the observational limit explained in Section~\ref{subsec:obs_limit} but also the X-ray flux limit in the XBACs sample ($f_{X}>5.0\times10^{12}$~[erg/cm$^2$/s]), where we have used the scaling relation reported in \citet{Yuan_2015ApJ...813...77Y} to translate the radio power at 1.4 GHz $P_{1.4}$ into the cluster X-ray luminosity $L_X$. The parameters listed in Table~\ref{tab:fRH_params} agree with the observed count (data points are adopted from \citet{Cassano_2012}), ensuring the compatibility between the observations and our models. In Section~\ref{sec:ASKAP}, we discuss the number count in future high-sensitivity survey.
\par

\subsection{Radio halo luminosity function\label{subsec:RHLF}}

\begin{figure*}
    \centering
    \plottwo{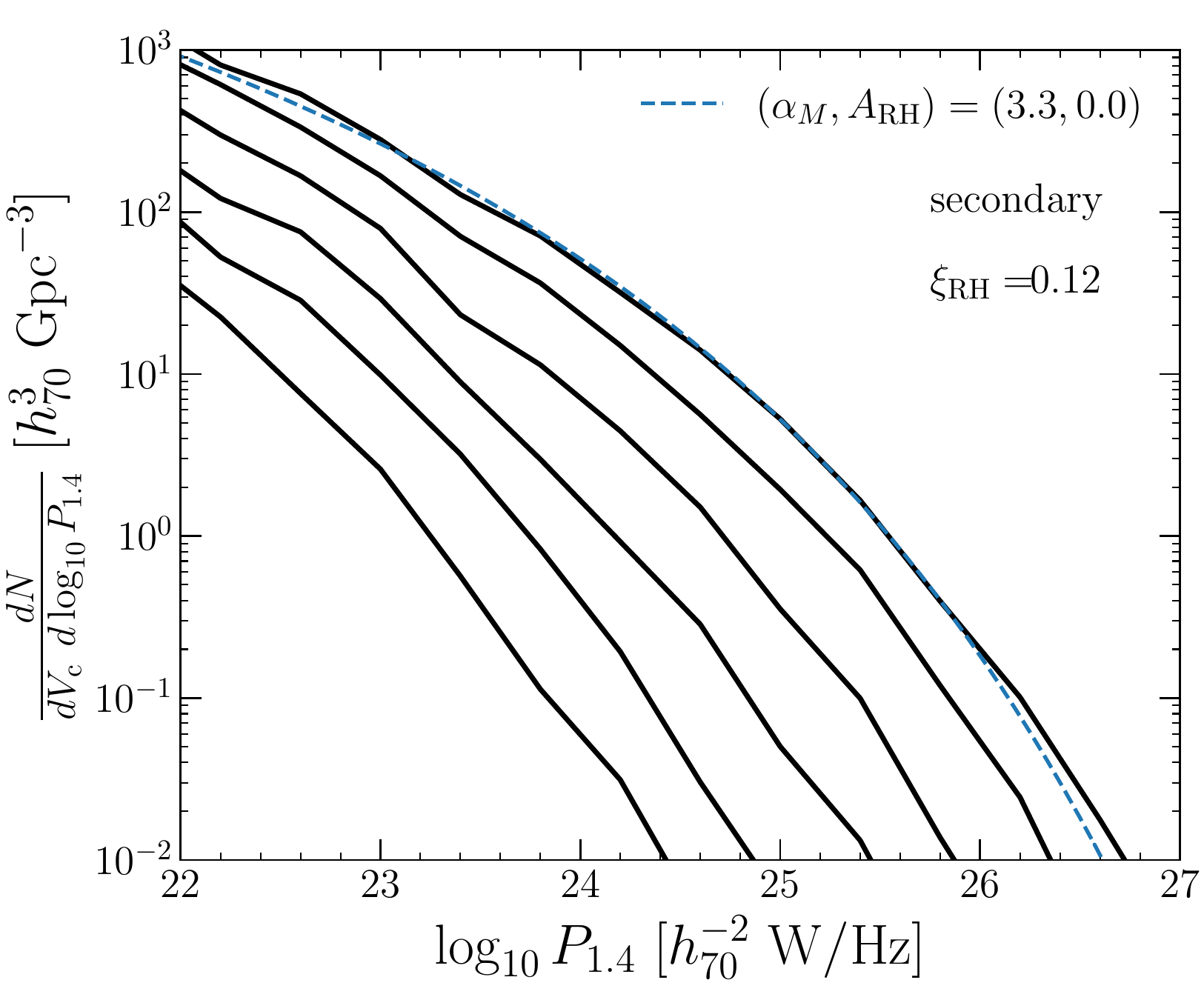}{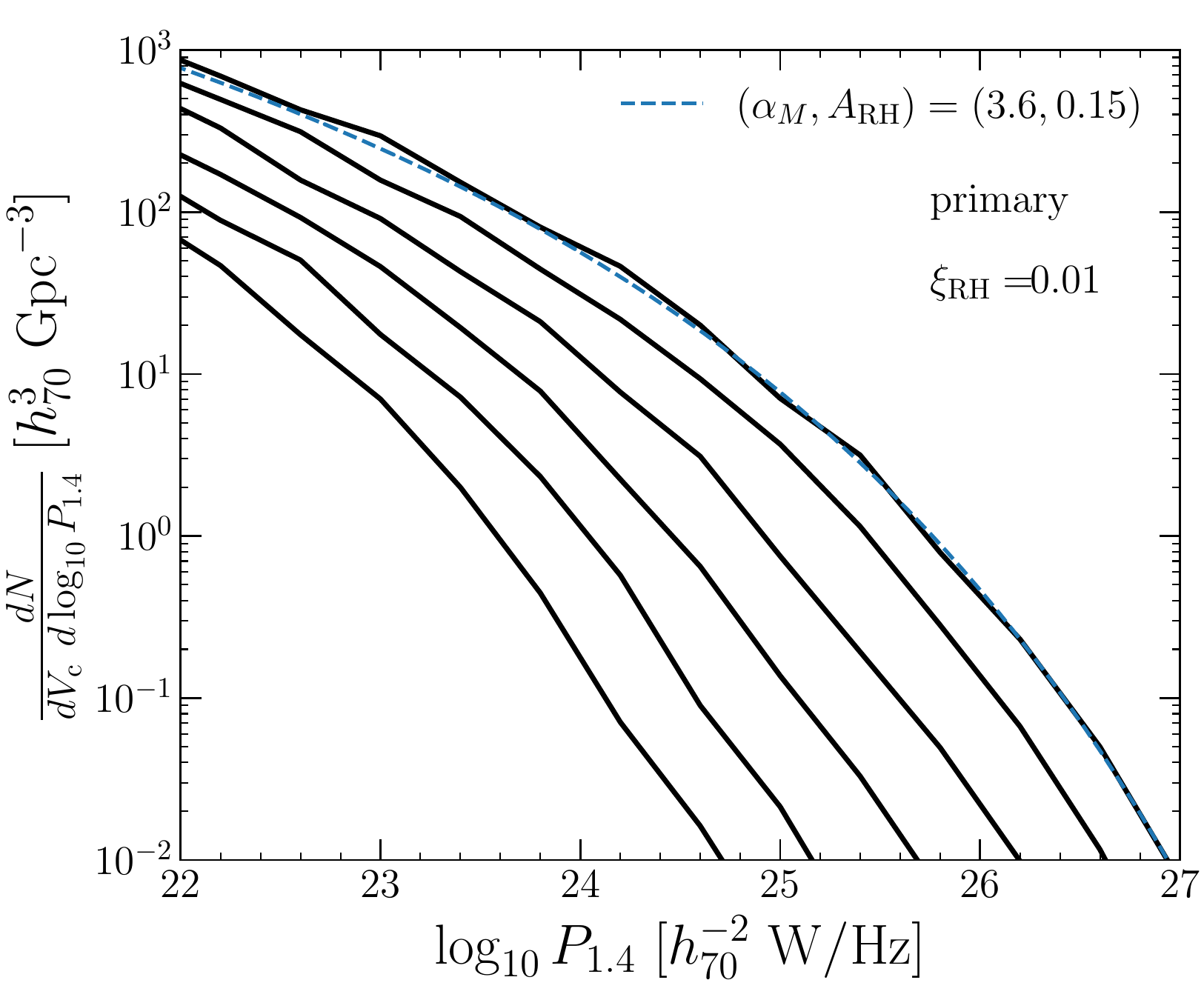}
    \caption{RHLF at different redshifts ($0\le z\le1.0$). The left and right panels show the results for the secondary and primary scenarios, respectively.} From top to bottom, results for $z=0.0,0.2,0.4,0.6,0.8$, and 1.0 are shown. Dashed lines are calculated with the HMF and the power-law scaling between $P_{1.4}$ and $M_{500}$, assuming the mean occurrence of $f_{\rm RH}=0.3$. 
    \label{fig:RHLF}
\end{figure*}

Figure \ref{fig:RHLF} shows total RHLF calculated in our model at different redshifts by directly counting the RH number in our MC merger tree. Any observational limits are not included in this figure. Because we do not impose any conditions concerning the mass for the RH onset, a large number of RHs with low luminosities are predicted. Both the secondary and primary models predict similar redshift evolution. The RHLF at $z=0$ is well approximated with that calculated from the HMF and the typical $P_{1.4}-M_{500}$ relation obtained from observations \citep[e.g.,][]{Cassano_2013ApJ...777..141C}, assuming mean occurrence of $f_{\rm RH}=0.3$ (dashed line) \citep[see also][]{Ensslin_2002}. The best fit value of $A_{\rm RH}$ for the RHLF, $A_{\rm RH}\approx0.0-0.1$, becomes smaller than those assumed for reproducing the peak luminosity in Table.~\ref{tab:fRH_params}, and it becomes similar to the fit to the observation \citep[e.g.,][]{2021A&A...647A..50C}. \par





\subsection{Caveats\label{subsec:caveat}}
We have assumed that all RHs follow the same luminosity evolution described with Eqs.~(\ref{eq:P14_sec}) and (\ref{eq:P14_pri}), regardless of $\xi$. The apparent difficulty in the primary scenario, such as the requirement of a small value for $\xi_{\rm RH}$ or relatively large value for $\eta_{\rm t}$, may be due to this simplification.  \par

In some numerical simulations of galaxy clusters, the turbulence is generated over a duration comparable to the crossing timescale of merger events \citep[e.g.,][]{Miniati_2014ApJ...782...21M,Vazza_2018MNRAS.481L.120V}. In such cases, the emission could be sustained over a longer duration than the lifetime we assumed in Section~\ref{subsec:radio_power}.  \par

Assuming a longer lifetime of RHs, one can obtain a larger $\xi_{\rm RH}$. To clarify this point, we have tested the primary scenario with a different assumption about its luminosity evolution. We introduce a decay time $t_{\rm decay}$ instead of using $t_{\rm cool}(z)$ in Eq.~(\ref{eq:P14_pri}). When $t_{\rm decay}>t_{\rm cool}(z)$, the decline of the luminosity after the peak becomes modest and the lifetime is effectively extended. Such a slow decline could be reproduced with a moderate acceleration efficiency, which is smaller than that we assumed in Section~\ref{subsec:Coma}. We find that the threshold mass ratio becomes $\xi_{\rm RH}=0.1$ for $t_{\rm decay}=1~{\rm Gyr}$, and $\xi_{\rm RH}=0.15$ for $t_{\rm decay}=2~{\rm Gyr}$, which is consistent with the estimate in \citet[][]{Cassano_2016}. Thus, the apparent difficulty in the primary scenario can be alleviated if the emission is sustained over $\sim$1 Gyr. \par
 Assuming that $t_{\rm decay}$ is comparable to the cascade timescale of the IK turbulence, we find that $\eta_{\rm t}\leq 0.1$ leads to $t_{\rm decay}>1~{\rm Gyr}$ for the major ($\xi\approx0.2$) merger of massive ($M_{500}\approx10^{15}~M_\odot$) clusters (Appendix~\ref{app:Dpp}).\par

We have simplified the luminosity evolution in a sequence of mergers, as we neglect the effect of mergers before the onset. However, the turbulent energy could be accumulated though the evolution of clusters \citep[e.g.,][]{Cassano_2005}. The time between mergers can be shorter than the cascade timescale of turbulence for minor mergers with $\xi<0.01$. Our method would underestimate the lifetime and the luminosity of RHs initiated by such mergers. \par

\section{correlation with merging systems \label{sec:X_dynamics}}
RHs are preferentially found in clusters showing the signatures of merger activities \citep[e.g.,][]{Schuecker_2001A&A...378..408S,Govoni_2004}. In this section, we quantify the fraction of RHs classified as a merging cluster in observations using our merger tree. We employ the same method used for the calculation of $f_{\rm RH}$ with the {\it mass-ratio condition} (Section~\ref{subsec:mass-ratio_condition}). \par

The RH lifetime in our secondary scenario is comparable to the loss timescale of CRPs, so it can exceed the dynamical timescale of cluster mergers. Thus, the secondary scenario predicts a fraction of RHs hosted in relaxed clusters. This scenario should be tested in the light of the observed correlation between the occurrence of RHs and the dynamical disturbance in X-ray morphology. \par

According to \citet{Cassano_2013ApJ...777..141C}, the fraction of all merging clusters, including both RHs and non-RHs, is about 60\% in both LM and HM bins. Firstly, we try to recover this fraction in our merger tree, using a similar method to that in Section~\ref{sec:Statis} \citep[see also][]{Cassano_2016}. We newly introduce two parameters, $\xi_{\rm mer}$ and $t_{\rm relax}$. The former one, $\xi_{\rm mer}$, is the threshold mass ratio, above which such a merger makes a descendant cluster disturbed enough to be classified as a merging system in the X-ray morphological analysis. The parameter $t_{\rm relax}$ is the timescale required for a merger system to recover a relaxed state, or the ``lifetime" of the merging system. \par

We do not model the detailed merger dynamics with parameters to diagnose if clusters are merging systems, such as the concentration parameter, $c$, the centroid shift, $w$, and the power ratio, $P_3/P_0$ \citep[e.g.,][]{Mohr_1993ApJ...413..492,Buote_2001,Santos_2008}. Those parameters generally depend on various factors other than the mass ratio, e.g., the projection effect and the impact parameter of the merger. \par


Unlike Section~\ref{sec:Statis}, we fix $\xi_{\rm mer}$ and find $t_{\rm relax}$ that can explain the merging fraction of $\approx60\%$, because the latter is less constrained from observations. As a reference value, we assume $\xi_{\rm mer}=0.1$ for the secondary scenario, which is close to the minimum mass ratio of observed RH-hosting clusters \citep[][]{Cassano_2016}. As in Section~\ref{subsec:occurrence}, we extract sub-sample of halos, which have masses in LM or HM bin within $0.088<z<0.33$. The redshift range is divided into 50 bins as before. A cluster is identified as a merging cluster only when the time interval between the merger and the observation is smaller than $t_{\rm relax}$, i.e., $t_{\rm R}\le t_{\rm relax}$. Under the above condition, we find that $t_{\rm relax}=3.6~{\rm Gyr}$ successfully explain the merging fraction of $\approx60\%$ \citep[see also][]{Cassano_2016}.\par


\begin{figure}
    \centering
    \plotone{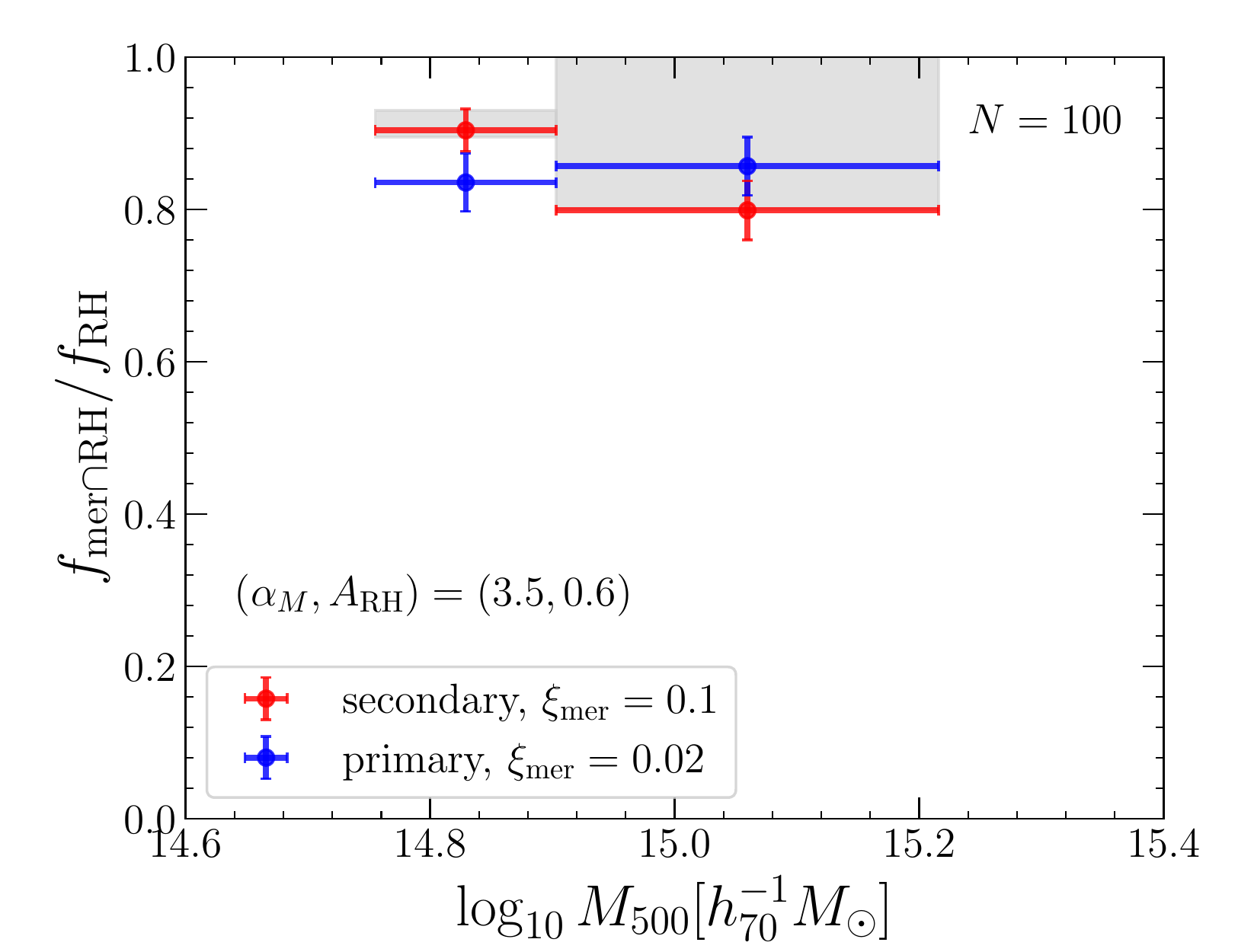}
    \caption{The conditional fraction of RHs in merging systems over the total fraction of RHs. The shaded region indicates the observed fraction in the sample of \citet{2021A&A...647A..50C}.  The flux limit for the GMRT sensitivity is assumed.
    }
    \label{fig:f_merger}
\end{figure}

In the next step, we calculate the fraction of RHs found in merging clusters over all RHs, $f_{\rm mer\cap RH}/f_{\rm RH}$, to test the correlation between the merging state and RHs. In the sample of \citet{2021A&A...647A..51C}, this fraction is 90\%-100\%. We use the same procedure as before to calculate that fraction in our merger tree; $f_{\rm mer\cap RH}$ is defined as the fraction of clusters that show the radio power larger than the observational limit and also satisfy $t_{\rm R}<t_{\rm relax}$ at the observed epoch.\par

Figure \ref{fig:f_merger} shows the results for models (i) and (ii). We find that the fraction is compatible with the observation. Thus, most of RHs in our merger tree should show disturbed morphology at the observed epoch and are classified as merging systems. In other words, the time interval between the onset and the observation, $t_{\rm R}$ (Eq.~(\ref{eq:P14_sec})), is typically shorter than $t_{\rm relax}$. \par



A similar discussion can be applicable for the primary scenario as well. The tight correlation between RHs and merger systems requires that $\xi_{\rm mer}$ should be similar to or smaller than $\xi_{\rm RH}$. Although $\xi_{\rm RH}$ can be lager with longer lifetime (Section~\ref{subsec:caveats}), $\xi_{\rm mer}\approx0.01$ is required in our fiducial primary scenario. 


We adopt $\xi_{\rm mer}=0.02$ for the primary scenario and find that $t_{\rm relax}=1.7~{\rm Gyr}$ leads to the merger fraction of $60\%$. With those $\xi_{\rm mer}$ and $t_{\rm relax}$, $f_{\rm mer\cap RH}/f_{\rm RH}\gtrsim0.8$ is well reproduced in this scenario (Figure~\ref{fig:f_merger}). However, that requirement for $\xi_{\rm mer}\simeq 0.01$ seems problematic, since the mass ratio estimated from optical or near infra-red observations is typically $\xi\gtrsim0.1$ for known merging clusters \citep[][]{Cassano_2016}.  \par

The above difficulty in the primary scenario may be due to our simplified treatment of the turbulent reacceleration. As discussed in Section~\ref{subsec:break-frequency_condition} this difficulty can be alleviated if the emission is sustained for a longer duration. \par




\section{Peak luminosity-mass relation \label{sec:origin_of_PM}}
In Section~\ref{subsec:radio_power}, we have introduced a simple power-law relation between the peak luminosity and the mass of RHs. In this section, we verify if such a steep relation could arise from the mass dependence of the reacceleration efficiency. Some previous attempts have succeeded in reproducing the relations similar to the observed ones \citep[e.g.,][]{Cassano_2007MNRAS.378.1565C,Zandanel2014b}. However, the relation between the steep mass dependence and the reacceleration parameters, such as $f_{\rm ep}$, $t_{\rm acc}$, or $t_{\rm R}$, has not yet been clarified. We examine this point, extending the model used in Section~\ref{sec:CR} for RHs with various masses. \par

In this section, we use a representative value for the acceleration timescale, $\tau_{\rm acc}$, which is the volume average of $t_{\rm acc}(r)=p^2/(4D_{pp}(r))$ within $r<0.5R_{500}$, where most of the radio emission is produced. As seen in \citet{paperI}, our turbulent reacceleration model is more sensitive to $t_{\rm R}$ or $\tau_{\rm acc}$ than the duration of the injection phase, and the energy injection from the reacceleration is at least an order of magnitude larger than that from the injection. The steep mass-dependence of the peak luminosity could be due to the steep dependence on the ratio $t_{\rm R}/\tau_{\rm acc}$. \par



\begin{figure}
    \centering
    \plotone{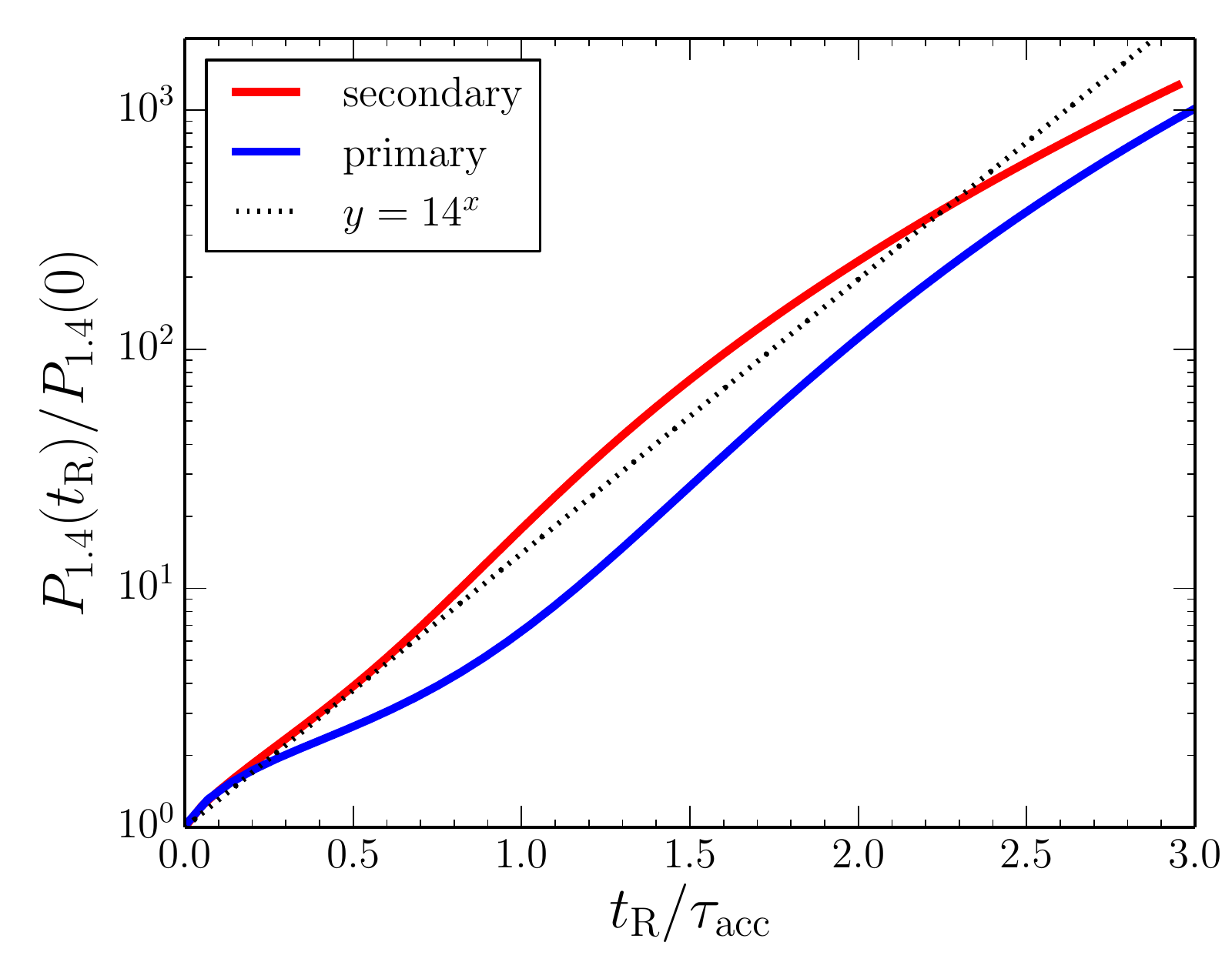}
    \caption{Time evolution of $P_{1.4}$ in the reacceleration phase. We adopt the same Coma-like model (Section~\ref{subsec:Coma}), and the radio power is shown as a function of the duration of the reacceleration phase, $t_{\rm R}$. The duration is normalized with the acceleration timescale of $\tau_{\rm acc}=253~{\rm Myr}$ and $231~{\rm Myr}$ for the secondary and the primary models, respectively. Those are the mean value within $r<0.5R_{500}$ for the Coma-like RH. }
    \label{fig:P14_t}
\end{figure}
\subsection{Luminosity evolution in the reacceleration phase}
First, we show the exponential increase of the radio luminosity in the reacceleration phase. We examine the Coma-like case as an example. We use the same model adopted in Section~\ref{subsec:Coma} with the same model parameters. The typical reacceleration timescale becomes $\tau_{\rm acc}=253~{\rm Myr}$ and $231~{\rm Myr}$ for the secondary and the primary scenarios, respectively.  \par


In Appendix~\ref{app:Dpp}, we roughly estimate the timescales, $\tau_{\rm acc}$ and $t_{\rm R}^{\rm peak}$ with the simple binary merger approximation. There, we introduce a parameter $\eta_{\rm t}$, which quantifies the fraction of the turbulent energy to the merger kinetic energy. A merger is parameterized with the mass ratio $\xi$ and the mass $M_{500}$. We find that the reacceleration with $\tau_{\rm acc}\approx 250~{\rm Myr}$ could be induced by a merger with $\xi\approx0.1$ and $M_{500}\approx10^{15}M_\odot$ when $\eta_{\rm t}\approx0.15$. The duration of $t_{\rm R}^{\rm peak}\approx500~{\rm Myr}$ assumed in Section~\ref{sec:lifetime} might correspond to the turbulent decay time at the injection scale but multiplied by a factor of $\approx$2. \par

In Figure~\ref{fig:P14_t}, we show the time evolution of $P_{1.4}$ during the reacceleration phase. The luminosity is normalized with the value at the beginning of the reacceleration phase, while $t_{\rm R}$ is normalized with $\tau_{\rm acc}$. The exponential increase of the luminosity can be seen in both the scenarios. The evolution in the secondary scenario can be approximated as $P_{1.4}\propto14^{x}$, where $x=t_{\rm R}/\tau_{\rm acc}$.

\begin{figure*}
    \centering
    \plottwo{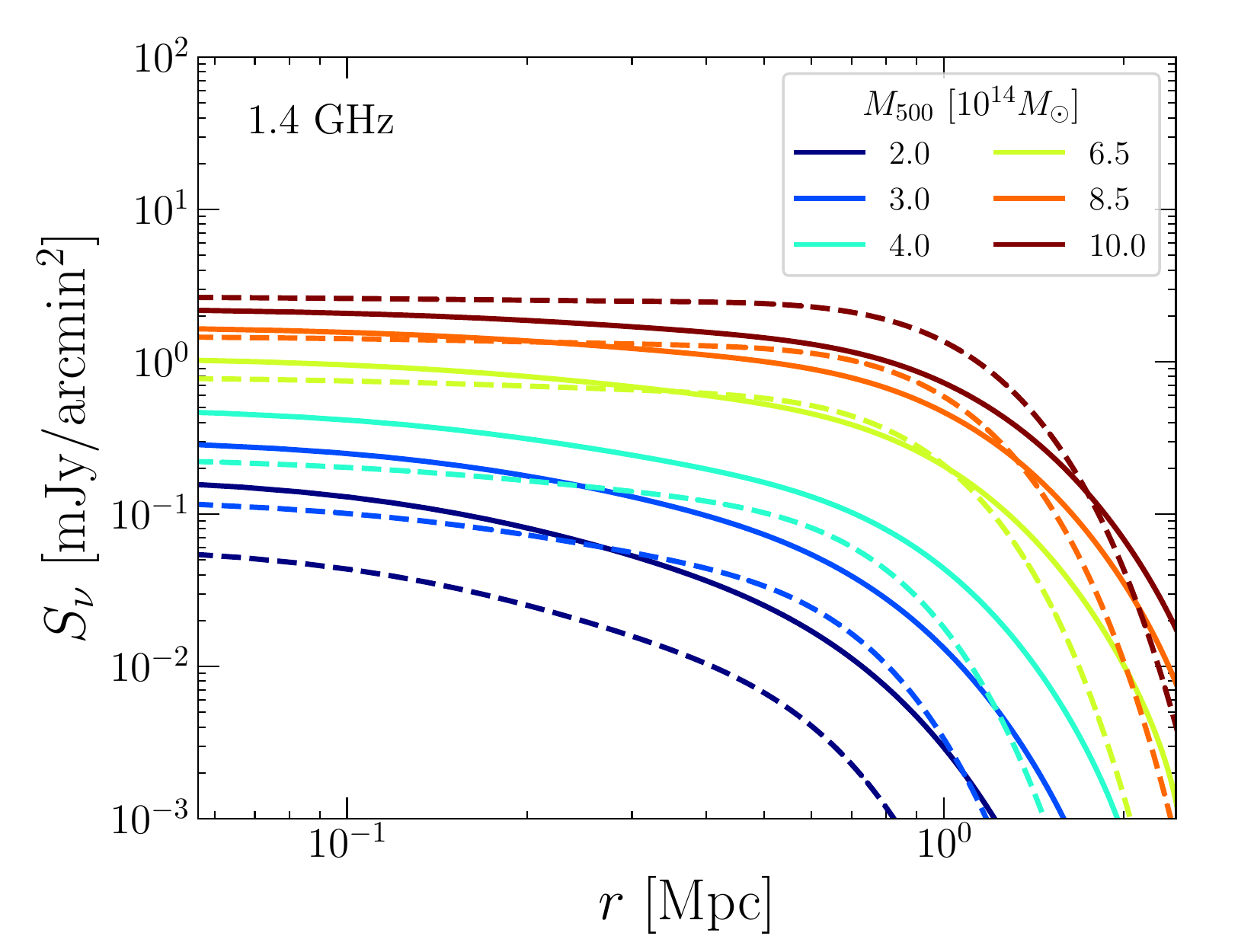}{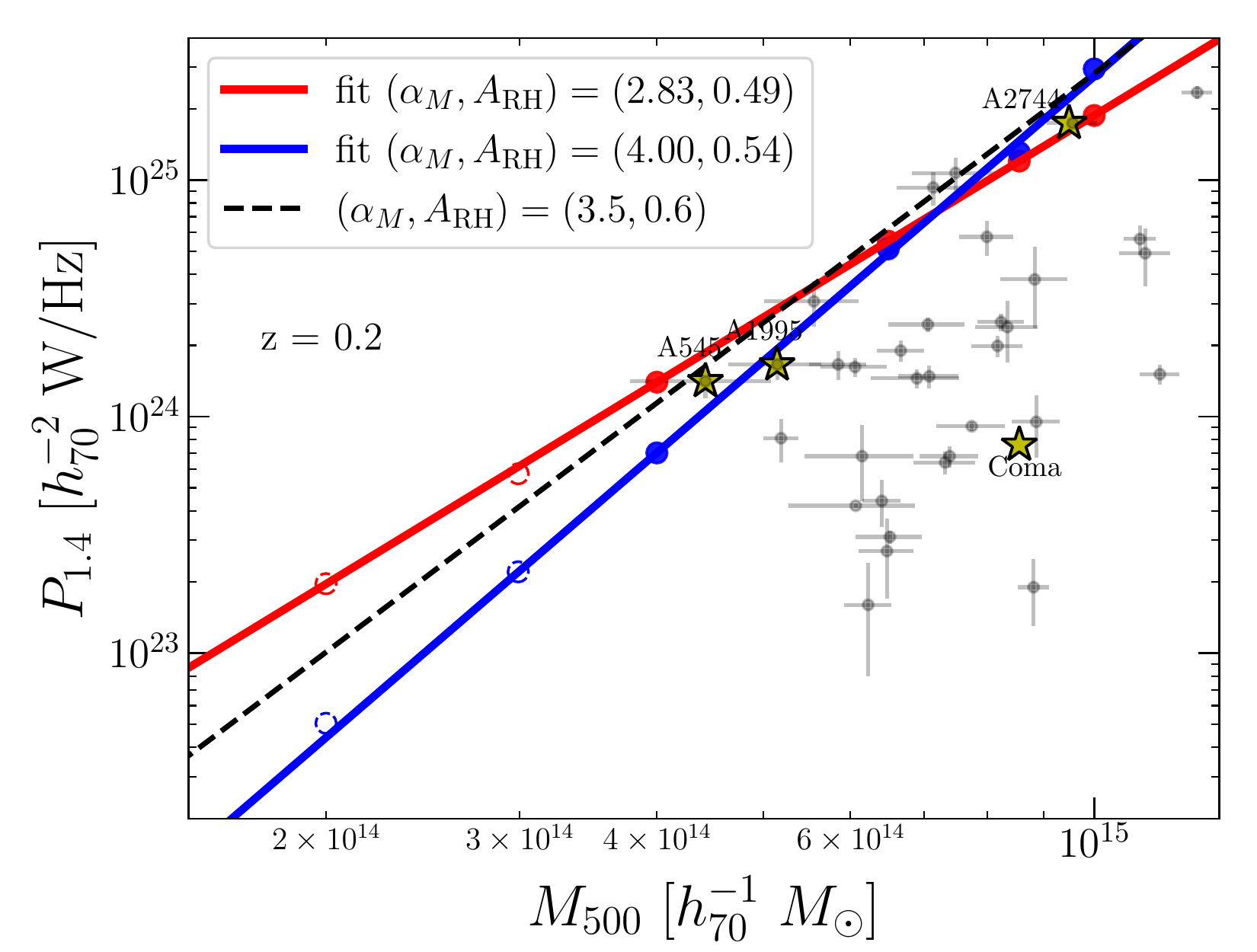}
    \caption{1.4 GHz brightness profile for various masses at the peak time (left) and the mass dependence of the peak luminosity (right) in our model calculations. In the left panel, we show RH profiles with masses of $M_{500} = 2.0,3.0,4.0,5.0,6.5,8.5, 10.0\times 10^{14}~M_\odot$. The results for different masses are distinguished by colors. The secondary and the primary scenarios are shown with solid and dashed lines, respectively. In the right panel, the red and blue circles show the calculated peak luminosity for the secondary and the primary scenarios, respectively. The clusters represented with open circles have masses smaller than the minimum mass in \citet{2021A&A...647A..50C}. We also plot the observed luminosities with thin black data points. Some of the RHs are highlighted with star symbols. We adopt $t_{\rm R}^{\rm peak} = 690~{\rm Myr}$ and $700~{\rm Myr}$ for the secondary and the primary scenarios, respectively. The solid lines show the fits to the red and blue points, while the dashed line is the relation assumed in Section~\ref{subsec:occurrence}.}
    \label{fig:P_M_fit}
\end{figure*}

As discussed in Section~\ref{sec:lifetime}, the synchrotron at high frequencies is mostly powered by the enhanced injection from reaccelerated CRPs. That discussion seems in line with the evolution seen in Figure~\ref{fig:P14_t}, as explained in the following. Since CRPs do not suffer from significant coolings, the hard-sphere type reacceleration causes the advection of the spectrum in the momentum space. At $t_{\rm R}=\tau_{\rm acc}$, the spectrum shifts to a higher energy by a factor of $e\approx2.718$. As we have assumed $\alpha_{\rm inj} = 2.45$ for the typical injection index, the production rate of secondary CREs, which is proportional to $N_p$, is increased by $e^{2.45}\approx11.59$,
slightly lower than 14. The slight deviation from this estimate would come from the spectral evolution of CREs or the definition of the typical acceleration timescale $\tau_{\rm acc}$ (Note that $D_{pp}$ depends on the radius). \par



The luminosity evolution in the primary scenario (blue line in Figure~\ref{fig:P14_t}) shows more complex features. The overall trend is slightly shallower than the secondary scenario, possibly because of the radiative cooling. It is characterised by the steepening around $t_{\rm R}/\tau_{\rm acc}\approx1$ and subsequent flattening around $2t_{\rm R}/\tau_{\rm acc}$. Those features should be related to the spectral evolution of CREs. As seen in \citet{paperI}, the CRE spectrum has a bump like-shape caused by the Coulomb cooling and the radiative cooling. The reacceleration shifts that bump towards higher energies. \par



\subsection{peak luminosity for various masses}
We calculate the peak luminosity of RHs with various masses, extending the model in Section~\ref{sec:lifetime}, which was adopted for the Coma RH (Section~\ref{subsec:Coma}). For this purpose, we introduce a scaling between the reacceleration parameter and the mass. As shown in Appendix~\ref{app:Dpp}, under the simple binary merger approximation with zero impact parameter and the IK scaling for the compressible turbulence, one can find $\tau_{\rm acc}\propto M_{500}^{-1/3}$. Thus, the reacceleration could be more efficient for larger masses. On the other hand, we assume that $t_{\rm R}^{\rm peak}$ is independent of the mass for simplicity. This may be justified if the duration is regulated by the turbulent cascade timescale (see Appendix~\ref{app:Dpp}). The values of $\tau_{\rm acc}$ and $t_{\rm R}^{\rm peak}$ will be normalized at the Coma mass, $M_{500}\approx8.5\times10^{14}~M_\odot$. \par

We neglect the $\xi$ dependence of the acceleration efficiency and the duration, although the acceleration efficiency is more sensitive to $\xi$ than $M_{500}$ (Figure~\ref{fig:Dpp} in Appendix~\ref{app:Dpp}). In other words, all RHs are assumed to be generated through the mergers with the threshold value $\xi\simeq\xi_{\rm RH}$, regardless of the mass. As noted in Section~\ref{subsec:radio_power}, the steep $\xi$ dependence of the accretion rate ensures that events with $\xi\gg\xi_{\rm RH}$ are negligible. More precisely, the $\xi$ dependence would cause a scatter in the peak luminosity-mass relation. The parameters we adopt here can be interpreted as the merger-rate weighted means for $\xi_{\rm RH}<\xi<1$.  \par


In addition to the above assumptions, we introduce following assumptions to make our model applicable to RHs with various masses:  
\begin{itemize}
    \item We adopt the beta-model of the ICM profile, where the core radius is scaled with the virial radius, $r_{\rm c}\approx0.2R_{500}$, and $\beta=0.75$ is taken to be constant. Under the assumption of a constant baryon fraction for all masses \citep[see, e.g.,][for the observational constraints]{Allen_2008}, this leads to a constant central density of $n_0\approx3\times10^{-3}~{\rm cm}^{-3}$. 
    \item We adopt the same Eq. (\ref{eq:B}), $B_0$, and $\eta_B$ for the profile of the magnetic field, without any dependence on mass or redshift.
    \item We adopt Eq. (\ref{eq:Q(r)}) with $\beta=0.75$ and $r_{\rm c}\approx0.2R_{500}$, and $\alpha_{\rm inj} = 2.45$ for the primary CR injection. The volumetric CR injection rate at the center ($r=0$) is scaled with the thermal energy density norimalized by the model for the Coma cluster.
   
    \item We adopt the same index for the turbulent profile, $\alpha_{\rm turb}$, as Coma.
    \item The ICM temperature follows the observed mass-temperature relation \citep[][]{Vikhlinin_2006ApJ...640..691}:
    \begin{eqnarray}\label{eq:TM}
    \left(\frac{T}{5~{\rm keV}}\right)=\left(\frac{E(z)M}{M_5}\right)^{k_T},
    \end{eqnarray}
    where $E(z)=H(z)/H_0$, $M_5 = 3.3\times10^{14}~h^{-1}M_\odot$ and $k_T = 1/1.47\approx0.68$.
    \item All clusters are observed at redshift $z=0.2$, which roughly corresponds to the mean redshift of the sample of \citet{2021A&A...647A..50C}.
    \item The duration of the injection phase is taken to be the time gap between $z=0.2$ and the epoch when the cluster mass was a half of that at the observation calculated with Eq.~(\ref{eq:medianMAR}).
\end{itemize}

The observations indicate that $\sim\mu{\rm G}$ magnetic field is ubiquitous in nearby merging clusters, while some cool-core clusters could have a central field of a few $\sim10\mu{\rm G}$ \citep[e.g.,][for review]{review_vanWeeren}. Recent low-frequency RH observation suggests that some distant clusters have similar magnetic field to those in nearby clusters \citep[][]{DiGennaro_2021NatAs...5..268D}. \par


With those assumptions, we calculate the peak luminosity of RHs with $M_{500}=2,3,4,5,6.5,8.5,10.0 \times 10^{14}~M_\odot$ at $z=0.2$. The acceleration timescale averaged within $r<0.5R_{500}$, $\tau_{\rm acc}$, is normalized at the Coma mass ($M_{500}=8.5\times10^{14}~M_\odot$) as $\tau_{\rm acc}=253~{\rm Myr}$ and $231~{\rm Myr}$ for the secondary and the primary scenarios, respectively. With the scaling of $\tau_{\rm acc}\propto M_{500}^{-1/3}$, $\tau_{\rm acc}$ spans $237~{\rm Myr}<\tau_{\rm acc}<405~{\rm Myr}$ for the above mass range in the case of the secondary scenario. We use the value of peak luminosity assumed in Section~\ref{subsec:occurrence} to derive the duration required to achieve the peak luminosity, $t_{\rm R}^{\rm peak}$. Adopting $(\alpha_{M},A_{\rm RH})=(3.5,0.6)$, the peak luminosity becomes $P_{1.4}\approx1\times10^{24}~{\rm W/Hz}$ at the Coma mass. We find that $t_{\rm R}^{\rm peak} = 690~{\rm Myr}$ and $t_{\rm R}^{\rm peak}=700~{\rm Myr}$ can reproduce that luminosity at the Coma mass for the secondary and the primary scenarios, respectively. As noted before, $t_{\rm R}^{\rm peak}$ is assumed to be constant with mass. With those assumptions, the peak luminosity increases with mass. \par

In the secondary scenario, the typical value of the average CR energy density in the RH volume ($r\leq0.5R_{500}$), $\varepsilon_{\rm CR}$, becomes $\approx5\%$ of the thermal energy density at the peak state. As discussed in Appendix~\ref{app:Dpp}, the acceleration timescale could be reproduced with $\eta_{\rm t}\approx0.15$, $\xi\approx0.2$ and $M_{500}\approx10^{15}~M_\odot$.  With those values, the energy density of the compressible turbulence induced by a merger  becomes $\approx40\%$ of the thermal one. Thus, $\eta_{\rm CR}\approx12.5\%$ of the turbulent energy should be consumed for the CR acceleration. \par

In the primary scenario ($f_{\rm ep}=0.01$), we find $\varepsilon_{\rm CR}\approx0.003\varepsilon_{\rm th}$ at the peak time. The energy density $\varepsilon_{\rm CR}$ is still dominated by CRPs even after the reacceleration. With the same values of $\eta_{\rm t}$, $\xi$, and $M_{500}$ as those in the secondary scenario above, only $\eta_{\rm CR}\approx0.75\%$ of the turbulent energy should be converted into the CR energy. \par

As seen in Section~\ref{sec:Statis}, however, $\xi_{\rm RH}\approx0.01$  rather than $0.2$ is required to explain the observed fraction of RHs in the primary scenario. When $\xi$ is taken to be $0.01$ as the typical value, the turbulent energy fraction is estimated as $\approx10\%$ of the thermal energy, then $\eta_{\rm CR}\approx3\%$ is required. On the other hand, a higher efficiency of the turbulent excitation ($\eta_{\rm t}\approx 0.4$) is required for this scenario (Appendix~\ref{app:Dpp}). \par


In the above discussion, the energy injection in the form of primary CR injection is not taken into account, since that is negligible compared to the energy injection by the reacceleration \citep[][]{paperI}.  \par


In the left panel of Figure~\ref{fig:P_M_fit}, we show the brightness profiles at the peak time. Since a larger $t_{\rm R}^{\rm peak}/\tau_{\rm acc}$ is assumed for higher masses, the overall brightness increases with mass. Note that all of those RHs are assumed to be at the same redshift, $z=0.2$. The typical size, or the cut-off radius of the profile, also increases with mass. We have assumed that the core radius of the ICM ($r_{\rm c}$) scales with $R_{500}$. The profile of the magnetic field, the CR injection, the turbulence for the reacceleration, and thus the synchrotron brightness also scale with $R_{500}$. The profiles in the primary scenario (dashed lines) are slightly flatter than the secondary scenario (solid lines), because of the flatter turbulent profile assumed in the primary scenario. In addition, the brightness in the primary scenario more steeply depends on the mass than the secondary scenario. \par

In the right panel of Figure~\ref{fig:P_M_fit}, the red and blue points show the peak luminosity at 1.4GHz for the secondary and the primary scenarios, respectively. Those points are fitted with a power-law function $\left(\frac{P_{1.4}}{10^{24.5}~{\rm W/Hz}}\right)=10^{A_{\rm RH}}\left(\frac{M_{500}}{10^{14.9}M_\odot}\right)^{\alpha_{M}}$. When the clusters with $M_{500}<4\times10^{14}M_\odot$ are excluded (only filled points), the best fit values are $(\alpha_M,A_{\rm RH}) = (2.83,0.49)$ (secondary) and $(4.00,0.54)$ (primary). We can reproduce the steep relation with $\alpha_{M}\approx3-4$ for both the scenarios. The slope of the relation reflects the steepness of the luminosity evolution around $t_{\rm R}/\tau_{\rm acc}\approx2$, shown in Figure~\ref{fig:P14_t}. Since the dependence on $t_{\rm R}/\tau_{\rm acc}$ is larger in the primary scenario, the fit result in Figure~\ref{fig:P_M_fit} is also steeper than that in the secondary scenario. 

\section{Gamma-ray limits in the secondary scenario \label{sec:gamma}}

As long as the Coma RH is assumed to be in the midst of the reacceleration phase or in the very early stage of the cooling phase, our model has not strong tension with the gamma-ray limit under the magnetic field constrained from RM (Section~\ref{subsec:Coma}). Recently, several papers on the analysis of Fermi data reported the detection of diffuse gamma-ray emission in the direction of Coma \citep[][]{Xi_2018PhRvD..98f3006X,Abdollahi_2020ApJS..247...33,2021A&A...648A..60A}. As noted by \citet{2021A&A...648A..60A}, the flux is comparable to the expectation in the secondary reacceleration models \citep[e.g.,][]{Brunetti2011}. \par

Although larger gamma-ray fluxes could be expected from brighter RHs, the current Fermi limits on the Coma cluster gives the most stringent constraint on the secondary model. For example, the RH in Abell 2744 is $\sim 23$ times more luminous than the Coma RH, but the expected gamma-ray flux becomes $\sim$100 times smaller due to the large distance to the source. \par

\section{number count in the ASKAP era \label{sec:ASKAP}}
We have focused on the RH population detectable with NVSS sensitivity. The RHLF in Section~\ref{subsec:RHLF} indicates a large number of RHs with low luminosities, which would be detectable with future survey with high sensitivity radio telescopes, such as Square Kilometre Array and its pathfinders. \par

The Australian SKA Pathfinder (ASKAP) is a next generation radio telescope array being built at Murchison Radio-astronomy Observatory in Western Australia \citep[][]{Johnston_2008ExA....22..151J}. It consists of 36 dish antennas, each in 12-m diameter, distributed over with baselines up to 6 km. Its array configuration balances the need for high sensitivity to extended structures with the need for high resolution for continuum projects such as EMU, the ''Evolutionary Map of the Universe \citep[][]{Norris_2011PASA...28..215N}. The short-spacing $uv$-coverage of the ASKAP array has higher sensitivity to extended structures such as cluster RHs. The EMU survey is expected to achieve the sky coverage (75\%) similar to the NVSS survey \citep[][]{Condon_1998AJ....115.1693C} but with 45 times better sensitivity. \par


\begin{figure*}
    \centering
    \plottwo{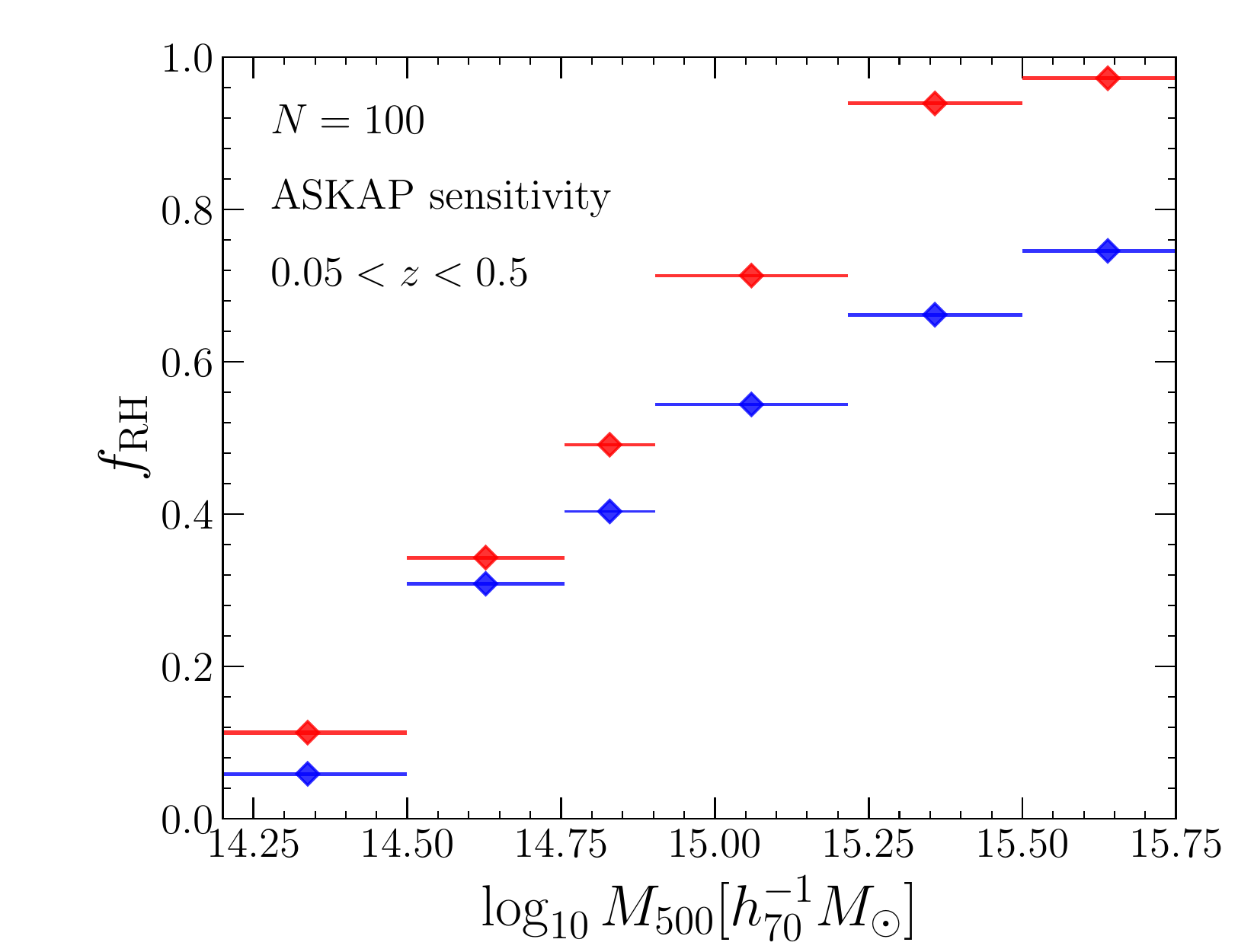}{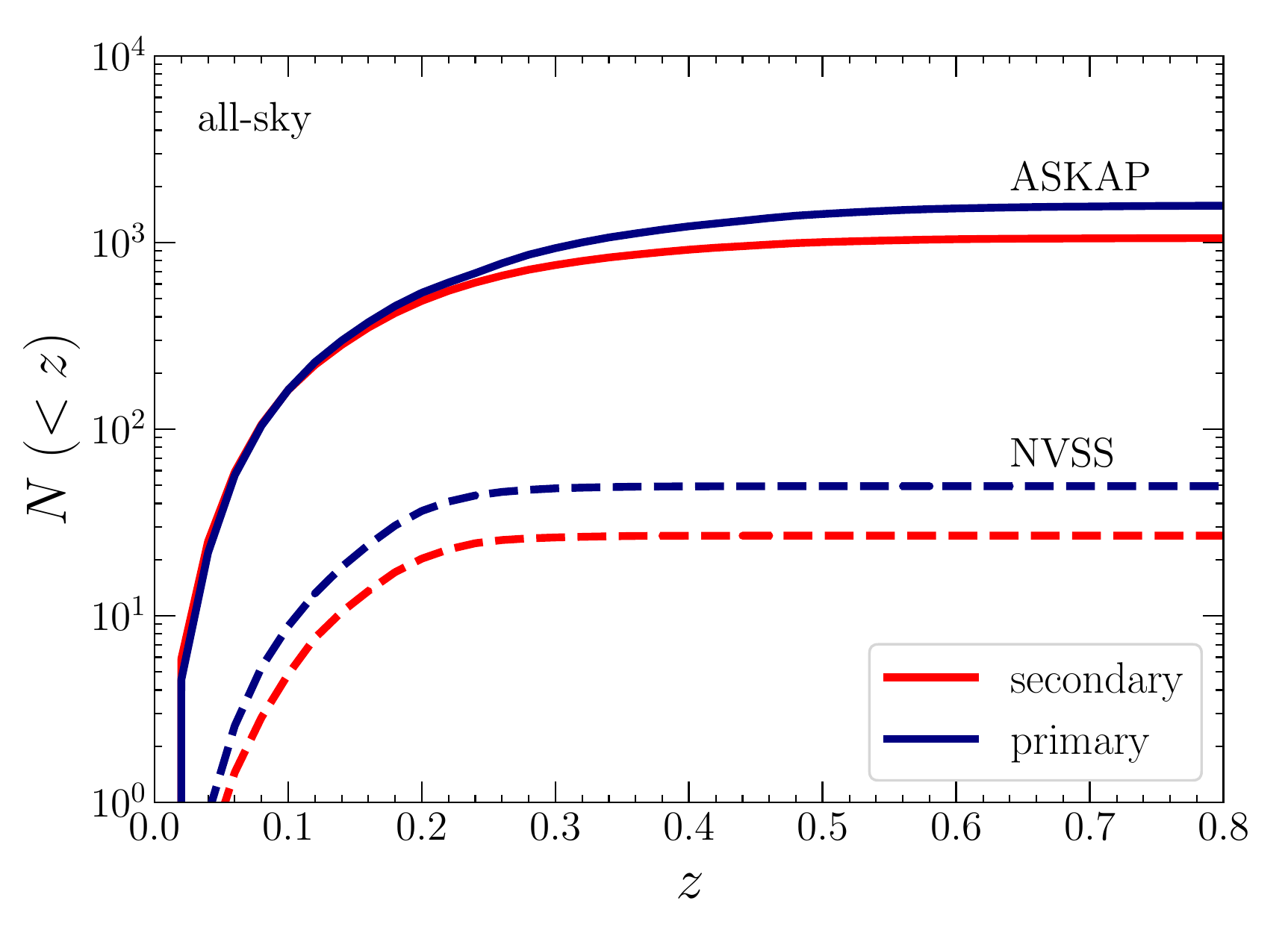}
    \caption{Left: RH fraction with the ASKAP sensitivity ($\theta_{\rm b}= 25~{\rm arcsec}$ and $F_{\rm rms}= 10~{\rm \mu Jy}$) within $0.05<z<0.5$. We adopt $\zeta_2 = 10$ and the parameters listed in Table~\ref{tab:fRH_params} for models (i) and (ii). Right: The all-sky number count of RHs observable with ASKAP (solid lines) and NVSS (dashed lines), accumulated with respect to redshift. In both the panels, the red and blue lines show the results for the secondary and the primary scenarios, respectively.}
    \label{fig:ASKAP}
\end{figure*}

We use the model explained in Section~\ref{sec:Statis} to calculate the RH population observable with the ASKAP sensitivity. Compared to Section~\ref{sec:Statis}, the only modification is the observational limit; we here use the flux-based criterion (Eq.~(\ref{eq:fmin_2})) with $\theta_{\rm b}= 25~{\rm arcsec}$, $F_{\rm rms}= 10~{\rm \mu Jy}$ and $\zeta_2 = 10$, following the previous study \citep[][]{Cassano_2012}. For the criteria for the RH onset, we here only discuss the {\it mass-ratio} condition with the parameters listed in Table~\ref{tab:fRH_params}. The luminosity evolution and its mass dependence are unchanged.\par

In the left panel of Figure~\ref{fig:ASKAP}, we show the mass trend of $f_{\rm RH}$ within $0.05<z<0.5$. The fraction $f_{\rm RH}$ at lower mass bins, where the radio power is relatively small, is larger than the values in Figure~\ref{fig:fRH}. The mass trend seems similar between the secondary (red) and the primary(blue) scenarios. In the right panel of Figure~\ref{fig:ASKAP}, we show the cumulative number count of RHs for models (i) and (ii). Most of RHs are expected to be found in $z<0.5$. The cluster mass corresponding to the minimum luminosity observable with the ASKAP sensitivity is $M_{500}\approx2\times10^{14}~M_\odot$. For $z>0.5$, the cut-off mass of the HMF becomes smaller than the above threshold mass. Thus, the RH detection at $z>0.5$ becomes inefficient. We can expect that the ASKAP survey will detect $10^3$ RHs, while only $\approx30$ RHs are available with NVSS. Since the RHLF (Figure~\ref{fig:RHLF}) is a steep function of the luminosity (Figure~\ref{fig:HMF}), the number count is dominated by the low-power systems ($P_{1.4}\lesssim3\times10^{23}~$ $h_{70}^{-2}~{\rm W/Hz}$). \par



Although the onset conditions are considerably different between models (i) and (ii), $f_{\rm RH}$ and the number count show similar trends. The results look very similar as in Figure~\ref{fig:RHLF}. This indicates that the survey at a single frequency is not enough to determine the CRE origin in RHs. A multi-band analysis ranging from $\approx100~{\rm MHz}$ to $\approx10~{\rm GHz}$ is important for further discussions. Especially at higher frequencies, where the CRE cooling is more efficient and the RH lifetime is expected even shorter in the primary scenario, the difference between those scenarios could be more apparent. In addition to the radio properties, the statistics of the merger mass ratio, which can be constrained from the X-ray and near-IR band observations, must be important, since it differs much between those scenarios (Section~\ref{sec:Statis}).


\section{Conclusions\label{sec:conclusion}}
An increasing number of RHs have been found in recent radio observations of galaxy clusters. The correlation between RHs and dynamically disturbed clusters supports the scenario where the RH emission is powered by the reacceleration of CRs by the merger-induced turbulence, but the origin of the seed CREs for the reacceleration is still poorly constrained. In this study, we compare two possibilities for the origin; the secondary scenario, where the seed population is provided as secondary particles injected via the pp collision, and the primary scenario, where most of the seed CREs are originated from the background thermal electrons and the secondary injection from the pp collision is negligible. \par

We have solved the FP equations to follow the evolution of the CR distribution in RHs. Our calculation method is basically the same as the one in \citet{paperI}, but we have newly included the radial dependence of the reacceleration efficiency in this paper. 
The radio and gamma-ray observations of the Coma cluster is used to constrain the model parameters. Assuming that the primary CRs are injected with the same radial profile as the thermal gas density, we have found that the radially increasing efficiency of the reacceleration reproduces both the radial profile and the spectrum of the Coma RH under the gamma-ray limit given by {\it Fermi}-LAT. \par 

Our model calibrated by the Coma RH is then used to study the lifetime of RHs in Section~\ref{sec:lifetime}. 
We have followed the evolution in the cooling phase, where the reacceleration ceases. 
The main results in this part are summarized as follows:
\begin{itemize}
    \item There are two factors that powers the radio emission: the reacceleration of the seed CREs 
    and the enhanced secondary injection from reaccelerated CRPs. 
    \item In the secondary scenario, a substantial part of the emission is produced by the injection from reaccelerated CRPs. This emission lasts much longer than the cooling timescale of CREs.
    \item In the primary scenario, the RH emission decays within the cooling timescale ($t_{\rm cool}\approx300~{\rm Myr}$).
    
    \item The reacceleration boosts the radio power almost two order of magnitude larger. Thus, we can clearly distinguish radio-loud and radio-quiet clusters.
\end{itemize}



Based on the merger tree provided in Section~\ref{sec:MT}, we have studied the occurrence conditions of RHs to satisfy the observed RH statistics (Section~\ref{sec:Statis}). The most important update from previous studies is the inclusion of the emission lifetime studied in Section~\ref{sec:lifetime}. We introduce the two conditions for the onset of RHs: {\it mass-ratio condition} and {\it break-frequency condition}. The former condition is expressed with the threshold mass ratio, $\xi_{\rm RH}$, while the latter has two parameters, i.e., the break frequency, $\nu_{\rm b}$, and the fraction of the turbulent energy to the merger kinetic energy, $\eta_{\rm t}$.

Our results in Sections~\ref{sec:Statis}-\ref{sec:ASKAP} can be summarized as follows:
\begin{itemize}
    \item The observed fraction and number counts of RHs could be explained in both the two scenarios, but the required threshold value for the merger mass ratio $\xi_{\rm RH}$ differs very much: $\xi_{\rm RH}=0.12$ and $\xi_{\rm RH}=0.01$ are required for the secondary and the primary scenarios, respectively.
    \item The present Coma RH should be in the middle of the luminosity growth due to the reacceleration. The Coma-like RH would achieve its peak luminosity of $P_{1.4}\approx10^{25}~{\rm W/Hz}$ at an elapsed time $\approx700$ Myr after the onset of the reacceleration in {\rm both} the secondary and the primary scenarios.
    \item A correlation between the halo mass and the peak radio luminosity is required. This correlation can be reproduced by the mass dependence of the acceleration timescale ($\tau_{\rm acc}\propto M^{-1/3}$) and a constant duration timescale of the reacceleration. Those parameter behaviors can be justified by a simple model as discussed in Appendix~\ref{app:Dpp}.
    \item In order to reproduce the tight correlation between the RHs and the dynamical disturbance seen in the X-ray morphology, the threshold mass ratio to induce the merger signature in X-ray observation should be close to $\xi_{\rm RH}$. For the primary scenario, this condition requires that even minor mergers with a mass ratio $\xi\approx0.01$ should produce observable global disturbances.
    \item The apparent difficulty in the primary scenario can be relaxed if the emission is  sustained for a few Gyrs by a gentle reacceleration.
    
    \item The RH fraction steeply decreases with decreasing halo mass. However, our model predicts a large number of RHs below the present observable flux limit. The future surveys such as ASKAP are expected to detect $\sim 1000$ RHs below redshift 0.5 for both the secondary and the primary models.
\end{itemize}

In summary, we have found that both the secondary and primary models for  the origins of the seed CREs are allowed from the current RH statistics. Both the models have both merits and demerits. The secondary scenario requires only major mergers for the RH onset, which is consistent with the fact that RH clusters tend to show observable morphological disturbance. However, for the Coma RH, the secondary model requires a steeper injection spectral index compared to the primary scenario (see Appendix \ref{app:ComaAlpha}). That requirement is even severer for USSRHs, and the CRP energy density possibly exceeds that of the thermal ICM for such RHs in the absence of the reacceleration \citep[e.g.,][]{Brunetti_2008Natur.455..944B}. On the other hand, the primary scenario provides a better fit to the Coma spectrum with a flatter injection index and potentially explains USSRHs without requiring an extremely high CR energy density. However, the short lifetime in this scenario requires frequent onsets with $\xi\approx0.01$ mergers, which is one order of magnitude smaller than the typical value estimated in RH clusters \citep[e.g.,][]{Cassano_2016}. The required efficiency for the turbulence excitation is relatively high in the primary model to explain the acceleration time of $\approx300~{\rm Myr}$ within the formulation of Appendix~\ref{app:Dpp}. Either an extended study dedicated for the RH population in lower ($\approx100{\rm MHz}$) and higher frequencies ($\approx10{\rm GHz}$), or a statistical analysis of the merger mass ratio of RHs would further constrain the origin of relativistic electrons in the ICM. \par




\acknowledgments
We thank the anonymous referee for the useful comments that greatly improved the presentation of the paper. K.N. acknowledges the support by the Forefront Physics and Mathematics Program to Drive Transformation (FoPM). This work is supported by the joint
research program of the Institute for Cosmic Ray Research
(ICRR), the University of Tokyo, and KAKENHI
No. 22K03684 (KA). 

\appendix
\section{Secondary scenario with a steeper index}
\label{app:ComaAlpha}
As discussed in Section~\ref{subsec:Coma}, our secondary model shows a slight tension with the observed radio flux of Coma above $1~{\rm GHz}$. In this section, we discuss the possibility that this tension could be relaxed when we adopt a steeper injection index $\alpha_{\rm inj}$.\par

\begin{figure}
    \centering
    \plotone{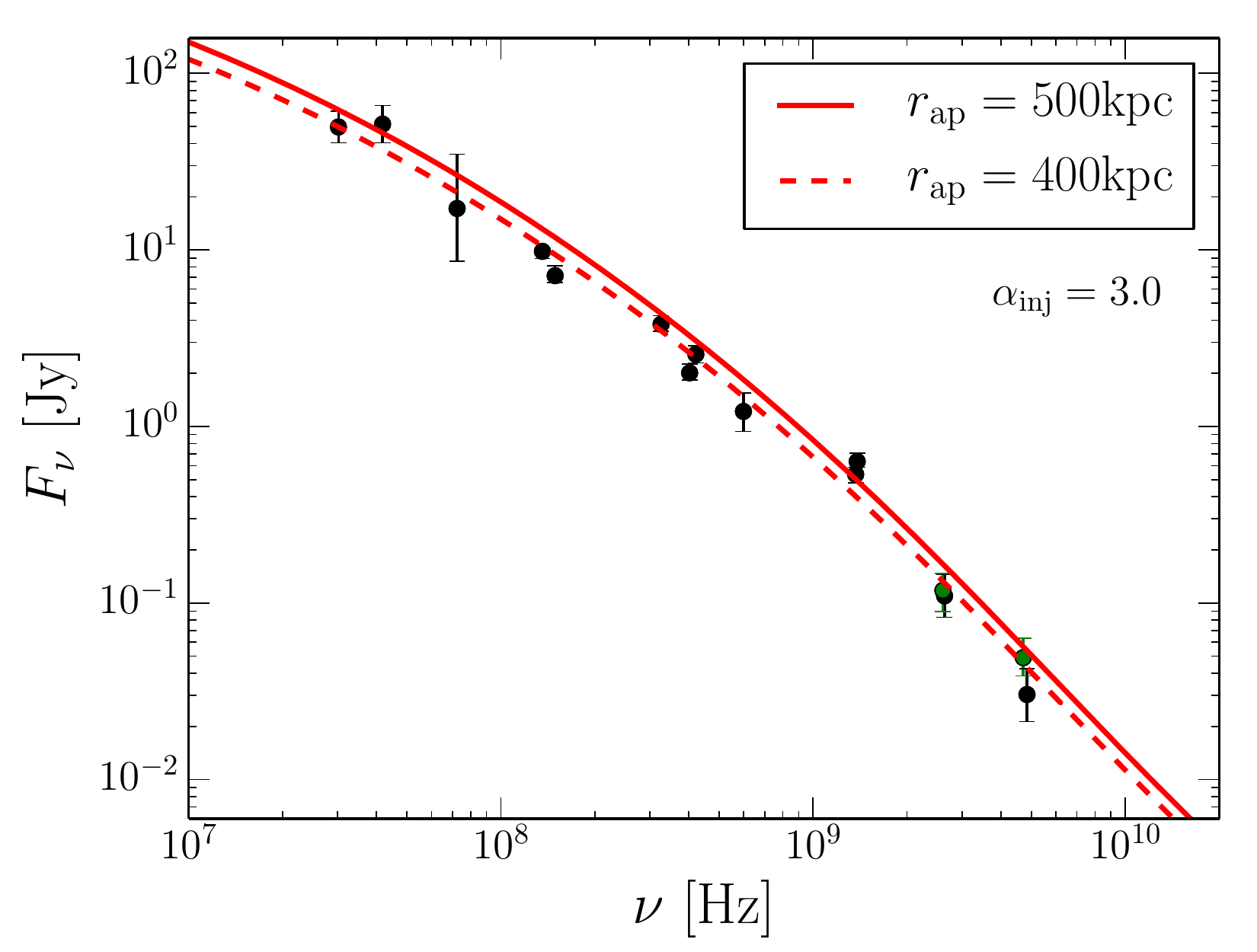}
    \caption{Synchrotron spectra of the Coma RH for the secondary scenario with $\alpha_{\rm inj} =3.0$ (solid: $r_{\rm ap} = 500~{\rm kpc}$, dashed: $r_{\rm ap} = 400~{\rm kpc}$).}
    \label{fig:Coma_a30}
\end{figure}

We have tested the case of $\alpha_{\rm inj} = 2.45$, 2.6, 2.8 and 3.0. Since the spectral shape of the synchrotron emission is mostly determined by $t_{\rm R}/t_{\rm acc}$ \citep[][]{paperI}, the parameter $t_{\rm acc}(r_{\rm c})$ was fixed to $310~{\rm Myr}$. From the gamma-ray upper limit, the lower limit of $t_{\rm R}$ is constrained as $t_{\rm R}>0.8t_{\rm acc}(r_{\rm c})$. For the turbulent profile, we firstly test the case of $\alpha_{\rm turb} = 0.28$, i.e. the same value as that in Section~\ref{subsec:Coma}. However, we find that this profile overproduces the 350 MHz intensity at $r>500~{\rm kpc}$ for $t_{\rm R}>0.8t_{\rm acc}(r_{\rm c})$. A model with a flatter profile of $\alpha_{\rm turb} = 0.35$ is compatible with both the radio profile and the gamma-ray limit. Finally, from the radio spectrum, we constrain $t_{\rm R} = 310$ Myr, .i.e., $t_{\rm R}/t_{\rm acc}(r_{\rm c})=1.0$. Figure~\ref{fig:Coma_a30} shows the synchrotron spectrum of the best fit model with $\alpha_{\rm inj} = 3.0$.
The observed spectrum is well fitted with the injection luminosity of $L_{\rm p}^{\rm inj} = 1.9\times10^{44}$ (solid line). The CR energy density in $r<0.5R_{500}$ becomes $\varepsilon_{\rm CR}/\varepsilon_{\rm th}\approx 0.013$, where $\varepsilon_{\rm th}$ is the thermal energy density averaged within $r<0.5R_{500}$. The fraction increases to $\varepsilon_{\rm CR}/\varepsilon_{\rm th}\approx 0.05$ at the peak state ($t_{\rm R}\approx500~{\rm Myr}$). \par

Throughout this paper, we have assumed $r_{\rm ap} = 500~{\rm kpc}$ for the aperture radius
to calculate the radio spectrum. With this assumption, the overall normalization of the 350 MHz profile becomes consistent with the flux around 350 MHz. In general, the fluxes at other frequencies are taken from different observations and thus originally measured with different aperture radii. We have not taken into account the systematic uncertainty in the re-scaling of those fluxes to $r_{\rm ap} = 500~{\rm kpc}$. \par

By assuming different $r_{\rm ap}$, we test the case when the normalization is not anchored to the flux at 350 MHz. Assuming $r_{\rm ap}=400~{\rm kpc}$, the calculated flux slightly shifts down and the fit to the data points at high frequency is slightly improved (dashed line). 

\section{reacceleration efficiency in the binary merger approximation \label{app:Dpp}}

In this section, we discuss the efficiency of the turbulent reacceleration triggered by a cluster merger. As explained in Section~\ref{sec:Statis}, we adopt the binary merger approximation. The mass of the descendant halo and the merger mass ratio are denoted as $M$ and $\xi=M_2/M_1$, respectively. The impact parameter is assumed to be zero for simplicity. We especially focus on the CR acceleration by TTD with compressible turbulence. As assumed in Section~\ref{subsec:acc_diff}, the interaction between turbulence and particles is assumed to be fully collisionless \citep[][]{Brunetti2011}. 
The discussion below is based on the one-zone approximation, unlike in Section~\ref{subsec:acc_diff}, where we have considered the radial dependence. \par

To quantify the turbulent energy induced by mergers, we introduce a parameter $\eta_{\rm t}$, which denotes the fraction of the turbulence energy dissipated from the merger kinetic energy. With the volume of the turbulent region $V_{\rm t}$, the turbulent energy density is written as $\varepsilon_{\rm t}=\eta_{\rm t}GM_1M_2/R_{\rm mer}/V_{\rm t}$, where $R_{\rm mer}$ is the distance between the centers at the impact. Following \citet{Cassano_2005}, we choose $R_{\rm mer}=R_1$, i.e., the physical radius of the major progenitor. Eq.~(\ref{eq:R_vir}) provides the radius of the cluster. \citet{Cassano_2005} evaluates $V_{\rm t}$ from the comparison between typical radius of the RHs and the stripping radius, where the ram pressure experienced by the minor progenitor becomes comparable to its static ICM pressure. We adopt a more simplified approach, $V_{\rm t}\approx\frac{4\pi}{3}(R_1R_2)^{3/2}$, assuming that $V_{\rm t}$ increases with both $M_1$ and $M_2$. The $\xi$ dependence of $\varepsilon_{\rm t}$ is affected by the choice of $R_{\rm mer}$ and $V_{\rm t}$, while $M$ dependence is not apparently affected, as long as both $R_{\rm mer}$ and $V_{\rm t}$ are expressed as functions of $R_1$ and $R_2$ (or $M_1$ and $M_2$).
\begin{figure*}
    \centering
    \plotone{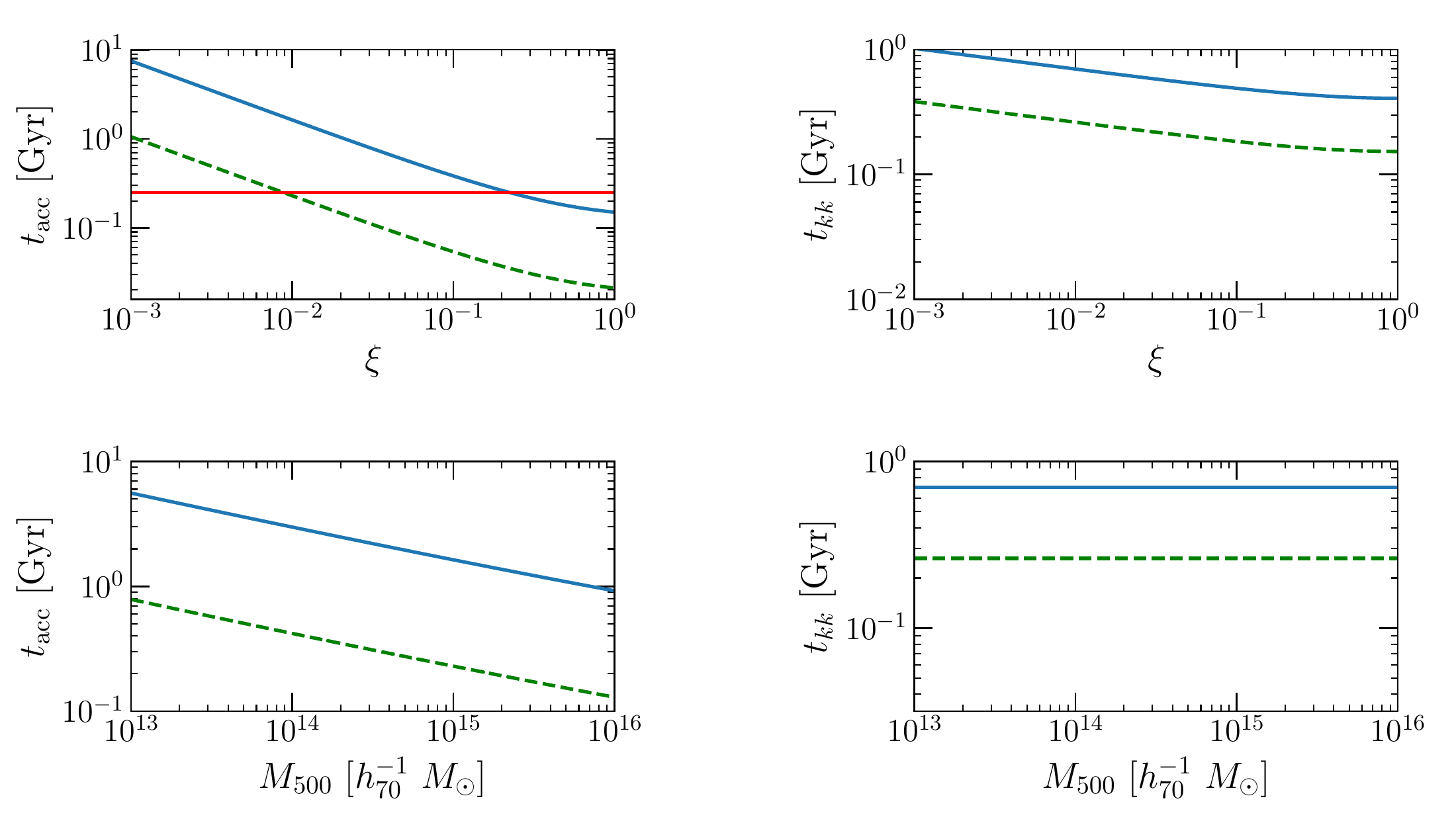}
    \caption{Timescales $t_{\rm acc}$ (left panels) and $t_{kk}$ (right panels) as functions of $M_{500}$ and $\xi$. The blue solid lines are results for $M_{500}=1.0\times10^{15}~M_\odot$ (upper panels) and $\xi=0.1$ (lower panels) with $\eta_{\rm t}=0.15$ and $n=10^{-3}~{\rm cm}^{-3}$. The green dashed lines show the results for $\eta_{\rm t}=0.40$ with the same values for the other parameters. The red horizontal line in the upper left panel shows the typical acceleration time scale for $M_{500}=1.0\times10^{15}~M_\odot$ assumed in Section~\ref{sec:origin_of_PM}.}
    \label{fig:Dpp}
\end{figure*}
We can evaluate the turbulent velocity at the scale of the injection from $\rho V_{L}^2\approx\varepsilon_{\rm t}$ as,
\begin{eqnarray}
V_{L} &\approx& 1.25\times10^3~{\rm km/s}~\frac{\xi^{\frac{1}{4}}}{(1+\xi)^{\frac{1}{3}}}\left(\frac{M}{10^{15}~M_\odot}\right)^{\frac{1}{3}} \nonumber \\
&&\times\left(\frac{\eta_{\rm t}}{0.1}\right)^{1/2}\left(\frac{\Delta}{500}\right)^{\frac{2}{3}}\left(\frac{n}{10^{-3}~{\rm cm}^{-3}}\right)^{-\frac{1}{2}},
\end{eqnarray}
where $\rho$ and $n$ are the mass and number densities of the ICM, respectively. The subscript $L$ denotes the injection scale of the turbulence. After a merger with $\xi=0.2$, the turbulent energy density in the cluster with $M_{500}=1.0\times10^{15}~M_\odot$ becomes $\approx40\%$ of the thermal energy density. This fraction is almost independent of the mass, since the mass dependence of the turbulent energy is similar to that of the temperature. \par
We adopt the IK scaling, and approximate the outer scale of the turbulence as the size of the minor progenitor, i.e. $L\approx R_2$. Since the ICM is a high-beta plasma, the sound speed $c_{\rm s}$ and $k_{\rm cut}$ characterize the timescale of TTD. We simply adopt the observed $M-T$ relation (Eq.~(\ref{eq:TM})) to calculate $c_{\rm s}$ for different masses. Because the observed relation is similar to the expectation from the virial theorem, $T\propto M^{2/3}$, the turbulent Mach number is nearly independent of the mass.  \par


We assume that the wave damping is dominated by the TTD interaction with non-relativistic thermal electrons in a high-beta plasma. In this case, the decay timescale of turbulence at scale $k$ is $\sim m_{\rm p}/(m_{\rm e}kv_{\rm th}^{\rm e})$, where $v_{\rm th}^{\rm e}$ is the thermal velocity of the electrons. The cut-off scale $k_{\rm cut}$ appears where the cascade timescale becomes comparable to the decay timescale. For the IK turbulence, the cascade timescale becomes $t_{kk}\approx\frac{2}{9}\frac{c_{\rm s}}{V_L^2}k_L^{-1/2}k^{-1/2}$, and $k_{\rm cut}$ is approximated as $k_{\rm cut}\approx 1.08\times10^4{\cal M}_{\rm s}^4k_L$, where ${\cal M}_{\rm s}=V_L/c_{\rm s}$ is the Mach number for the turbulent velocity at the injection scale \citep[e.g.,][]{Brunetti2007}.
Since $k_{\rm cut}$ is much larger than $k_{L}$, $D_{pp}$ (Eq.~(\ref{eq:Dpp_integral})) can be written as
\begin{eqnarray}
\frac{D_{pp}}{p^2}&\approx&\frac{\pi}{4c}I_\theta (x)k_{\rm cut}^2{\cal W}(k_{\rm cut}), \\
&\propto& I_\theta (x){\cal M}_{\rm s}^4c^2_{\rm s}k_{L}.
\end{eqnarray}
In the ICM, $x=c_{\rm s}/c\ll1$ so $I_\theta(x)=\frac{x^4}{4}+x^2-(1+2x^2)\ln x-\frac{5}{4}$ has only a log-dependence on $x$. Neglecting this dependence, $D_{pp}$ scales as $D_{pp}\propto \xi^{2/3}(1+\xi)^{-1}M^{1/3}$. The acceleration timescale, $t_{\rm acc}=p^2/4D_{pp}$, is plotted in Figure~\ref{fig:Dpp} (left panels). The mass dependence is slightly weaker than $M^{1/3}$, since the integral $I_\theta(c_{\rm s}/c)$ (Eq.~(\ref{eq:Dpp_integral})) has a weak dependence on $c_{\rm s}$. The typical acceleration timescale assumed in Section~\ref{sec:origin_of_PM}, $t_{\rm acc}\approx 250~{\rm Myr}$, could be reproduced with $\eta_{\rm t}=0.15$ and $\xi=0.2$ (solid line in the upper left panel). To achieve a similar acceleration efficiency with $\xi=0.01$ mergers, the efficiency of turbulent excitation should be as large as $\eta_{\rm t}\approx0.4$ (green dashed line).  \par

In Section~\ref{sec:origin_of_PM}, we have shown that the steep mass dependence in the observed $P_{1.4}-M_{500}$ relation could be explained by the reacceleration with a constant duration. One possible explanation for this model is that the duration is regulated by the timescale of the turbulent cascade, $t_{kk}$, rather than that of the dynamical motion that injects turbulence. The cascade timescale at the injection scale can be expressed as \citep[][]{Brunetti2007},
\begin{eqnarray}
t_{kk} & = & \left. \frac{k^3}{\frac{\partial}{\partial k}(k^2D_{kk})}\right|_{k=k_L}, \nonumber \\
\approx&& 0.45~{\rm Gyr}\left(\frac{L}{0.5~{\rm Mpc}}\right)\left(\frac{V_L}{10^3~{\rm km/s}}\right)^{-1}\left(\frac{M_s}{0.5}\right)^{-1},
\end{eqnarray}
where $D_{kk}$ is the wave-wave diffusion coefficient \citep[e.g.,][]{Brunetti2007}. Adopting the same models for $L$ and $V_L$ as that adopted for $t_{\rm acc}$, one can find the scaling of $t_{kk}\propto \xi^{-1/6}(1+\xi)^{-1}M^{0}$. The duration adopted in Section~\ref{sec:origin_of_PM}, $t_{\rm R}\approx 600~{\rm Myr}$, corresponds to1.5$\times t_{kk}$ for $\xi=0.1$.

\bibliography{Lifetime}{}
\bibliographystyle{aasjournal}

\end{document}